\documentclass[12pt, draftclsnofoot, onecolumn]{IEEEtran}

\normalsize

\ifCLASSINFOpdf

\else

\fi
\usepackage[cmex10]{amsmath}
\usepackage{amssymb}
\usepackage{cite}
\usepackage{graphicx}
\usepackage{array,color}
\usepackage{amsmath,amsthm}
\usepackage{stfloats}
\usepackage{graphicx}
\usepackage{subfigure}
\usepackage{tabularx}
\usepackage{epsfig,epsf,color,balance,cite}
\usepackage{algorithmic}
\usepackage{algorithm}
\usepackage{bm}
\usepackage{textcomp}

\usepackage{epstopdf}
\usepackage{graphicx}
\usepackage{epstopdf}
\usepackage{hyperref}

\newtheorem{lemma}{\textbf{Lemma}}

\begin{document}
\title{Outage Constrained Robust Transmission Design for IRS-aided Secure Communications
with Direct Communication Links}
\author{ Sheng Hong, Cunhua Pan, Gui Zhou, Hong Ren, and Kezhi Wang
\thanks{S. Hong is with Information Engineering School of Nanchang University, Nanchang 330031, China. (email: shenghong@ncu.edu.cn). C. Pan and H. Ren are with the National Mobile
	Communications Research Laboratory, Southeast University, Nanjing
	210096, China. (cpan, hren@seu.edu.cn). G. Zhou are with the Institute for Digital Communications, Friedrich-Alexander-University Erlangen-N\"{u}rnberg (FAU), 91054 Erlangen, Germany (e-mail: gui.zhou@fau.de). K. Wang is with Department of Computer and Information Sciences, Northumbria University, UK. (email: kezhi.wang@northumbria.ac.uk). (Corresponding author: Cunhua Pan)
}
}

\markboth{IEEE Transactions on Communications}%
{Submitted paper}
%

\maketitle
\vspace{-2cm}
\begin{abstract}
This paper considers the outage constrained robust transmission design for an intelligent reflecting surface (IRS) aided secure communication with direct communication links. We assume a scenario of multiple-input-single-output wiretap channels where a legitimate
receiver (Bob) served by a base station (BS) is overheard by multiple eavesdroppers (Eves), meanwhile the artificial noise (AN) is incorporated to confuse the Eves. In particular, we aim to jointly optimize the transmit beamforming, the AN spatial distribution at the BS, and the passive beamforming at the IRS, where the Eve-related channels are only partially available due to the channel estimation errors and their concealment. Two scenarios with partial channel state information (CSI) error of only cascaded BS-IRS-Eve channel and full CSI errors of both cascaded BS-IRS-Eve channel and direct BS-Eve channel are investigated under the statistical CSI error model. A robust power minimization problem constrained by the minimum data rate requirement of Bob and the outage probability of maximum data rate limitation of Eves is investigated. To solve it, the Bernstein-type inequality and alternating optimization strategy are exploited, and artful mathematical manipulations are performed to facilitate the optimization on the phase shifts of the IRS. Simulation results confirm the performance advantages of the proposed algorithm.
\end{abstract}
\begin{IEEEkeywords}
Intelligent Reflecting Surface (IRS), CSI errors, robust transmission design, secrecy, outage probability, direct communication link.
\end{IEEEkeywords}

\IEEEpeerreviewmaketitle

\section{Introduction}
\IEEEPARstart{T}{he} intelligent reflecting surface (IRS) is a promising technique in future six generation (6G) communication networks \cite{wu2019towards,pan2020reconfigurable}. The IRS consists of a large number of reflecting units, each of which can reflect the incident signal passively \cite{cui2014coding}. By properly tuning the phase shifts of reflecting units, the reflected signals can be added constructively or destructively \cite{wu2019towards}. Thus an IRS can intelligently configure the wireless environment to help the transmissions between the sender and the receiver. Since the reflecting unit of the IRS operates in a passive mode without any radio frequency (RF) chain, deploying an IRS costs much less than deploying a relay\cite{di2020reconfigurable}. The IRS can be readily installed with low cost due to its light weight and compact size. Moreover, it can be easily integrated into traditional communication systems with only minor modifications. Therefore, the IRS-aided wireless communications have received extensive research attention in multicell networks \cite{pan2019intelligent2}, mobile edge computing \cite{9133107}, multigroup multicast communication\cite{zhou2020intelligent}, cognitive radio system \cite{zhang2020intelligent}, and wireless power transfer design \cite{pan2020intelligent}.

In view of the great potential, the IRS has recently been exploited to enhance the physical layer security in wireless communications \cite{9014322,chen2019intelligent,8972406}. The effectiveness of the IRS is more prominent in some tough scenarios, where the non-zero secrecy rate is difficult to achieve without the IRS. In \cite{cui2019secure}, an IRS was applied to tackle the challenging scenarios where the channel of the legitimate communication link and that of the eavesdropping link were highly correlated. The authors in \cite{guan2020intelligent} showed that the advantages of joint use of an IRS and artificial noise (AN). A scenario where the eavesdroppers (Eves) are closer to the base station (BS) than the desired users was investigated by employing an IRS in \cite{9201173,chu2019intelligent}. The work in \cite{xu2019resource} verified the performance enhancement by using an IRS when the direct BS-user links were blocked.

However, all the above-mentioned papers assumed that the channel state information (CSI) associated with all involved channels is perfectly known at the BS, which is too idealistic. Since Eves are usually unregistered users and the IRS is passive, the Eves' channel, especially the IRS-related channel cannot be perfect. In IRS-aided systems, the direct channels from BS to users are first estimated by turning off the IRS \cite{wei2021channel}. Then, the IRS is turned on, and the IRS-related channels are estimated. Currently, there are roughly two approaches for estimating the IRS-related channels: 1) separately estimating the channels of BS-IRS link and IRS-user link \cite{taha2019enabling}. The main idea is to install some active channel estimators at the IRS, and then estimate the two kinds of channels individually, and finally send the estimated channels to the BS; 2) directly estimating the cascaded channels \cite{9241029,wang2020channel,wang2019compressed,chen2019channel}, which is the composite channel of the BS-IRS link and IRS-user link, and can be exploited to achieve the optimal beamforming design\cite{zhang2020capacity,zhou2020intelligent}. The second channel estimation (CE) approach is more attractive than the first approach, because no additional active hardware is required, no additional power is consumed, and the channel training and feedback overhead is reduced.

In spite of these two feasible CE approaches for IRS-related channels, the CSI error is still unavoidable. Naturally, such CSI error will heavily deteriorate the system performance. Thus, a few recent works have addressed the robust design problem for IRS-aided communications. The earliest research on robust design for the IRS-aided communication relies on the first CE approach, which was proposed first. A worst-case robust design was investigated in \cite{zhou2020robust} with CSI errors on IRS-user channels in a multiuser MISO communication system. Then, a robust design algorithm in IRS-aided secure communications was proposed in \cite{yu2020robust}, where the IRS-Eve link has bounded CSI errors, and the direct communication links were blocked. Recently, the research on robust design mainly focuses on the the second CE approach, which is more appealing in practice. The robust design based on the second approach was firstly proposed in \cite{zhou2020framework} for IRS-aided MISO communications, where the cascaded BS-IRS-user channels were imperfect with both the bounded and statistical channel errors. Similarly, the robust design for an IRS-aided cognitive radio system was investigated with imperfect cascaded CSI on primary user (PU)-related channels in \cite{zhang2020robust}. For the bounded CSI errors, a worst-case robust transmit power minimization problem was investigated for the IRS-aided green MISO communications based on the second CE method \cite{yu2020irs}, and a robust sum-rate maximization problem was solved for multiuser MISO systmes with self-sustainable IRS \cite{hu_robust_2021}. For the statistical CSI errors, a robust probabilistic-constrained transmit power minimization problem was investigated for IRS-aided MISO communications without direct links \cite{9167248}. The outage-constrained robust beamforming problem was transformed into a outage
probability minimization problem in \cite{zhao2020outage}. We investigated the robust transmission design in IRS-assisted secure communications in \cite{hong2020robust} by assuming that the cascaded BS-IRS-Eve channels have statistic CSI errors, and the direct links are blocked. From above research, we find that most robust designs with statistical CSI errors cannot apply to the case with correlated CSI errors, e.g., the methods in \cite{zhou2020framework,zhang2020robust}. Moreover, the direct communication links were usually assumed to be blocked, e.g., the methods in \cite{9167248,hong2020robust}, which is only applicable for some special scenarios.

In this paper, we investigate an outage constrained robust power minimization (OCR-PM) problem in an IRS-aided secure communication system by jointly optimizing transmit beamformer, the AN spatial distribution, and the phase shifts at the IRS. The Eves' channels are assumed to be partially available, and the formulated problem is subject to the outage constraint of maximum information leakage to Eves. In the robust design, we consider the direct communication links from the BS to all users. An outage constrained robust (OCR) design algorithm catering for both uncorrelated or correlated CSI errors is proposed. The contributions of this work are summarized as follows:
\begin{enumerate}
\item In contrast to existing literature \cite{yu2020robust,hong2020robust} which assumes that the direct BS-users links were blocked, we investigate the robust design in a more general and practical scenario with both direct links and IRS reflecting links. The existence of direct communication link makes the optimization of phase shifts at the IRS much more challenging since almost all constraints become nonconvex. To address this issue,
we propose a series of artful mathematical manipulations to transform these constraints into convex
ones, thus provide a robust design framework for the case with direct link.

\item In the robust design problem, we consider two kinds of scenarios where only the CSI errors exist for cascaded channels (i.e. partial CSI errors) as well as for both the cascaded channels and direct channels (i.e. full CSI errors). To solve the formulated problem, the alternation optimization (AO) strategy is leveraged to decouple the optimization variables. The Bernstein-type inequality (BTI) \cite{wang2014outage} is utilized to tackle the probability constraints. The penalty convex-concave procedure (CCP) \cite{zhou2020framework} is explored to handle the nonconvex unit modulus constraints of IRS phase shifts.

\item In contrast to existing research \cite{zhou2020framework,zhang2020robust} where the covariance matrix of CSI errors is simplified into an identity matrix to facilitate the algorithm design, our proposed algorithm can be applied with more general forms of covariance matrices, which can describe both uncorrelated and correlated CSI errors.

\item Simulation results verify the effectiveness of our proposed algorithm under both scenarios of partial CSI errors and full CSI errors, and reveal that the proposed algorithm outperforms the maximum ratio transmission (MRT) and isotropic AN based baseline schemes.
\end{enumerate}
\begin{figure}[b]
	\centering
	\includegraphics[width=3in]{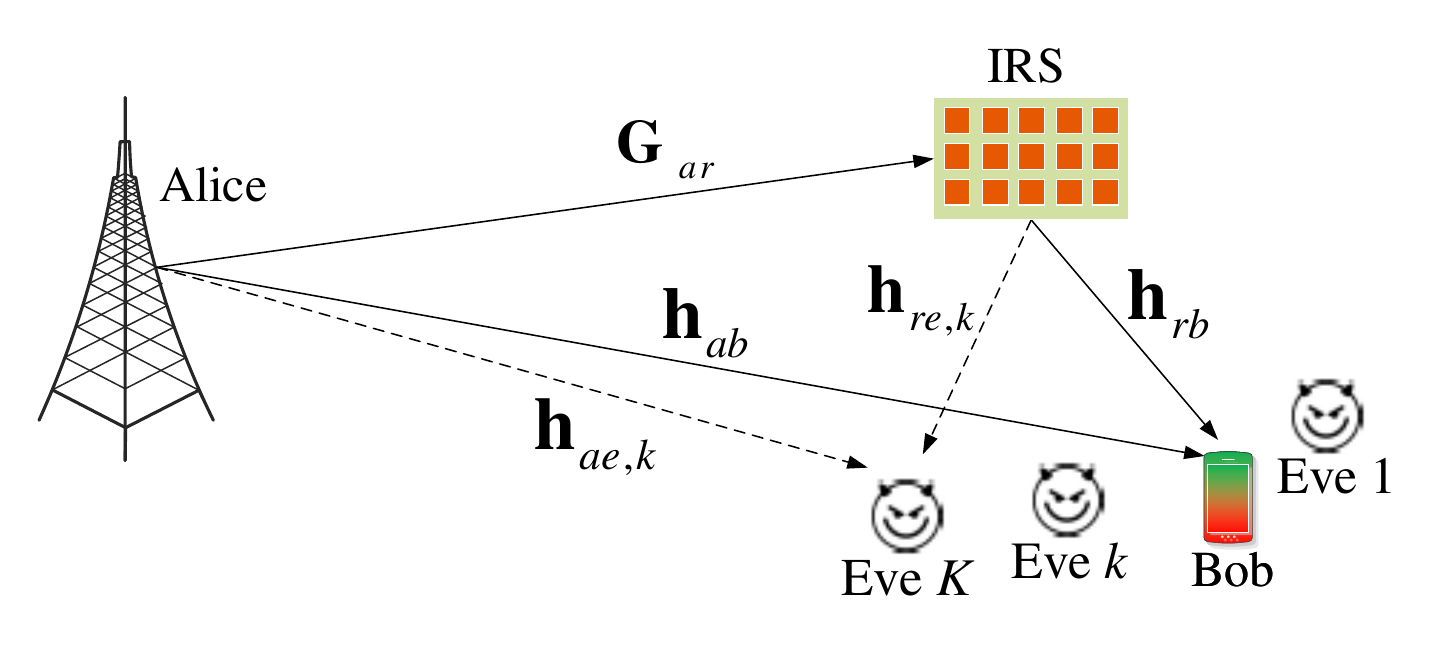}
	\caption{An IRS-aided secure communication with one Bob and multiple Eves under MISO wiretap channels.}\vspace{-0.5cm}
	\label{figsystemmodel}
\end{figure}
\emph{Notations}: Boldface lowercase and uppercase letters are used to represent vectors and matrices, respectively. The superscripts ${\left( \cdot \right)^{\rm{T}}}$, ${\left(  \cdot \right)^{\rm{H}}}$, and ${\left( \cdot \right)^{\rm{\ast}}}$ stand for the transpose, Hermitian, and conjugate operators, respectively. ${{\mathbb{ C}}^{M \times N}}$ represents the set of all $M \times N$ complex matrices, and $\mathbb{H}^{n}$ denotes the set of all $n \times n$ Hermitian matrices. ${\rm{Tr}}\left( \cdot \right)$, ${\rm{Re}}\{\cdot\}$, and ${\rm{diag}}\{\cdot\}$ denote the trace, the real part of a complex value, and a diagonal matrix. $\textrm{vec}(\mathbf{A})$ denotes the vectorization operation on the matrix $\mathbf{A}$. $\sigma_{\textrm{max}}\text{\{}\mathbf{A}\}$
represents the maximum singular value of matrix $\mathbf{A}$. $\lambda_{\textrm{max}}\text{\{}\mathbf{A}\}$ and $\lambda_{\textrm{min}}\{\mathbf{A}\}$ denote the maximum and minimum eigen value of matrix $\mathbf{A}$, respectively. ${\cal C}{\cal N}({\bm{\mu}},{\bf{Z}})$ represents a circularly symmetric complex gaussian (CSCG) distribution with a mean vector ${\bm{\mu}}$ and covariance matrix ${\bf{Z}}$.

\section{System model}
\subsection{Signal Transmission Model}
As illustrated in Fig. 1, we consider a wireless downlink scenario, where a single-antenna legitimate receiver (referred as Bob), is
	overheard by multiple single-antenna eavesdroppers (referred as Eves).  Eves are assumed to be participating users, thus the transmitter (referred as Alice) has Eves’
	channel state information (CSI) to some extend (but is not accurate). The Alice is equipped with $N_{t}$ antennas, and the
IRS is equipped with $M$ reflection units, thus the spatial degrees of freedom (DoFs) at
the transmitter and IRS are utilized to degrade the Eves' interceptions. By considering the AN-aided transmit beamforming, the transmit signal vector at Alice can be represented as
\begin{equation}
\mathbf{x}=\mathbf{s}+\mathbf{z}=\mathbf{w}s+\mathbf{z},
\end{equation}
where $s$ is the data symbol intended for Bob, and $\mathbf{z}$ is the AN generated by Alice to confuse Eves. $\mathbf{w}\in\mathbb{C}^{N_{t}\times1}$
is the transmit beamforming vector. We assume that the transmit signal vector
$\mathbf{s}$ and the noise vector $\mathbf{z}$ follow the complex Gaussian distributions of $\mathcal{CN}(\mathbf{0},\mathbf{W})$ and $\mathcal{CN}(\mathbf{0},\mathbf{Z})$, respectively, where $\mathbf{W=}\mathbf{w}\mathbf{w}^{H}$. Obviously, both $\mathbf{W}$ and $\mathbf{Z}$ are positive semidefinite matrices, and we have $\textrm{rank}(\mathbf{W})=1$.

Then the received signals at Bob and the $k$th Eve can be respectively expressed as \begin{subequations}
\begin{alignat}{1}
y_{b} & =\hat{\mathbf{h}}_{b}^{H}\mathbf{x}+n_{b}=(\mathbf{h}_{ab}^{H}+\mathbf{\mathbf{h}}_{rb}^{H}\mathbf{\Phi}\mathbf{G}_{ar})\mathbf{x}+n_{b},\\
y_{e,k} & =\mathbf{\hat{h}}_{e,k}^{H}\mathbf{x}\!+\!n_{e,k}=(\mathbf{h}_{ae,k}^{H}\!+\!\mathbf{h}_{re,k}^{H}\mathbf{\Phi}\mathbf{G}_{ar})\mathbf{x}+n_{e,k},
\end{alignat}
\end{subequations} where $\forall k\in\mathcal{K}=\{1,2,\cdots,K\}$. The channels of direct communication links are $\mathbf{h}_{ab}\in\mathbb{\mathbb{C}}^{N_{t}\times1}$ and $\mathbf{h}_{ae,k}\in\mathbb{\mathbb{C}}^{N_{t}\times1}$, which respectively denote the Alice-Bob link and Alice-Eve link. The IRS can provide reflecting links to enhance the communication for Bob, and interfere the communication for Eves. The channel from Alice to IRS is modeled by $\mathbf{\mathbf{G}}_{ar}\in\mathbb{\mathbb{C}}^{M\times N_{t}}$. The channels from IRS to Bob and from IRS to Eves are modeled by $\mathbf{h}_{rb}\in\mathbb{\mathbb{C}}^{M\times1}$ and $\mathbf{h}_{re,k}\in\mathbb{\mathbb{C}}^{M\times1}$, respectively. Let $\phi_{m}=e^{j\theta_{m}}$ and define the reflection coefficients matrix of the IRS by $\mathbf{\Phi}=\text{diag}\{\phi_{1},\cdots,\phi_{m},\cdots,\phi_{M}\}$, where $\theta_{m}\in[0,2\pi)$
denotes the phase shift of the $m$th unit of the IRS. Then the equivalent channel of the composite Alice-IRS-Bob link can be defined by $\hat{\mathbf{h}}_{b}\triangleq\mathbf{h}_{ab}+\mathbf{G}_{ar}^{H}\mathbf{\mathbf{\Phi}}^{H}\mathbf{h}_{rb}$,
$\hat{\mathbf{h}}_{b}\in\mathbb{C}^{N_{t}\times1}$, while the equivalent channel of the composite Alice-IRS-Eve link can be defined by $\mathbf{\hat{h}}_{e,k}\triangleq\mathbf{h}_{ae,k}+\mathbf{G}_{ar}^{H}\mathbf{\mathbf{\Phi}}^{H}\mathbf{h}_{re,k}$,
$\mathbf{\hat{h}}_{e,k}\in\mathbb{C}^{N_{t}\times1}$. $n_{b}\sim\mathcal{CN}(0,\sigma_{b}^{2})$ denotes the additive white Gaussian noise (AWGN) received at Bob, while $n_{e,k}\sim\mathcal{CN}(0,\sigma_{e,k}^{2})$ denotes the AWGN received at the $k$th Eve.

By defining the vector $\bm{\phi}=[\phi_{1},\cdots,\phi_{m},\cdots,\phi_{M}]^{T}$, the equivalent channel $\hat{\mathbf{h}}_{b}$ and $\mathbf{\hat{h}}_{e,k}$
can be respectively reexpressed as \begin{subequations}
\begin{alignat}{1}
\hat{\mathbf{h}}_{b}^{H} & \triangleq\mathbf{h}_{ab}^{H}+\bm{\phi}^{T}\mathbf{G}_{cb},\label{BobChnnlextnd}\\
\hat{\mathbf{h}}_{e,k}^{H} & \triangleq\mathbf{h}_{ae,k}^{H}+\bm{\phi}^{T}\mathbf{G}_{ce,k},\forall k\in\mathcal{K},\label{EveChnnlextnd}
\end{alignat}
\end{subequations} where the $\mathbf{G}_{cb}\triangleq\textrm{diag(\ensuremath{\mathbf{\mathbf{h}}_{rb}^{H}})}\mathbf{G}_{ar}\in\mathbb{C}^{M\times N_{t}}$
is defined as the cascaded Alice-IRS-Bob channel, and
$\mathbf{G}_{ce,k}\triangleq\textrm{diag(\ensuremath{\mathbf{\mathbf{h}}_{re,k}^{H}})}\mathbf{G}_{ar}\in\mathbb{C}^{M\times N_{t}}$ is defined as the cascaded Alice-IRS-Eve channel.

Based on above channel models, the achievable data rates in (bit/s/Hz) of Bob and the $k$th Eve are
\begin{subequations}
\begin{align}
C_{b}(\mathbf{W},\mathbf{Z},{\bf \Phi}) & =\log\left(1+\frac{\hat{\mathbf{h}}_{b}^{H}\mathbf{W}\hat{\mathbf{h}}_{b}}{\sigma_{b}^{2}+\hat{\mathbf{h}}_{b}^{H}\mathbf{Z}\hat{\mathbf{h}}_{b}}\right).\label{eq:Cbdef} \\
C_{e,k}(\mathbf{W},\mathbf{Z},{\bf \Phi}) & =\log\left(1+\frac{\hat{\mathbf{h}}_{e,k}^{H}\mathbf{W}\hat{\mathbf{h}}_{e,k}}{\sigma_{e,k}^{2}+\hat{\mathbf{h}}_{e,k}^{H}\mathbf{Z}\hat{\mathbf{h}}_{e,k}}\right).\label{eq:Cekdef}
\end{align}
\end{subequations}
Then the achievable secrecy rate \cite{liao2010qos} can be written as
\begin{alignat}{1}
{{R}_{s}}(\mathbf{W},\mathbf{Z},\mathbf{\Phi})\text{ }={{[\underset{k\in\mathcal{K}}{\mathop{\min}}\,\left\{ {{C}_{b}}(\mathbf{W},\mathbf{Z},\mathbf{\Phi})-{{C}_{e,k}}(\mathbf{W},\mathbf{Z},\mathbf{\Phi})\right\} ]}^{+}},\label{eq:SRdefnt}
\end{alignat}
where ${{\left[a\right]}^{+}}\triangleq\text{max}\left(a,0\right)$.

\subsection{Two CSI Error Scenarios}

Since Bob is a registered user, we assume that Bob's CSI is perfect while Eves' CSI is imperfect. Different from the communication system without IRS, there
are two types of channels from Alice to Eves, which are the direct
Alice-Eve channel $\mathbf{h}_{ae,k}$ and the cascaded Alice-IRS-Eve
channel $\mathbf{G}_{ce,k}$. Thus, we consider two scenarios of CSI errors next.

\subsubsection{Scenario 1: Partial CSI Errors}
The CSI of the cascaded Alice-IRS-Eve
link at the transmitter is much more challenging to obtain than
the CSI of the direct Alice-Eve link at the transmitter according to the passive nature of the IRS. Hence,
in this scenario, we assume the direct channel $\mathbf{h}_{ae,k}$
is perfect, while the cascaded channel $\mathbf{G}_{ce,k}$ is imperfect.
The imperfect cascaded Alice-IRS-Eve channel can be presented as
\begin{alignat}{1}
\mathbf{G}_{ce,k} & =\bar{\mathbf{G}}_{ce,k}+\triangle\mathbf{G}_{ce,k},\forall k\in\mathcal{K},\label{eq:CascdChnnlerror}
\end{alignat}
where $\bar{\mathbf{G}}_{ce,k}$ is the estimated value of $\mathbf{G}_{ce,k}$ which is known to Alice,
and $\triangle\mathbf{G}_{ce,k}$ denotes the corresponding CSI error.

\subsubsection{Scenario 2: Full CSI Errors}
Due to the concealment of Eves, we consider full CSI errors on both the direct channels and reflection channels for Eves in Scenario 2. Other than the imperfect cascaded channel $\mathbf{G}_{ce,k}$ (i.e., reflection channel) in \eqref{eq:CascdChnnlerror}, the imperfect direct channel
$\mathbf{h}_{ae,k}$ can be described as
\begin{alignat}{1}
\mathbf{h}_{ae,k} & =\bar{\mathbf{h}}_{ae,k}+\triangle\mathbf{h}_{ae,k},\forall k\in\mathcal{K},\label{eq:DirctChnnlerror}
\end{alignat}
where $\bar{\mathbf{h}}_{ae,k}$ is the estimated value of $\mathbf{h}_{ae,k}$ which is known to Alice, and $\triangle\mathbf{h}_{ae,k}$ denotes the corresponding CSI error.

The statistical
CSI error usually describes the channel estimation error, and is mainly considered here. Specifically, the CSI errors of $\mathbf{h}_{ae,k}$
and $\mathbf{G}_{ce,k}$ are assumed to be random and follow a CSCG distribution known a priori, i.e., \begin{subequations}\label{StatisticalCSI}
\begin{alignat}{1}
\mathbf{g}_{he,k}\triangleq\triangle\mathbf{h}_{ae,k} & \thicksim\mathcal{CN}(\mathbf{0},\mathbf{\Sigma}_{he,k}),\mathbf{\Sigma}_{he,k}\succeq\mathbf{0},\label{StatisticalCSI_g_hek}\\
\mathbf{g}_{ge,k}\triangleq\textrm{vec}(\triangle\mathbf{G}_{ce,k}) & \thicksim\mathcal{CN}(\mathbf{0},\mathbf{\Sigma}_{ge,k}),\mathbf{\Sigma}_{ge,k}\succeq\mathbf{0}, \label{StatisticalCSI_g_gek}
\end{alignat}
\end{subequations} where $\forall k\in\mathcal{K}$, and $\mathbf{\Sigma}_{he,k}\in\mathbb{C}^{N_{t}\times N_{t}}$
and $\mathbf{\Sigma}_{ge,k}\in\mathbb{C}^{MN_{t}\times MN_{t}}$ are
positive semidefinite covariance matrices of the CSI error. In addition,
$\mathbf{g}_{he,k}$ is independent of $\mathbf{g}_{he,j}$ for
any $k\neq j$, and $\mathbf{g}_{ge,k}$ is independent of $\mathbf{g}_{ge,j}$
for any $k\neq j$.
\section{The OCR Transmission Design}
By considering Eves' CSI errors, it is very necessary to obtain a robust design of the transmit beamformer $\mathbf{w}$, AN spatial covariance $\mathbf{Z}$, and IRS phase shifts ${\bf \Phi}$ to ensure the system security. Thus, $\mathbf{w}$, $\mathbf{Z}$ and ${\bf \Phi}$ are jointly optimized by an AO
algorithm, where the BTI, semi-definite relaxation (SDR) technique, and
penalty CCP method\cite{zhou2020framework} are leveraged.
\vspace{-0.7cm}
\subsection{Scenario 1: Partial CSI Error} \label{OCR-Partial}

\subsubsection{Problem Formulation}
When the direct channel $\mathbf{h}_{ae,k}$ of the Alice-Eve link is perfect, and the cascaded
channel $\triangle\mathbf{G}_{ce,k}$ of the Alice-IRS-Eve link is
imperfect, we formulate an OCR-PM problem to minimize the transmit power as \begin{subequations}\label{OCR_PM_Partialorgn}
\begin{alignat}{1}
&\underset{\mathbf{W},\mathbf{Z},{\bf \Phi}}{\min}  {\rm Tr(}{\bf W}{\rm +}{\bf Z}{\rm )}\label{OCR_PM_Partialorgn_obj}\\
&\textrm{s.t.}\,\,  C_{b}(\mathbf{W},\mathbf{Z},{\bf \Phi})\geq\log\gamma,\label{OCR_PM_Partialorgn_bob}\\
& \quad\,\,\,\textrm{Pr}_{\mathbf{g}_{ge,k}}\!\!\left\{ C_{e,k}(\mathbf{W},\mathbf{Z},{\bf \Phi})\!\leq\!\log\beta\right\} \!\geq\!1\!-\!\rho_{k},\forall k\in\mathcal{K},\label{OCR_PM_Partialorgn_eve}\\
 &\quad\,\,\, {\bf Z}\succeq0,\label{OCR_PM_Partialorgn_Z0}\\
 & \quad\,\,\,{\bf W}\succeq0,\label{OCR_PM_Partialorgn_W0}\\
 & \quad\,\,\,\textrm{rank}({\bf W})=1,\label{OCR_PM_Partialorgn_Wrank}\\
 & \quad\,\, \left|\phi_{m}\right|=1,m=1,\cdots,M,\label{OCR_PM_Partialorgn_fai1}
\end{alignat}
\end{subequations} where $\gamma\geq1$ and $\beta\geq1$ are constant values, and $\gamma\geq\beta$ is imposed to ensure a non-negative
secrecy rate. $\rho_{k}\in\left(0,1\right]$ denotes the rate outage probability for the $k$th Eve. The chance constraint \eqref{OCR_PM_Partialorgn_eve} combined with (9b) guarantees that the probability for the secrecy rate to be larger than $\log\gamma-\log\beta$ is no less than $1-{{\rho}_{k}}$ in the presence of random CSI errors. By substituting \eqref{eq:Cbdef}
and \eqref{eq:Cekdef} into \eqref{OCR_PM_Partialorgn}, we can transform
Problem \eqref{OCR_PM_Partialorgn} into\begin{subequations}\label{OCR_PM_Partialextnd}
\begin{alignat}{1}
\underset{\mathbf{W},\mathbf{Z},\bm{\phi}}{\min} & {\rm Tr(}{\bf W}{\rm +}{\bf Z}{\rm )}\label{OCR_PM_Partialextnd_obj}\\
\textrm{s.t.}\;\, & \hat{\mathbf{h}}_{b}^{H}[\mathbf{W}-(\gamma-1)\mathbf{Z}]\hat{\mathbf{h}}_{b}\geq(\gamma-1)\sigma_{b}^{2},\label{OCR_PM_Partialextnd_bob}\\
\textrm{} & \textrm{Pr}_{\mathbf{g}_{ge,k}}\{ \hat{\mathbf{h}}_{e,k}^{H}[\mathbf{W}-(\beta-1)\mathbf{Z}]\hat{\mathbf{h}}_{e,k}\leq(\beta-1)\sigma_{e,k}^{2}\} \geq1-\rho_{k},\forall k\in\mathcal{K},\label{OCR_PM_Partialextnd_eve}\\
 & \eqref{OCR_PM_Partialorgn_Z0},\eqref{OCR_PM_Partialorgn_W0},\eqref{OCR_PM_Partialorgn_Wrank},\eqref{OCR_PM_Partialorgn_fai1}.\label{OCR_PM_Partialextnd_ZWrankfai}
\end{alignat}
\end{subequations}

The main challenge for solving Problem \eqref{OCR_PM_Partialextnd} lies in the rate outage probability constraints \eqref{OCR_PM_Partialextnd_eve}. To tackle it,
we develop computable upper bounds for \eqref{OCR_PM_Partialextnd_eve} by using the BTI given in Lemma \ref{Lemma_BIT}.
 \begin{lemma}
\label{Lemma_BIT} (Bernstein-Type
Inequality \cite{wang2014outage}) For any $(\mathbf{A},\mathbf{u},c)\in\mathbb{H}^{n}\times\mathbb{C}^{n}\times\mathbb{R}$, $\mathbf{v}\sim\mathcal{CN}(\mathbf{0},\mathbf{I}_{n})$
and $\rho\in\text{(0,1]}$, the following implication holds:
\begin{flalign}
 & \textrm{Pr}_{\mathbf{v}}\left\{ \mathbf{v}^{H}\mathbf{A}\mathbf{v}+2\textrm{Re}\{\mathbf{u}^{H}\mathbf{v}\}+c\geq0\right\} \geq1-\rho, \label{Lemma1A}\\
\Longleftarrow & \begin{cases}
\textrm{Tr}(\mathbf{A})-\sqrt{-2\ln(\rho)}\cdot x+\ln(\rho)\cdot y+c\geq0,\\
\left\Vert \left[\begin{array}{c}
\textrm{vec}(\mathbf{A})\\
\sqrt{2}\mathbf{u}
\end{array}\right]\right\Vert _{2}\leq x,\\
y\mathbf{I}_{n}+\mathbf{A}\succeq\mathbf{0},y\geq0,
\end{cases} \label{Lemma1B}
\end{flalign}
where $x$, $y$ are the introduced slack variables. \end{lemma}
As illustrated in \cite{wang2014outage}, the \eqref{Lemma1B} is a safe approximation of \eqref{Lemma1A}, and the tightness is achieved when the equalities in \eqref{Lemma1B} hold.

\subsubsection{Reformulation}

To apply Lemma \ref{Lemma_BIT}, we transform the probability constraints \eqref{OCR_PM_Partialextnd_eve} into the form of \eqref{Lemma1A} as follows. By substituting \eqref{eq:CascdChnnlerror} into \eqref{OCR_PM_Partialextnd_eve},
and defining $\mathbf{\Xi}_{e}\triangleq(\beta-1)\mathbf{Z}-\mathbf{W}$
and $\tilde{\sigma}_{e,k}^{2}\triangleq(\beta-1)\sigma_{e,k}^{2}$, the probability of Eves' information leakage can be represented as
\begin{subequations}\label{eq:evedatarate}
	\begin{align}
		& \textrm{Pr}_{\mathbf{g}_{ge,k}}\{\hat{\mathbf{h}}_{e,k}^{H}[\mathbf{W}-(\beta-1)\mathbf{Z}]\hat{\mathbf{h}}_{e,k}\leq\tilde{\sigma}_{e,k}^{2}\}\nonumber \\
		= & \textrm{Pr}_{\mathbf{g}_{ge,k}}\{[\mathbf{h}_{ae,k}^{H}+\bm{\phi}^{T}(\mathbf{\bar{\mathbf{G}}}_{ce,k}+\triangle\mathbf{G}_{ce,k})]\mathbf{\Xi}_{e}[\mathbf{h}_{ae,k}^{H}+\bm{\phi}^{T}(\mathbf{\bar{\mathbf{G}}}_{ce,k}+\triangle\mathbf{G}_{ce,k})]^{H}+\tilde{\sigma}_{e,k}^{2}\geq0\}\label{eq:evedatarate_a}\\
		= & \textrm{Pr}_{\mathbf{g}_{ge,k}}\{\underset{f_{1,k}^{\textrm{Partial}}}{\underbrace{\bm{\phi}^{T}\triangle\mathbf{G}_{ce,k}\mathbf{\Xi}_{e}\triangle\mathbf{G}_{ce,k}^{H}\bm{\phi}^{*}}}+2\textrm{Re}[\underset{f_{2,k}^{\textrm{Partial}}}{\underbrace{\bm{\phi}^{T}\triangle\mathbf{G}_{ce,k}\mathbf{\Xi}_{e}(\mathbf{h}_{ae,k}+\mathbf{\bar{\mathbf{G}}}_{ce,k}^{H}\bm{\phi}^{*})}]}\nonumber \\
		& +\underset{c_{k}^{\textrm{Partial}}}{\underbrace{(\mathbf{h}_{ae,k}^{H}+\bm{\phi}^{T}\mathbf{\bar{\mathbf{G}}}_{ce,k})\mathbf{\Xi}_{e}(\mathbf{h}_{ae,k}^{H}+\bm{\phi}^{T}\mathbf{\bar{\mathbf{G}}}_{ce,k})^{H}+\tilde{\sigma}_{e,k}^{2}}}\geq0\}.\label{eq:evedatarate_b}
	\end{align}
\end{subequations}

%

The vectored CSI error $\mathbf{g}_{ge,k}$ in \eqref{StatisticalCSI_g_gek} can be represented by
$\mathbf{g}_{ge,k}=\mathbf{\Sigma}_{ge,k}^{1/2}\mathbf{v}_{ge,k}$,
where $\mathbf{v}_{ge,k}$ is a Gaussian random vector, i.e., $\mathbf{v}_{ge,k}\sim\mathcal{CN}(\mathbf{0},\mathbf{I}_{MN_{t}})$,
and $\mathbf{\Sigma}_{ge,k}=\mathbf{\Sigma}_{ge,k}^{1/2}\mathbf{\Sigma}_{ge,k}^{1/2}$.
Since $\mathbf{\Sigma}_{ge,k}$ is a positive semidefinite matrix, we have
$(\mathbf{\Sigma}_{ge,k}^{1/2})^{H}=\mathbf{\Sigma}_{ge,k}^{1/2}$ and
$(\mathbf{\Sigma}_{ge,k}^{1/2})^{T}=(\mathbf{\Sigma}_{ge,k}^{1/2})^{*}$. The
expression of $f_{1,k}^{\textrm{Partial}}$ in \eqref{eq:evedatarate_b} can be reformulated as\begin{alignat}{1}
f_{1,k}^{\textrm{Partial}} & =\textrm{Tr}(\triangle\mathbf{G}_{ce,k}\mathbf{\Xi}_{e}\triangle\mathbf{G}_{ce,k}^{H}\bm{\phi}^{*}\bm{\phi}^{T})=\textrm{Tr}(\triangle\mathbf{G}_{ce,k}^{H}\mathbf{E}\triangle\mathbf{G}_{ce,k}\mathbf{\Xi}_{e})\nonumber \\
 & \stackrel{(a)}{=}\textrm{vec}^{H}(\triangle\mathbf{G}_{ce,k})(\mathbf{\Xi}_{e}^{T}\otimes\mathbf{E})\textrm{vec}(\triangle\mathbf{G}_{ce,k})\triangleq\mathbf{v}_{ge,k}^{H}\mathbf{A}_{e,k}\mathbf{v}_{ge,k},\label{eq:f1}
\end{alignat}
where $\mathbf{E}\triangleq\bm{\phi}^{*}\bm{\phi}^{T}$, $\mathbf{A}_{e,k}\triangleq\mathbf{\Sigma}_{ge,k}^{1/2}(\mathbf{\Xi}_{e}^{T}\otimes\mathbf{E})\mathbf{\Sigma}_{ge,k}^{1/2}$,
and $(a)$ is obtained due to $\textrm{Tr}(\mathbf{A}^{H}\mathbf{BCD})=\textrm{vec}^{H}(\mathbf{A})(\mathbf{D}^{T}\otimes\mathbf{B})\textrm{vec}(\mathbf{C})$.
The expression of $f_{2,k}^{\textrm{Partial}}$ in \eqref{eq:evedatarate_b} can be reformulated as
\begin{alignat}{1}
f_{2,k}^{\textrm{Partial}} & =\textrm{Tr}(\triangle\mathbf{G}_{ce,k}\mathbf{\Xi}_{e}(\mathbf{h}_{ae,k}\bm{\phi}^{T}+\mathbf{\bar{\mathbf{G}}}_{ce,k}^{H}\mathbf{E}))\overset{(b)}{=}\textrm{vec}^{H}(\bm{\phi}^{*}\mathbf{h}_{ae,k}^{H}+\mathbf{E}\mathbf{\bar{\mathbf{G}}}_{ce,k})(\mathbf{\Xi}_{e}^{T}\otimes\mathbf{I}_{M})\textrm{vec}(\triangle\mathbf{G}_{ce,k})\nonumber \\
 & \triangleq\mathbf{u}_{e,k}^{H}\mathbf{v}_{ge,k},\label{eq:f2}
\end{alignat}
where $\mathbf{u}_{e,k}\triangleq\mathbf{\Sigma}_{ge,k}^{1/2}(\mathbf{\Xi}_{e}^{*}\otimes\mathbf{I}_{M})\textrm{vec}(\bm{\phi}^{*}\mathbf{h}_{ae,k}^{H}+\mathbf{E}\mathbf{\bar{\mathbf{G}}}_{ce,k})$,
and $(b)$ is obtained due to $\textrm{Tr}(\mathbf{A}\mathbf{B}\mathbf{C}^{H})=\textrm{vec}^{H}(\mathbf{C})(\mathbf{B}^{T}\otimes\mathbf{I})\textrm{vec}(\mathbf{A})$.

By substituting \eqref{eq:f1} and \eqref{eq:f2} into \eqref{eq:evedatarate_b},
the leakage data rate outage constraints \eqref{OCR_PM_Partialextnd_eve}
for Eves become
\begin{alignat}{2}
\eqref{OCR_PM_Partialextnd_eve} \Leftrightarrow & \textrm{Pr}_{\mathbf{v}_{ce,k}}\left\{ \mathbf{v}_{ge,k}^{H}\mathbf{A}_{e,k}\mathbf{v}_{ge,k}+2\textrm{Re}\{\mathbf{u}_{e,k}^{H}\mathbf{v}_{ge,k}\}+c_{k}^{\textrm{Partial}}\geq0\right\} \geq1-\rho_{k},\forall k\in\mathcal{K}.\label{eq:Gussianvectoreql}
\end{alignat}

In \eqref{eq:Gussianvectoreql}, the outage
probability w.r.t. the random CSI error $\mathbf{g}_{ge,k}$ in \eqref{OCR_PM_Partialextnd_eve} is equivalently transformed into the outage probability of a real Gaussian quadratic form w.r.t. $\mathbf{v}_{ge,k}$, which facilitates the application of Lemma \ref{Lemma_BIT}. It is also seen that Lemma \ref{Lemma_BIT} can be applied as long as the CSI error $\mathbf{g}_{ge,k}$ follows the CSCG distribution and regardless of the correlation among its elements. Then the chance constraint in \eqref{OCR_PM_Partialextnd_eve} can be conservatively approximated and replaced by computable constraints according to Lemma \ref{Lemma_BIT}, and Problem \eqref{OCR_PM_Partialextnd} becomes \begin{subequations}\label{BITOCRPM}
\begin{alignat}{1}
\underset{\mathbf{W},\mathbf{Z},{\rm \bm{\phi}},\mathbf{A}_{e},\mathbf{x},\mathbf{y}}{\min} & {\rm Tr(}{\bf W}{\rm +}{\bf Z}{\rm )}\label{eq:BITOCRPM_a}\\
\textrm{s.t.}\quad & \textrm{Tr}(\mathbf{A}_{e,k})-\sqrt{-2\ln(\rho_{k})}\cdot x_{k}+\ln(\rho_{k})\cdot y_{k}+c_{k}^{\textrm{Partial}}\geq0,\label{eq:BITOCRPM_b}\\
 & \left\Vert \left[\begin{array}{c}
\textrm{vec}(\mathbf{A}_{e,k})\\
\sqrt{2}\mathbf{u}_{e,k}
\end{array}\right]\right\Vert _{2}\leq x_{k},\label{eq:BITOCRPM_c}\\
 & y_{k}\mathbf{I}_{MN_{t}}+\mathbf{A}_{e,k}\succeq\mathbf{0},y_{k}\geq0,\label{eq:BITOCRPM_d}\\
 & \eqref{OCR_PM_Partialextnd_bob},\eqref{OCR_PM_Partialorgn_Z0},\eqref{OCR_PM_Partialorgn_W0},\eqref{OCR_PM_Partialorgn_Wrank},\eqref{OCR_PM_Partialorgn_fai1},
\end{alignat}
\end{subequations} where $\mathbf{A}_{e}=[\mathbf{A}_{e,1},\mathbf{A}_{e,2},\cdots,\mathbf{A}_{e,K}]$, $\mathbf{x}=[x_{1},x_{2},\cdots,x_{K}]^{T}$ and $\mathbf{y}=[y_{1},y_{2},\cdots,y_{K}]^{T}$ are introduced slack variables.
We further simplify the $\left\Vert \textrm{vec}(\mathbf{A}_{e,k})\right\Vert ^{2}$ in constraint \eqref{eq:BITOCRPM_c} by
\begin{alignat}{1}
\left\Vert \textrm{vec}(\mathbf{A}_{e,k})\right\Vert ^{2}  &=\left\Vert \mathbf{A}_{e,k}\right\Vert _{F}^{2}=\textrm{Tr}[\mathbf{A}_{e,k}\mathbf{A}_{e,k}^{H}] =\textrm{Tr}[(\mathbf{\Xi}_{e}^{T}\otimes\mathbf{E})^{H}\mathbf{\Sigma}_{ge,k}(\mathbf{\Xi}_{e}^{T}\otimes\mathbf{E})\mathbf{\Sigma}_{ge,k}]\nonumber \\
 & \overset{(c)}{=}\textrm{vec}^{H}(\mathbf{\Xi}_{e}^{T}\otimes\mathbf{E})(\mathbf{\Sigma}_{ge,k}^{T}\otimes\mathbf{\Sigma}_{ge,k})\textrm{vec}(\mathbf{\Xi}_{e}^{T}\otimes\mathbf{E})\nonumber \\
 & =\textrm{vec}^{H}(\mathbf{\Xi}_{e}^{T}\otimes\mathbf{E})[(\mathbf{\Sigma}_{ge,k}^{1/2T}\otimes\mathbf{\Sigma}_{ge,k}^{1/2})^{H}(\mathbf{\Sigma}_{ge,k}^{1/2T}\otimes\mathbf{\Sigma}_{ge,k}^{1/2})]\textrm{vec}(\mathbf{\Xi}_{e}^{T}\otimes\mathbf{E})\nonumber \\
 & =\left\Vert (\mathbf{\Sigma}_{ge,k}^{1/2T}\otimes\mathbf{\Sigma}_{ge,k}^{1/2})\textrm{vec}(\mathbf{\Xi}_{e}^{T}\otimes\mathbf{E})\right\Vert ^{2}.\label{eq:vecAkforoptZW}
\end{alignat}
where $(c)$ is obtained by invoking the identity $\textrm{Tr}(\mathbf{A}^{H}\mathbf{BCD})=\textrm{vec}^{H}(\mathbf{A})(\mathbf{D}^{T}\otimes\mathbf{B})\textrm{vec}(\mathbf{C})$.

By
substituting the expressions of \eqref{eq:vecAkforoptZW} into \eqref{eq:BITOCRPM_c}, we have the OCR-PM Problem in
\eqref{BITOCRPM} written more explicitly as \begin{subequations}\label{BITOCRPMsimplfy}
\begin{alignat}{1}
&\underset{\mathbf{W},\mathbf{Z},{\bf \bm{\phi}},\mathbf{x},\mathbf{y}}{\min}  {\rm Tr(}{\bf W}{\rm +}{\bf Z}{\rm )}\label{eq:BITOCRPMsimplfypow}\\
&\textrm{s.t.}\ \ \textrm{Tr}(\mathbf{\Sigma}_{ge,k}^{1/2}(\mathbf{\Xi}_{e}^{T}\otimes\mathbf{E})\mathbf{\Sigma}_{ge,k}^{1/2})-\sqrt{-2\ln(\rho_{k})}\cdot x_{k}+\ln(\rho_{k})\cdot y_{k}+c_{k}^{\textrm{Partial}}\geq0,\label{eq:BITOCRPMsimplfyLMI}\\
 & \left\Vert \!\left[\!\!\!\begin{array}{c}
(\mathbf{\Sigma}_{ge,k}^{1/2T}\otimes\mathbf{\Sigma}_{ge,k}^{1/2})\textrm{vec}(\mathbf{\Xi}_{e}^{T}\otimes\mathbf{E})\\
\sqrt{2}\mathbf{\Sigma}_{ge,k}^{1/2}(\mathbf{\Xi}_{e}^{*}\otimes\mathbf{I}_{M})\textrm{vec}(\bm{\phi}^{*}\mathbf{h}_{ae,k}^{H}\!\!+\!\!\mathbf{E}\mathbf{\bar{\mathbf{G}}}_{ce,k})
\end{array}\!\!\!\right]\!\right\Vert _{2}\!\!\leq \!\!x_{k},\label{eq:BITOCRPMsimplfySOC}\\
 & y_{k}\mathbf{I}_{MN_{t}}+\mathbf{\Sigma}_{ge,k}^{1/2}(\mathbf{\Xi}_{e}^{T}\otimes\mathbf{E})\mathbf{\Sigma}_{ge,k}^{1/2}\succeq\mathbf{0},y_{k}\geq0,\label{eq:BITOCRPMsimplfyeig}\\
 & \eqref{OCR_PM_Partialextnd_bob},\eqref{OCR_PM_Partialorgn_Z0},\eqref{OCR_PM_Partialorgn_W0},\eqref{OCR_PM_Partialorgn_Wrank},\eqref{OCR_PM_Partialorgn_fai1}.
\end{alignat}
\end{subequations}
The chance constraints are removed in Problem \eqref{BITOCRPMsimplfy}. However, the optimization variables $\left\{ \mathbf{W},\mathbf{Z}\right\}$ and $\bm{\phi}$ are coupled. We propose to tackle the coupling via the AO method. The variables $\left\{ \mathbf{W},\mathbf{Z}\right\}$ and $\mathbf{\bm{\phi}}$ are updated alternately.
\subsubsection{Optimization of Transmit Beamforming and AN}
Obviously, when ${\bf \bm{\phi}}$ is fixed, the $c_{k}^{\textrm{Partial}}$,
	$\textrm{Tr}(\mathbf{A}_{k})$, $\left\Vert \textrm{vec}(\mathbf{A}_{k})\right\Vert ^{2}$
	and $\left\Vert \mathbf{u}_{ce,k}\right\Vert $ are all convex functions
	of $\mathbf{\Xi}_{e}$. Then, Problem \eqref{BITOCRPMsimplfy} cannot be solved efficiently only due to the nonconvexity of $\textrm{rank}({\bf W})=1$ in \eqref{OCR_PM_Partialorgn_Wrank}. By removing \eqref{OCR_PM_Partialorgn_Wrank}, the $\left\{ \mathbf{W},\mathbf{Z}\right\} $ can be solved by the SDR, and the corresponding optimization problem is
\begin{subequations}\label{BITOCRPMsimplfyWZxy}
\begin{alignat}{1}
\underset{\mathbf{W},\mathbf{Z},\mathbf{x},\mathbf{y}}{\min} & {\rm Tr(}{\bf W}{\rm +}{\bf Z}{\rm )}\\
\textrm{s.t.}\quad & \eqref{eq:BITOCRPMsimplfyLMI},\eqref{eq:BITOCRPMsimplfySOC},\eqref{eq:BITOCRPMsimplfyeig},\eqref{OCR_PM_Partialextnd_bob},\eqref{OCR_PM_Partialorgn_Z0},\eqref{OCR_PM_Partialorgn_W0}.
\end{alignat}
\end{subequations}

The SDR in Problem \eqref{BITOCRPMsimplfyWZxy} is tight, which means that the solved $\mathbf{W}$ always satisfies $\textrm{rank}({\bf W})=1$. Then the beamforming vector $\mathbf{w}$ can be
recovered from $\mathbf{W}=\mathbf{w}{\mathbf{w}}^H$ by performing the Cholesky decomposition. The tightness of the SDR is proved in the following
Theorem.
\newtheorem{mytheorem}{Theorem}
\begin{mytheorem}\label{Theorem1}
	Assume that the optimal solution of Problem \eqref{BITOCRPMsimplfyWZxy} is $(\mathbf{W}^{\star},\mathbf{Z}^{\star},\mathbf{x}^{\star},\mathbf{y}^{\star})$,
	where $\textrm{rank}(\mathbf{W}^{\star})\geq1$. Then there always
	exists another optimal solution of Problem \eqref{BITOCRPMsimplfyWZxy}, denoted as $(\tilde{\mathbf{W}}^{\star},\tilde{\mathbf{Z}}^{\star},\tilde{\mathbf{x}}^{\star},\tilde{\mathbf{y}}^{\star})$,
	which not only has the same objective value as $(\mathbf{W}^{\star},\mathbf{Z}^{\star},\mathbf{x}^{\star},\mathbf{y}^{\star})$,
	but also satisfies the rank-one constraint, i.e., $\textrm{rank}(\tilde{\mathbf{W}}^{\star})=1$.
\end{mytheorem}
\vspace{-0.7cm}
\begin{proof}[Proof]
	The proving process is the same as that in Appendix A of \cite{hong2020robust} after replacing the channel vector ${{\mathbf{\hat{h}}}_{b}}$ in \cite{hong2020robust} by $\hat{\mathbf{h}}_{b}^{H}\triangleq\mathbf{h}_{ab}^{H}+\bm{\phi}^{T}\mathbf{G}_{cb}$ defined in \eqref{BobChnnlextnd}.
\end{proof}
\subsubsection{Optimization of Phase Shifts at the IRS} \label{OCR-Partial-Optphi}
For given $\mathbf{W}$ and $\mathbf{Z}$, the objective function of Problem \eqref{BITOCRPMsimplfy} is irrelevant with $\mathbf{\bm{\phi}}$. To achieve better convergence, slack variables are introduced, and the data rate constraints in \eqref{OCR_PM_Partialextnd_bob}
and in \eqref{eq:evedatarate_a} can be modified respectively as \begin{subequations}
\begin{align}
 & \hat{\mathbf{h}}_{b}^{H}[\mathbf{W}-(\gamma-1)\mathbf{Z}]\hat{\mathbf{h}}_{b}\geq\tilde{\sigma}_{b}^{2}+\delta_{0},\delta_{0}\geq0,\label{eq:bobraterelaxed}\\
 & [\mathbf{h}_{ae,k}^{H}\!\!+\!\!\bm{\phi}^{T}\!(\mathbf{\bar{\mathbf{G}}}_{ce,k}\!\!+\!\!\triangle\mathbf{G}_{ce,k})]\mathbf{\Xi}_{e}[\mathbf{h}_{ae,k}^{H}\!\!+\!\!\bm{\phi}^{T}\!(\mathbf{\bar{\mathbf{G}}}_{ce,k}\!\!+\!\!\triangle\mathbf{G}_{ce,k})]^{H}+\tilde{\sigma}_{e,k}^{2}-\delta_{k}\geq0,\delta_{k}\geq0.\label{eq:everaterelaxed}
\end{align}
\end{subequations}
where $\tilde{\sigma}_{b}^{2}\triangleq(\gamma-1)\sigma_{b}^{2}$, and $\delta_{k}$ is a slack variable.

Then, the outage probability of \eqref{eq:everaterelaxed} can be safely approximated again by invoking Lemma \ref{Lemma_BIT},
and the optimization problem for $\bm{\phi}$ can be written as \begin{subequations}\label{Findfainonconvx}
\begin{alignat}{1}
\underset{\bm{\phi},\bm{\delta},\mathbf{x},\mathbf{y}}{\textrm{max}} & \ \ \sum_{k=0}^K{{\delta}_k}\\
\textrm{s.t.}\quad & \textrm{Tr}(\mathbf{\Sigma}_{ge,k}^{1/2}(\mathbf{\Xi}_{e}^{T}\otimes\mathbf{E})\mathbf{\Sigma}_{ge,k}^{1/2})-\sqrt{-2\ln(\rho_{k})}\cdot x_{k}+\ln(\rho_{k})\cdot y_{k}+c_{k}^{\textrm{Partial}}-\delta_{k}\geq0,\label{FindfainonconvxLMI}\\
 & \eqref{eq:BITOCRPMsimplfySOC},\eqref{eq:BITOCRPMsimplfyeig},\eqref{OCR_PM_Partialorgn_fai1},\label{FindfainonconvxSOC}\\
 & \hat{\mathbf{h}}_{b}^{H}[\mathbf{W}-(\gamma-1)\mathbf{Z}]\hat{\mathbf{h}}_{b}\geq\tilde{\sigma}_{b}^{2}+\delta_{0},\label{Findfainonconvx_bob}\\
 & \mathbf{\bm{\delta}}\geq\mathbf{0},\label{Findfainonconvx_delta}
\end{alignat}
\end{subequations} where $\mathbf{\bm{\delta}}=[\delta_{0},\delta_{1},\delta_{2},\cdots,\delta_{K}]^{T}$ are slack variables.

From Problem \eqref{Findfainonconvx}, it is seen that the constraints cannot be transformed into functions of only $\mathbf{E}=\bm{\phi}^{*}\bm{\phi}^{T}$ as in \cite{hong2020robust} where the direct communication link is obstructed, but are functions of both $\mathbf{E}$ and $\bm{\phi}$ here. Thus, the optimization variable can only be $\bm{\phi}$ instead of $\mathbf{E}$. To transform Problem \eqref{Findfainonconvx} into a convex problem w.r.t. $\bm{\phi}$, we propose novel mathematical manipulations
on these constraints in four steps as follows.

\textbf{Step 1}: Transform the constraint of \eqref{FindfainonconvxLMI} into convex ones.

\textbf{(1)} We note that the $c_{k}^{\textrm{Partial}}$ in \eqref{FindfainonconvxLMI}
is non-concave w.r.t. $\bm{\phi}$ due to the fact that $\mathbf{\bar{\mathbf{G}}}_{ce,k}\mathbf{\Xi}_{e}\mathbf{\bar{\mathbf{G}}}_{ce,k}^{H}$
is a non-negative semidefinite matrix. To address this issue, the term
$c_{k}^{\textrm{Partial}}$ can be equivalently rewritten as
\begin{alignat}{1}
&\!\!c_{k}^{\textrm{Partial}}(\bm{\phi})  \!\!=\!\!\bm{\phi}^{T}(\mathbf{\bar{\mathbf{G}}}_{ce,k}\mathbf{\Xi}_{e}\mathbf{\bar{\mathbf{G}}}_{ce,k}^{H}\!-\!c_{c,k}\mathbf{I}_{M})\bm{\phi}^{*}\!\!+\!\!2\textrm{Re}\{\bm{\phi}^{T}\mathbf{\bar{\mathbf{G}}}_{ce,k}\mathbf{\Xi}_{e}\mathbf{h}_{ae,k}\}\!\!+\!\!\mathbf{h}_{ae,k}^{H}\mathbf{\Xi}_{e}\mathbf{h}_{ae,k}\!\!+\!\!Mc_{c,k}\!\!+\!\!\tilde{\sigma}_{e,k}^{2},\label{ck_partial_Concave}
\end{alignat}
where the unit-modulus property $\bm{\phi}^{T}\bm{\phi}^{*}=M$ is
utilized. $c_{c,k}$ is chosen as $c_{c,k}=\lambda_{\textrm{max}}\text{\{\ensuremath{\mathbf{\bar{\mathbf{G}}}_{ce,k}}}(\beta-1)\mathbf{Z}\mathbf{\bar{\mathbf{G}}}_{ce,k}^{H}\}$.
Since $\mathbf{\Xi}_{e}=(\beta-1)\mathbf{Z}-\mathbf{W}$, we have
$\mathbf{\bar{\mathbf{G}}}_{ce,k}\mathbf{\Xi}_{e}\mathbf{\bar{\mathbf{G}}}_{ce,k}^{H}-c_{c,k}\mathbf{I}_{M}\preceq\mathbf{0}$,
and the expression of $c_{k}^{\textrm{Partial}}(\bm{\phi})$ in \eqref{ck_partial_Concave}
becomes concave w.r.t. $\bm{\phi}$.

\textbf{(2)} The term $\textrm{Tr}(\mathbf{\Sigma}_{ge,k}^{1/2}(\mathbf{\Xi}_{e}^{T}\otimes\mathbf{E})\mathbf{\Sigma}_{ge,k}^{1/2})$
in \eqref{FindfainonconvxLMI} is equivalently transformed as
\begin{alignat}{1}
&\textrm{Tr}(\mathbf{\Sigma}_{ge,k}^{1/2}(\mathbf{\Xi}_{e}^{T}\otimes\mathbf{E})\mathbf{\Sigma}_{ge,k}^{1/2}) \overset{(d)}{=}\textrm{vec}^{H}(\mathbf{\Sigma}_{ge,k}^{1/2})(\mathbf{I}_{MN_{t}}\otimes(\mathbf{\Xi}_{e}^{T}\otimes(\bm{\phi}^{*}\bm{\phi}^{T})))\textrm{vec}(\mathbf{\Sigma}_{ge,k}^{1/2})\nonumber \\
 & =\textrm{vec}^{H}(\mathbf{\Sigma}_{gx,k})(\mathbf{I}_{MN_{t}}\otimes\mathbf{\Xi}_{e}^{T}\otimes(\bm{\phi}^{*}\bm{\phi}^{T}))\textrm{vec}(\mathbf{\Sigma}_{gx,k}) \overset{(e)}{=}\bm{\phi}^{T}\mathbf{\Sigma}_{gx,k}(\mathbf{I}_{MN_{t}}\otimes\mathbf{\Xi}_{e})\mathbf{\Sigma}_{gx,k}^{H}\bm{\phi}^{*},\label{eq:TrAkquadratic}
\end{alignat}
where $(d)$ is obtained by invoking $\textrm{Tr}(\mathbf{A}^{H}\mathbf{BCD})=\textrm{vec}^{H}(\mathbf{A})(\mathbf{D}^{T}\otimes\mathbf{B})\textrm{vec}(\mathbf{C})$, $(e)$ follows from the equality $\textrm{vec}(\mathbf{X})^{H}(\mathbf{B}^{T}\otimes\mathbf{c}\mathbf{a}^{H})\textrm{vec}(\mathbf{X})=\mathbf{a}^{H}\mathbf{X}\mathbf{B}\mathbf{X}^{H}\mathbf{c}$,
where $\mathbf{B}^{T}=\mathbf{I}_{MN_{t}}\otimes\mathbf{\Xi}_{e}^{T}$,
$\mathbf{c}=\bm{\phi}^{*}$, $\mathbf{a}^{H}=\bm{\phi}^{T}$, and
$\mathbf{X}=\mathbf{\Sigma}_{gx,k}$. When using this identity, the
matrix dimension must be matched. Thus, $\mathbf{\Sigma}_{gx,k}\in\mathbb{C}^{M\times MN_{t}^{2}}$
is obtained by remaping the column vector $\textrm{vec}(\mathbf{\Sigma}_{ge,k}^{1/2})\in\mathbb{C}^{(MN_{t})^{2}\times1}$
into an $M\times MN_{t}^{2}$ dimensional matrix, which can be expressed
as $\mathbf{\Sigma}_{gx,k}=\textrm{unvec}[\textrm{vec}(\mathbf{\Sigma}_{ge,k}^{1/2})]$.

It is also noted that \eqref{eq:TrAkquadratic} is also non-concave
w.r.t. $\bm{\phi}$.
Similarly, we reformulate \eqref{eq:TrAkquadratic} into a concave
form as
\begin{alignat}{1}
&\textrm{Tr}(\mathbf{\Sigma}_{ge,k}^{1/2}(\mathbf{\Xi}_{e}^{T}\otimes\mathbf{E})\mathbf{\Sigma}_{ge,k}^{1/2})=\bm{\phi}^{T}[\mathbf{\Sigma}_{gx,k}(\mathbf{I}_{MN_{t}}\otimes\mathbf{\Xi}_{e})\mathbf{\Sigma}_{gx,k}^{H}-c_{t,k}\mathbf{I}_{M}]\bm{\phi}^{*}\!+\!Mc_{t,k},\label{eq:TrAkConcave}
\end{alignat}
where $c_{t,k}=\lambda_{\textrm{max}}\text{\{}\mathbf{\Sigma}_{gx,k}(\mathbf{I}_{MN_{t}}\otimes((\beta-1)\mathbf{Z}))\mathbf{\Sigma}_{gx,k}^{H}\}$.

\textbf{(3)} By substituting \eqref{ck_partial_Concave} and \eqref{eq:TrAkConcave}
into \eqref{FindfainonconvxLMI}, we have the following convex constraint:
\begin{alignat}{1}
\eqref{FindfainonconvxLMI}\Leftrightarrow & c_{tc,k}^{\textrm{partial}}(\bm{\phi})\geq0,\label{findfaiLMIconvx}
\end{alignat}
where
\begin{alignat}{1}
&c_{tc,k}^{\textrm{partial}}(\bm{\phi})\triangleq  \bm{\phi}^{T}[\mathbf{\Sigma}_{gx,k}(\mathbf{I}_{MN_{t}}\otimes\mathbf{\Xi}_{e})\mathbf{\Sigma}_{gx,k}^{H}-c_{t,k}\mathbf{I}_{M}]\bm{\phi}^{*}+Mc_{t,k}-\sqrt{-2\ln(\rho_{k})}\cdot x_{k}+\ln(\rho_{k})\cdot y_{k}\nonumber \\
 & \!+\!\bm{\phi}^{T}(\mathbf{\bar{\mathbf{G}}}_{ce,k}\mathbf{\Xi}_{e}\mathbf{\bar{\mathbf{G}}}_{ce,k}^{H}\!-\!c_{c,k}\mathbf{I}_{M})\bm{\phi}^{*}\!+\!2\textrm{Re}\{\bm{\phi}^{T}\mathbf{\bar{\mathbf{G}}}_{ce,k}\mathbf{\Xi}_{e}\mathbf{h}_{ae,k}\}\!+\!\mathbf{h}_{ae,k}^{H}\mathbf{\Xi}_{e}\mathbf{h}_{ae,k}\!+\!Mc_{c,k}\!+\!\tilde{\sigma}_{e,k}^{2}\!-\!\delta_{k}.\label{findfaiLMIconvxExprssn}
\end{alignat}
As a result, the non-convex constraint in \eqref{FindfainonconvxLMI} is transformed into a convex one in \eqref{findfaiLMIconvx}.

\textbf{Step 2}: Transform the constraint in \eqref{eq:BITOCRPMsimplfySOC}
to be convex.

\textbf{(1)} In constraint \eqref{eq:BITOCRPMsimplfySOC}, we can find an upper
bound of $(\mathbf{\Sigma}_{ge,k}^{1/2T}\otimes\mathbf{\Sigma}_{ge,k}^{1/2})\textrm{vec}(\mathbf{\Xi}_{e}^{T}\otimes\mathbf{E})$
as \begin{subequations}
\begin{alignat}{1}
  \left\Vert (\mathbf{\Sigma}_{ge,k}^{1/2T}\otimes\mathbf{\Sigma}_{ge,k}^{1/2})\textrm{vec}(\mathbf{\Xi}_{e}^{T}\otimes\mathbf{E})\right\Vert&\overset{(f)}{\leq}\left\Vert (\mathbf{\Sigma}_{ge,k}^{1/2T}\otimes\mathbf{\Sigma}_{ge,k}^{1/2})\right\Vert _{2}\left\Vert \textrm{vec}(\mathbf{\Xi}_{e}^{T}\otimes\mathbf{E})\right\Vert _{2}, \label{norminequalitypre_a}\\
\Leftrightarrow &  \overset{(g)}{=}\lambda_{\textrm{max}}(\mathbf{\Sigma}_{ge,k}^{1/2T}\otimes\mathbf{\Sigma}_{ge,k}^{1/2})M\left\Vert \mathbf{\Xi}_{e}\right\Vert _{F},\label{norminequalitypre}
\end{alignat}
\end{subequations} where the step $(f)$ is obtained due to $\left\Vert \mathbf{A}\mathbf{x}\right\Vert _{2}\leq\left\Vert \mathbf{A}\right\Vert _{2}\left\Vert \mathbf{x}\right\Vert _{2}$,
$\left\Vert \mathbf{A}\right\Vert _{2}=\sigma_{\textrm{max}}\text{\{}\mathbf{A}\}$
denotes the spectral norm of matrix $\mathbf{A}$. The equality
$\left\Vert \mathbf{A}\mathbf{x}\right\Vert _{2}=\left\Vert \mathbf{A}\right\Vert _{2}\left\Vert \mathbf{x}\right\Vert _{2}$
holds when the matrix $\mathbf{A}$ is an unitary matrix. The
step $(g)$ is obtained due to $\sigma_{\textrm{max}}\text{\{}\mathbf{A}\}=\lambda_{\textrm{max}}\text{\{}\mathbf{A}\}$
if $\mathbf{A}\in\mathbb{H}^{n}$, and the following
property:
\begin{align}
\left\Vert \!\textrm{vec}(\mathbf{\Xi}_{e}^{T}\!\otimes\!\mathbf{E})\!\right\Vert ^{2}  \!\!&=\!\textrm{Tr}[(\mathbf{\Xi}_{e}^{T}\!\otimes\!\mathbf{E})(\mathbf{\Xi}_{e}^{*}\!\otimes\!\mathbf{E})]\!=\!\!\textrm{Tr}[(\mathbf{\Xi}_{e}^{T}\mathbf{\Xi}_{e}^{*})\!\otimes\!(\mathbf{E}\mathbf{E})]\nonumber\\
&=\textrm{Tr}[(\mathbf{\Xi}_{e}^{T}\mathbf{\Xi}_{e}^{*})]\textrm{Tr}[(\mathbf{E}\mathbf{E})]\!=\!M^{2}\left\Vert \mathbf{\Xi}_{e}\right\Vert _{F}^{2}.\label{vecKronMproprty}
\end{align}

It is observed that when $\mathbf{\Sigma}_{ge,k}^{1/2T}\otimes\mathbf{\Sigma}_{ge,k}^{1/2}$ is an unitary matrix, e.g., $\mathbf{\Sigma}_{ge,k}=\mathbf{I}_{MN_{t}}$,
the inequality in \eqref{norminequalitypre_a} holds with equality.

\textbf{(2)} In \eqref{eq:BITOCRPMsimplfySOC}, we can find an upper
bound of $\mathbf{\Sigma}_{ge,k}^{1/2}(\mathbf{\Xi}_{e}^{*}\otimes\mathbf{I}_{M})\textrm{vec}(\bm{\phi}^{*}\mathbf{h}_{ae,k}^{H}+\mathbf{E}\mathbf{\bar{\mathbf{G}}}_{ce,k})$
as \begin{subequations}
\begin{alignat}{1}
\!\!\!\!\!\left\Vert\!  \mathbf{\Sigma}_{ge,k}^{1/2}(\mathbf{\Xi}_{e}^{*}\!\otimes\!\mathbf{I}_{M}\!)\textrm{vec}[\bm{\phi}^{*}(\mathbf{h}_{ae,k}^{H}\!+\!\bm{\phi}^{T}\mathbf{\bar{\mathbf{G}}}_{ce,k})]\right\Vert&  \!\!\overset{(h)}{\leq}\!\lambda_{\textrm{max}}(\mathbf{\Sigma}_{ge,k}^{1/2}(\mathbf{\Xi}_{e}^{*}\otimes\mathbf{I}_{M}))\!\left\Vert\! (\mathbf{h}_{ae,k}^{*}\!\!+\!\!\mathbf{\bar{\mathbf{G}}}_{ce,k}^{T}\bm{\phi})\otimes\bm{\phi}^{*}\!\right\Vert ,\label{norminequalityaftr_a}\\
 & \!\!\overset{(i)}{=}\!\lambda_{\textrm{max}}(\mathbf{\Sigma}_{ge,k}^{1/2}(\mathbf{\Xi}_{e}^{*}\otimes\mathbf{I}_{M}))\sqrt{M}\!\left\Vert (\mathbf{h}_{ae,k}^{*}\!+\!\mathbf{\bar{\mathbf{G}}}_{ce,k}^{T}\bm{\phi})\right\Vert ,\label{norminequalityaftr}
\end{alignat}
\end{subequations} where $(h)$ is obtained due to $\left\Vert \mathbf{A}\mathbf{x}\right\Vert _{2}\leq\left\Vert \mathbf{A}\right\Vert _{2}\left\Vert \mathbf{x}\right\Vert _{2}$
and $\textrm{vec}(\mathbf{a}\mathbf{b}^{T})=\mathbf{b}\otimes\mathbf{a}$.
The step in $(i)$ is obtained by using Lemma \ref{Lemma_KronckrNorm}
as follows. It is observed that when $\mathbf{\Sigma}_{ge,k}^{1/2}(\mathbf{\Xi}_{e}^{*}\otimes\mathbf{I}_{M})$ is an unitary matrix,
the inequality in \eqref{norminequalityaftr_a} becomes an equality.
\begin{lemma}\label{Lemma_KronckrNorm} For any two vectors $\mathbf{b}$
and $\mathbf{c}$, we have $\left\Vert \mathbf{b}\otimes\mathbf{c}\right\Vert =\left\Vert \mathbf{b}\right\Vert \left\Vert \mathbf{c}\right\Vert $.
In particular, we have $\left\Vert \mathbf{a}\otimes\mathbf{a}\right\Vert =\mathbf{a}^{H}\mathbf{a}$.
\end{lemma}

\begin{proof}[\textbf{Proof}]
Please refer to Appendix \ref{Appendix_C}.
\end{proof}

Thus, by utilizing the upper bounds in \eqref{norminequalitypre}
and \eqref{norminequalityaftr}, the constraint in \eqref{eq:BITOCRPMsimplfySOC}
can be approximated by
\begin{alignat}{1}
&\left\Vert\!\!\! \begin{array}{c}
\lambda_{\textrm{max}}(\mathbf{\Sigma}_{ge,k}^{1/2T}\otimes\mathbf{\Sigma}_{ge,k}^{1/2})M\textrm{vec}(\mathbf{\Xi}_{e})\\
\sqrt{2M}\lambda_{\textrm{max}}(\mathbf{\Sigma}_{ge,k}^{1/2}(\mathbf{\Xi}_{e}^{*}\otimes\mathbf{I}_{M}))(\mathbf{h}_{ae,k}^{*}\!+\!\mathbf{\bar{\mathbf{G}}}_{ce,k}^{T}\bm{\phi})
\end{array}\!\!\!\right\Vert \!\!  \leq \!x_{k}.\label{FindfaiSOCconvx}
\end{alignat}
where $\eqref{eq:BITOCRPMsimplfySOC}\Leftarrow\eqref{FindfaiSOCconvx}$, and the constraint \eqref{FindfaiSOCconvx} is convex. When both the
$(\mathbf{\Sigma}_{ge,k}^{1/2T}\otimes\mathbf{\Sigma}_{ge,k}^{1/2})$
and the $\mathbf{\Sigma}_{ge,k}^{1/2}(\mathbf{\Xi}_{e}^{*}\otimes\mathbf{I}_{M})$
are unitary matrices, the inequality in \eqref{FindfaiSOCconvx} becomes an equality.

\textbf{Step 3}: Transform the constraint of \eqref{eq:BITOCRPMsimplfyeig}
to be convex.

Since $\beta\geq1$, we note that $c_{g}=\lambda_{\textrm{max}}\{(\beta-1)\mathbf{Z}\}\geq0$.
By substituting $\mathbf{\Xi}_{e}=(\beta-1)\mathbf{Z}-\mathbf{W}$
into \eqref{eq:BITOCRPMsimplfyeig}, and adding $\mathbf{\Sigma}_{ge,k}^{1/2}((c_{g}\mathbf{I}_{N_{t}})^{T}\otimes\mathbf{E})\mathbf{\Sigma}_{ge,k}^{1/2}$
on both sides of \eqref{eq:BITOCRPMsimplfyeig}, we have \begin{subequations}
\begin{alignat}{1}
\!\!\!\!\!\eqref{eq:BITOCRPMsimplfyeig}\!\Leftrightarrow& y_{k}\mathbf{I}_{N_{t}M}+\mathbf{\Sigma}_{ge,k}^{1/2}((c_{g}\mathbf{I}_{N_{t}})^{T}\otimes\mathbf{E})\mathbf{\Sigma}_{ge,k}^{1/2}\succeq  \mathbf{\Sigma}_{ge,k}^{1/2}(\mathbf{W}^{T}\otimes\mathbf{E})\mathbf{\Sigma}_{ge,k}^{1/2}\nonumber \\
 &\qquad \qquad \qquad \qquad \qquad \qquad \qquad \qquad +\mathbf{\Sigma}_{ge,k}^{1/2}[(c_{g}\mathbf{I}_{N_{t}}\!-\!(\beta\!-\!1)\mathbf{Z})^{T}\otimes\mathbf{E}]\mathbf{\Sigma}_{ge,k}^{1/2}\\
\Leftrightarrow& y_{k}\mathbf{I}_{N_{t}M}+\mathbf{\Sigma}_{ge,k}^{1/2}((c_{g}\mathbf{I}_{N_{t}})^{T}\otimes\mathbf{E})\mathbf{\Sigma}_{ge,k}^{1/2}\succeq \mathbf{\Sigma}_{ge,k}^{1/2}[(\mathbf{W}+\mathbf{Z}_{g})^{T}\otimes\mathbf{E}]\mathbf{\Sigma}_{ge,k}^{1/2}\\
\Leftrightarrow& y_{k}+\lambda_{\textrm{min}}\{\mathbf{\Sigma}_{ge,k}^{1/2}((c_{g}\mathbf{I}_{N_{t}})^{T}\otimes\mathbf{E})\mathbf{\Sigma}_{ge,k}^{1/2}\}\geq \! \lambda_{\textrm{max}}\{\mathbf{\Sigma}_{ge,k}^{1/2}(\mathbf{V}_{wgz}^{*}\!\otimes\!\bm{\phi}^{*})(\mathbf{V}_{wgz}^{T}\!\otimes\!\bm{\phi}^{T})\mathbf{\Sigma}_{ge,k}^{1/2}\}\\
\Leftrightarrow& y_{k}\overset{(j)}{\geq}  \left\Vert \mathbf{\Sigma}_{ge,k}^{1/2}(\mathbf{V}_{wgz}^{*}\otimes\bm{\phi}^{*})\right\Vert _{2}^{2},\label{FindfaiEIGconvx}
\end{alignat}
\end{subequations} where $\mathbf{Z}_{g}\triangleq c_{g}\mathbf{I}_{N_{t}}-(\beta-1)\mathbf{Z}$,
$\mathbf{W}+\mathbf{Z}_{g}\triangleq\mathbf{V}_{wgz}\mathbf{V}_{wgz}^{H}$,
$\mathbf{Z}_{g}\succeq\mathbf{0}$ and $\mathbf{W}+\mathbf{Z}_{g}\succeq\mathbf{0}$.
The step in $(j)$ is obtained because $\lambda_{\textrm{min}}\{\mathbf{\Sigma}_{ge,k}^{1/2}((c_{g}\mathbf{I}_{N_{t}})^{T}\otimes\mathbf{E})\mathbf{\Sigma}_{ge,k}^{1/2}\}=0$,
and the proof is given in Appendix \ref{Appendix_A}. The constraint
\eqref{FindfaiEIGconvx} is convex, and can be equivalently written
as
\begin{alignat}{1}
\eqref{eq:BITOCRPMsimplfyeig}\Leftrightarrow\left\Vert(\mathbf{\Sigma}_{ge,k}^{1/2}(\mathbf{V}_{wgz}^{*}\otimes\bm{\phi}^{*}))\right\Vert _{2}-\sqrt{y_{k}}\leq0,\label{FindfaiEIG_CVX}
\end{alignat}
which is convex.

\textbf{Step 4}: Transform the constraint of \eqref{Findfainonconvx_bob} to
be convex.

By substituting \eqref{BobChnnlextnd} into the data rate requirement
for Bob in \eqref{Findfainonconvx_bob}, and defining $\mathbf{\Xi}_{b}\triangleq(\gamma-1)\mathbf{Z}-\mathbf{W}$,
the constraint \eqref{Findfainonconvx_bob} can be recast as\begin{subequations}
\begin{alignat}{1}
\eqref{Findfainonconvx_bob}\Leftrightarrow & (\mathbf{h}_{ab}^{H}+\bm{\phi}^{T}\mathbf{G}_{cb})[\mathbf{W}-(\gamma-1)\mathbf{Z}](\mathbf{h}_{ab}+\mathbf{G}_{cb}^{H}\bm{\phi}^{*})\geq\tilde{\sigma}_{b}^{2}+\delta_{0},\\
\Leftrightarrow & \bm{\phi}^{T}\mathbf{G}_{cb}\mathbf{\Xi}_{b}\mathbf{G}_{cb}^{H}\bm{\phi}^{*}+2\textrm{Re}\{\bm{\phi}^{T}\mathbf{G}_{cb}\mathbf{\Xi}_{b}\mathbf{h}_{ab}\}+\mathbf{h}_{ab}^{H}\mathbf{\Xi}_{b}\mathbf{h}_{ab}+\tilde{\sigma}_{b}^{2}+\delta_{0}\leq0.\label{Bobdatarate}
\end{alignat}
\end{subequations} To build the convexity of \eqref{Bobdatarate},
it is reformulated as
\begin{alignat}{1}
\eqref{Findfainonconvx_bob}\Leftrightarrow & b(\bm{\phi})\leq0,\label{BobdatarateRelaxConvx}
\end{alignat}
where $c_{b}=\lambda_{\textrm{min}}\{\mathbf{G}_{cb}\mathbf{\Xi}_{b}\mathbf{G}_{cb}^{H}\}$ and
\begin{alignat}{1}
b(\bm{\phi})\triangleq & \bm{\phi}^{T}[\mathbf{G}_{cb}\mathbf{\Xi}_{b}\mathbf{G}_{cb}^{H}-c_{b}\mathbf{I}_{M}]\bm{\phi}^{*}+2\textrm{Re}\{\bm{\phi}^{T}\mathbf{G}_{cb}\mathbf{\Xi}_{b}\mathbf{h}_{ab}\}+\mathbf{h}_{ab}^{H}\mathbf{\Xi}_{b}\mathbf{h}_{ab}+\tilde{\sigma}_{b}^{2}+\delta_{0}+Mc_{b}.\label{BobdatarateRelaxConvxExprssn}
\end{alignat}

Finally, based on the mathematical manipulations of \textbf{Steps} \textbf{1}-\textbf{4} above, Problem \eqref{Findfainonconvx} is reformulated into \begin{subequations}\label{FindfaiConvx}
\begin{alignat}{1}
\underset{\bm{\phi},\bm{\delta},\mathbf{x},\mathbf{y}}{\textrm{max}} & \ \ \sum_{k=0}^K{{\delta}_k}\\
\textrm{s.t.}\;\, & \eqref{findfaiLMIconvx},\eqref{FindfaiSOCconvx},\eqref{FindfaiEIG_CVX},\eqref{BobdatarateRelaxConvx},\eqref{Findfainonconvx_delta},\label{FindfaiConvx_A}\\
 & \left|\phi_{m}\right|=1,m=1,\cdots,M.\label{FindfaiConvxunitmodulus}
\end{alignat}
\end{subequations}

Problem \eqref{FindfaiConvx} is nonconvex due to
the unit-modulus constraint. To tackle it, the penalty CCP algorithm
is leveraged. The constraint of \eqref{FindfaiConvxunitmodulus} is
equivalent to $\left|\phi_{m}\right|^{2}\leq1$ and $\left|\phi_{m}\right|^{2}\geq1$,
where the former is convex and the latter is not convex. By using the first-order Taylor expansion, the nonconvex constraint $\left|\phi_{m}\right|^{2}\geq1$
is transformed into an affine constraint. Then, Problem \eqref{FindfaiConvx} can be transformed into an iterative optimization process,
and the optimization problem at the $(n+1)$th iteration is \begin{subequations}\label{FindfaiConvxCCP}
\begin{alignat}{1}
&\underset{\bm{\phi},\bm{\delta},\mathbf{b},\mathbf{x},\mathbf{y}}{\textrm{max}}  \ \ \sum_{k=0}^K{{\delta}_k}-\lambda^{(n+1)}\sum_{l=1}^{2M}{{b}_l}\\
&\textrm{s.t.}\;\,  \eqref{findfaiLMIconvx},\eqref{FindfaiSOCconvx},\eqref{FindfaiEIG_CVX},\eqref{BobdatarateRelaxConvx},\eqref{Findfainonconvx_delta},\label{FindfaiConvxCCP_A}\\
 &\quad  \ \left|\phi_{m}^{(n)}\right|^{2}\!-\!2\textrm{Re}\{\phi_{m}^{*}\phi_{m}^{(n)}\}\leq b_{m}\!-\!1,m=1,\cdots,M,\label{FindfaiConvxCCPunitmodulus_apprx}\\
 &\quad  \ \left|\phi_{m}\right|^{2}\leq1+b_{M+m},m=1,\cdots,M,\label{FindfaiConvxCCPunitmodulus_noapprx}\\
 & \quad \ \mathbf{b}\succeq\mathbf{0},\label{FindfaiConvxCCPunitmodulus_bvctr}
\end{alignat}
\end{subequations} where $\bm{\phi}_{m}^{(n)}$ is the phase shift of last
iteration ($n$th iteration), $\mathbf{b}=[b_{1},\cdots,b_{2M}]^{T}$ are
the introduced slack variables, $\sum_{l=1}^{2M}{{b}_l}$
is added into the objective function as a penalty term, and $\lambda^{(n+1)}$
is employed as the regularization factor to control the feasibility
of the constraints.

\subsection{Scenario 2: Full CSI Errors}\label{OCR-Full}
\subsubsection{Problem Reformulation}
Considering the full
statistical CSI error model in \eqref{StatisticalCSI}, the outage probability constraints in \eqref{OCR_PM_Partialextnd_eve} for Eves' leaked data rate can
be extended to
\begin{align}
\textrm{Pr}_{\mathbf{g}_{he,k},\mathbf{g}_{ge,k}}&\left\{ \hat{\mathbf{h}}_{e,k}^{H}[\mathbf{W}-(\beta-1)\mathbf{Z}]\hat{\mathbf{h}}_{e,k}\leq(\beta-1)\sigma_{e,k}^{2}\right\} \geq1-\rho_{k},\forall k\in\mathcal{K}.\label{OCR_PM_Fullextnd_eve}
\end{align}

To apply Lemma \ref{Lemma_BIT}, we transform the probability constraints \eqref{OCR_PM_Fullextnd_eve} into the form of \eqref{Lemma1A} as follows. By substituting \eqref{eq:CascdChnnlerror} and \eqref{eq:DirctChnnlerror} into \eqref{OCR_PM_Fullextnd_eve}, it can be reformulated in \eqref{eq:evedatarate_FULL}.
\begin{subequations}\label{eq:evedatarate_FULL}
	\begin{align}
		& \textrm{Pr}_{\mathbf{g}_{he,k},\mathbf{g}_{ge,k}}\{[\mathbf{h}_{ae,k}^{H}+\triangle\mathbf{h}_{ae,k}^{H}+\bm{\phi}^{T}(\mathbf{\bar{\mathbf{G}}}_{ce,k}+\triangle\mathbf{G}_{ce,k})]\mathbf{\Xi}_{e}\nonumber \\
		&\qquad \qquad\qquad \qquad \qquad  \qquad [\mathbf{h}_{ae,k}^{H}+\triangle\mathbf{h}_{ae,k}^{H}+\bm{\phi}^{T}(\mathbf{\bar{\mathbf{G}}}_{ce,k}+\triangle\mathbf{G}_{ce,k})]^{H}+\tilde{\sigma}_{e,k}^{2}\geq0\}\label{eq:evedatarate_aFULL}\\
		 &= \textrm{Pr}_{\mathbf{g}_{he,k},\mathbf{g}_{ge,k}}\{\underset{f_{1,k}^{\textrm{full}}}{\underbrace{(\triangle\mathbf{h}_{ae,k}^{H}+\bm{\phi}^{T}\triangle\mathbf{G}_{ce,k})\mathbf{\Xi}_{e}(\triangle\mathbf{h}_{ae,k}+\triangle\mathbf{G}_{ce,k}^{H}\bm{\phi}^{*})}}\nonumber \\
		&\quad+2\textrm{Re}[\underset{f_{2,k}^{\textrm{full}}}{\underbrace{(\mathbf{h}_{ae,k}^{H}+\bm{\phi}^{T}\mathbf{\bar{\mathbf{G}}}_{ce,k})\mathbf{\Xi}_{e}(\triangle\mathbf{h}_{ae,k}+\triangle\mathbf{G}_{ce,k}^{H}\bm{\phi}^{*})}}]\nonumber \\
		&\quad +\underset{c_{k}^{\textrm{full}}}{\underbrace{(\mathbf{h}_{ae,k}^{H}+\bm{\phi}^{T}\mathbf{\bar{\mathbf{G}}}_{ce,k})\mathbf{\Xi}_{e}(\mathbf{h}_{ae,k}^{H}+\bm{\phi}^{T}\mathbf{\bar{\mathbf{G}}}_{ce,k})^{H}+\tilde{\sigma}_{e,k}^{2}}}\geq0\}.\label{eq:evedatarate_bFULL}
	\end{align}
\end{subequations}
%

In addition to the CSI error vector $\mathbf{g}_{ge,k}$ for the cascaded
channel, we denote the CSI error vector for the direct channel by
$\mathbf{g}_{he,k}\triangleq\triangle\mathbf{h}_{ae,k}=\mathbf{\Sigma}_{he,k}^{1/2}\mathbf{v}_{he,k}$, where $\mathbf{v}_{he,k}\sim\mathcal{CN}(\mathbf{0},\mathbf{I}_{N_{t}})$ and $\mathbf{\Sigma}_{he,k}=\mathbf{\Sigma}_{he,k}^{1/2}\mathbf{\Sigma}_{he,k}^{1/2}$.
Since $\mathbf{\Sigma}_{he,k}$ is a semidefinite matrix, we have
$(\mathbf{\Sigma}_{he,k}^{1/2})^{H}=\mathbf{\Sigma}_{he,k}^{1/2}$ and
$(\mathbf{\Sigma}_{he,k}^{1/2})^{T}=(\mathbf{\Sigma}_{he,k}^{1/2})^{*}$. By
invoking the identity $\textrm{Tr}(\mathbf{A}\mathbf{BCD})=\textrm{vec}^{T}(\mathbf{D})(\mathbf{A}\otimes\mathbf{C}^{T})\textrm{vec}(\mathbf{B}^{T})$,
the expression of $f_{2,k}^{\textrm{full}}$ in \eqref{eq:evedatarate_bFULL} can be rewritten as
\begin{alignat}{1}
f_{2,k}^{\textrm{full}}
=&(\mathbf{h}_{ae,k}^{H}+\bm{\phi}^{T}\mathbf{\bar{\mathbf{G}}}_{ce,k})\mathbf{\Xi}_{e}\mathbf{g}_{he,k}+\textrm{Tr}[\mathbf{\Xi}_{e}\triangle\mathbf{G}_{ce,k}^{H}\bm{\phi}^{*}(\mathbf{h}_{ae,k}^{H}+\bm{\phi}^{T}\mathbf{\bar{\mathbf{G}}}_{ce,k})]\nonumber \\
=&(\mathbf{h}_{ae,k}^{H}+\bm{\phi}^{T}\mathbf{\bar{\mathbf{G}}}_{ce,k})\mathbf{\Xi}_{e}\mathbf{g}_{he,k}+\textrm{vec}^{T}(\mathbf{h}_{ae,k}^{H}+\bm{\phi}^{T}\mathbf{\bar{\mathbf{G}}}_{ce,k})(\mathbf{\Xi}_{e}\otimes\bm{\phi}^{H})\textrm{vec}(\triangle\mathbf{G}_{ce,k}^{*})\nonumber \\
=&(\mathbf{h}_{ae,k}^{H}+\bm{\phi}^{T}\mathbf{\bar{\mathbf{G}}}_{ce,k})\mathbf{\Xi}_{e}\mathbf{\Sigma}_{he,k}^{1/2}\mathbf{v}_{he,k}+(\mathbf{h}_{ae,k}^{H}+\bm{\phi}^{T}\mathbf{\bar{\mathbf{G}}}_{ce,k})(\mathbf{\Xi}_{e}\otimes\bm{\phi}^{H})\mathbf{\Sigma}_{ge,k}^{1/2*}\mathbf{v}_{ge,k}^{*}\nonumber \\
 \triangleq& \tilde{\mathbf{u}}_{e,k}^{H}\tilde{\mathbf{v}}_{e,k},\label{fFull2exprssn}
\end{alignat}
where $\tilde{\mathbf{v}}_{e,k}=[\mathbf{v}_{he,k}^{H},\mathbf{v}_{ge,k}^{T}]^{H}$,
and
\begin{alignat}{1}
\tilde{\mathbf{u}}_{e,k} & =\left[\begin{array}{c}
\mathbf{\Sigma}_{he,k}^{1/2}\mathbf{\Xi}_{e}(\mathbf{h}_{ae,k}+\mathbf{\bar{\mathbf{G}}}_{ce,k}^{H}\bm{\phi}^{*})\\
\mathbf{\Sigma}_{ge,k}^{1/2T}(\mathbf{\Xi}_{e}\otimes\bm{\phi})(\mathbf{h}_{ae,k}+\mathbf{\bar{\mathbf{G}}}_{ce,k}^{H}\bm{\phi}^{*})
\end{array}\right] \overset{(k)}{=}\left[\begin{array}{c}
\mathbf{\Sigma}_{he,k}^{1/2}\mathbf{\Xi}_{e}(\mathbf{h}_{ae,k}+\mathbf{\bar{\mathbf{G}}}_{ce,k}^{H}\bm{\phi}^{*})\\
\mathbf{\Sigma}_{ge,k}^{1/2T}([\mathbf{\Xi}_{e}(\mathbf{h}_{ae,k}+\mathbf{\bar{\mathbf{G}}}_{ce,k}^{H}\bm{\phi}^{*})]\otimes(\mathbf{I}_{M}\bm{\phi}))
\end{array}\right]\nonumber \\
 & =\left[\!\!\begin{array}{c}
\mathbf{\Sigma}_{he,k}^{1/2}\mathbf{\Xi}_{e}(\mathbf{h}_{ae,k}\!+\!\mathbf{\bar{\mathbf{G}}}_{ce,k}^{H}\bm{\phi}^{*})\\
\mathbf{\Sigma}_{ge,k}^{1/2T}(\mathbf{\Xi}_{e}\otimes\mathbf{I}_{M})((\mathbf{h}_{ae,k}\!+\!\mathbf{\bar{\mathbf{G}}}_{ce,k}^{H}\bm{\phi}^{*})\otimes\bm{\phi})
\end{array}\!\!\right].\label{uekExprssn}
\end{alignat}

The the step in $(k)$ in \eqref{uekExprssn} is obtained by invoking the property
$(\mathbf{A}\otimes\mathbf{b})\mathbf{c}=(\mathbf{A}\mathbf{c})\otimes\mathbf{b}$, when $\mathbf{b}$ and $\mathbf{c}$ are column vectors.
The expression of $f_{1,k}^{\textrm{full}}$ can be rewritten as
\begin{alignat}{1}
&f_{1,k}^{\textrm{full}}
  =\triangle\mathbf{h}_{ae,k}^{H}\mathbf{\Xi}_{e}\triangle\mathbf{h}_{ae,k}+\textrm{Tr}(\mathbf{\Xi}_{e}\triangle\mathbf{G}_{ce,k}^{H}\mathbf{E}\triangle\mathbf{G}_{ce,k})+2\textrm{Re}\{\textrm{Tr}(\mathbf{\Xi}_{e}\triangle\mathbf{G}_{ce,k}^{H}\bm{\phi}^{*}\triangle\mathbf{h}_{ae,k}^{H})\}\nonumber \\
  &\overset{(l)}{=}\triangle\mathbf{h}_{ae,k}^{H}\mathbf{\Xi}_{e}\triangle\mathbf{h}_{ae,k}+\textrm{vec}^{T}(\triangle\mathbf{G}_{ce,k})(\mathbf{\Xi}_{e}\otimes\mathbf{E}^{T})\textrm{vec}(\triangle\mathbf{G}_{ce,k}^{*})\nonumber \\
 & +2\textrm{Re}\{\textrm{vec}^{T}(\triangle\mathbf{h}_{ae,k}^{H})(\mathbf{\Xi}_{e}\otimes\bm{\phi}^{H})\textrm{vec}(\triangle\mathbf{G}_{ce,k}^{*})\}\nonumber \\
  &=\!\mathbf{v}_{he,k}^{H}\mathbf{\Sigma}_{he,k}^{1/2}\mathbf{\Xi}_{e}\mathbf{\Sigma}_{he,k}^{1/2}\mathbf{v}_{he,k}\!\!+\!\!\mathbf{v}_{ge,k}^{T}\mathbf{\Sigma}_{ge,k}^{1/2T}(\mathbf{\Xi}_{e}\!\otimes\!\mathbf{E}^{T})\mathbf{\Sigma}_{ge,k}^{1/2*}\mathbf{v}_{ge,k}^{*} \!\!+\!\!2\textrm{Re}\{\!\mathbf{v}_{he,k}^{H}\mathbf{\Sigma}_{he,k}^{1/2}(\mathbf{\Xi}_{e}\!\otimes\!\bm{\phi}^{H})\mathbf{\Sigma}_{ge,k}^{1/2*}\mathbf{v}_{ge,k}^{*}\!\}\nonumber \\
  &\triangleq\tilde{\mathbf{v}}_{e,k}^{H}\tilde{\mathbf{A}}_{e,k}\tilde{\mathbf{v}}_{e,k},\label{fFull1exprssn}
\end{alignat}
where $(l)$ is obtained by invoking $\textrm{Tr}(\mathbf{A}\mathbf{BCD})=\textrm{vec}^{T}(\mathbf{D})(\mathbf{A}\otimes\mathbf{C}^{T})\textrm{vec}(\mathbf{B}^{T})$,
and
\begin{alignat}{1}
\!\!\!\!\!\tilde{\mathbf{A}}_{e,k} &\!\! \triangleq\!\!\left[\!\!\begin{array}{cc}\!\!
\mathbf{\Sigma}_{he,k}^{1/2}\mathbf{\Xi}_{e}\mathbf{\Sigma}_{he,k}^{1/2} \!\!& \!\!\mathbf{\Sigma}_{he,k}^{1/2}(\mathbf{\Xi}_{e}\otimes\bm{\phi}^{H})\mathbf{\Sigma}_{ge,k}^{1/2*}\\
\!\!\mathbf{\Sigma}_{ge,k}^{1/2T}(\mathbf{\Xi}_{e}\otimes\bm{\phi})\mathbf{\Sigma}_{he,k}^{1/2} \!\!& \!\!\mathbf{\Sigma}_{ge,k}^{1/2T}(\mathbf{\Xi}_{e}\otimes\mathbf{E}^{T})\mathbf{\Sigma}_{ge,k}^{1/2*}
\end{array}\!\!\!\!\right].\label{AekExprssn}
\end{alignat}

Then, the outage probability constraint \eqref{eq:evedatarate_bFULL} is recast
as
\begin{alignat}{1}
\textrm{Pr}_{\mathbf{g}_{he,k},\mathbf{g}_{ge,k}}&\left\{ \tilde{\mathbf{v}}_{e,k}^{H}\tilde{\mathbf{A}}_{e,k}\tilde{\mathbf{v}}_{e,k}+2\textrm{Re}\{\tilde{\mathbf{u}}_{e,k}^{H}\tilde{\mathbf{v}}_{e,k}\}+c_{k}^{\textrm{full}}\geq0\right\} \geq1-\rho_{k},\forall k\in\mathcal{K},\label{OCR_PMfull_extnd_eve}
\end{alignat}

It is seen from \eqref{OCR_PMfull_extnd_eve} that Lemma \ref{Lemma_BIT} can be applied as long as the CSI error $\mathbf{g}_{ge,k}$ and $\mathbf{g}_{he,k}$ follow the CSCG distribution and regardless of the correlation among their elements. By leveraging Lemma \ref{Lemma_BIT} again, the chance constraints in \eqref{OCR_PM_Fullextnd_eve} are approximated by computable constraints, which results the equivalent OCR-PM problem with full CSI errors as \begin{subequations}\label{BITOCRPMfull}
\begin{alignat}{1}
\underset{\mathbf{W},\mathbf{Z},{\rm \bm{\phi}},\mathbf{\tilde{A}},\mathbf{\tilde{x}},\mathbf{\tilde{y}}}{\min} & {\rm Tr(}{\bf W}{\rm +}{\bf Z}{\rm )}\label{eq:BITOCRPMfull_a}\\
\textrm{s.t.}\quad & \textrm{Tr}(\tilde{\mathbf{A}}_{e,k})-\sqrt{-2\ln(\rho_{k})}\cdot\tilde{x}_{k}+\ln(\rho_{k})\cdot\tilde{y}_{k}+c_{k}^{\textrm{full}}\geq0,\label{eq:BITOCRPMfull_LMI}\\
 & \left\Vert \left[\begin{array}{c}
\textrm{vec}(\mathbf{\tilde{A}}_{e,k})\\
\sqrt{2}\tilde{\mathbf{u}}_{e,k}
\end{array}\right]\right\Vert _{2}\leq\tilde{x}_{k},\label{eq:BITOCRPMfull_SOCP}\\
 & \tilde{y}_{k}\mathbf{I}_{MN_{t}+N_{t}}+\mathbf{\tilde{A}}_{k}\succeq\mathbf{0},\tilde{y}_{k}\geq0,\label{eq:BITOCRPMfull_EIG}\\
 & \eqref{OCR_PM_Partialextnd_bob},\eqref{OCR_PM_Partialorgn_Z0},\eqref{OCR_PM_Partialorgn_W0},\eqref{OCR_PM_Partialorgn_Wrank},\eqref{OCR_PM_Partialorgn_fai1},
\end{alignat}
\end{subequations} where $\mathbf{\tilde{A}}_{e}=[\mathbf{\tilde{A}}_{e,1},\mathbf{\tilde{A}}_{e,2},\cdots,\mathbf{\tilde{A}}_{e,K}]$, and $\tilde{\mathbf{x}}=[\tilde{x}_{1},\tilde{x}_{2},\cdots,\tilde{x}_{K}]^{T}$
and $\tilde{\mathbf{y}}=[\tilde{y}_{1},\tilde{y}_{2},\cdots,\tilde{y}_{K}]^{T}$ are introduced variables.
\subsubsection{Optimization of Transmit Beamforming and AN}
The AO method is also utilized to decouple the variables. When ${\rm \bm{\phi}}$
is fixed, the $\tilde{\mathbf{A}}_{e,k}$ and $\tilde{\mathbf{u}}_{e,k}$
are linear functions of the ${\bf W}$ and ${\bf Z}$. Then, the ${\bf W}$
and ${\bf Z}$ can be obtained by the SDR technique, and the $\{{\bf W}^{(n)},\mathbf{Z}^{(n)}\}$
obtained at the $n$th iteration are given as \begin{subequations}\label{OCR_PMfull_WZ}
\begin{alignat}{1}
\{{\bf W}^{(n)},\mathbf{Z}^{(n)}\} & =\arg\underset{\mathbf{W},\mathbf{Z},\mathbf{x},\mathbf{y}}{\min}{\rm Tr(}{\bf W}{\rm +}{\bf Z}{\rm )}\\
  \textrm{s.t.}&\ \eqref{eq:BITOCRPMfull_LMI},\eqref{eq:BITOCRPMfull_SOCP},\eqref{eq:BITOCRPMfull_EIG},\eqref{OCR_PM_Partialextnd_bob},\eqref{OCR_PM_Partialorgn_Z0},\eqref{OCR_PM_Partialorgn_W0}.
\end{alignat}
\end{subequations}
\subsubsection{Optimization of Phase Shifts at the IRS} \label{OCR-Full-Optphi}
When ${\bf W}$ and ${\bf Z}$ are fixed, the optimization problem for $\bm{\phi}$ becomes the feasibility check problem. The slack variables $\mathbf{\bm{\delta}}=[\delta_{0},\delta_{1},\delta_{2},\cdots,\delta_{K}]^{T}$
are introduced to improve the convergence of the AO algorithm. Due to the existence of direct communication link, the optimization variable can only be $\bm{\phi}$ instead of $\mathbf{E}$, and all constraints are nonconvex w.r.t. $\bm{\phi}$. Thus, we propose novel mathematical manipulations to transform the constraints of Problem \eqref{BITOCRPMfull} into convex forms in the following three steps. This manipulations are different but inherited from those in the scenario of partial CSI errors.

\textbf{Step 1}: Transform the constraint in \eqref{eq:BITOCRPMfull_LMI} to
be convex.

We find that $c_{k}^{\textrm{full}}=c_{k}^{\textrm{partial}}$, thus
the process on $c_{k}^{\textrm{full}}$ is the same as the process
on $c_{k}^{\textrm{partial}}$ under the partial CSI error.
Then we have $c_{k}^{\textrm{full}}=\eqref{ck_partial_Concave}$,
the expression of which is concave w.r.t. $\bm{\phi}$. The $\textrm{Tr}[\tilde{\mathbf{A}}_{e,k}]$
are transformed as
\begin{alignat}{1}
\textrm{Tr}[\tilde{\mathbf{A}}_{e,k}]  &=\textrm{Tr}\text{[\ensuremath{\mathbf{\Sigma}_{he,k}^{1/2}\mathbf{\Xi}_{e}\mathbf{\Sigma}_{he,k}^{1/2}}]}\!+\!\left\{\! \textrm{Tr}[\mathbf{\Sigma}_{ge,k}^{1/2T}(\mathbf{\Xi}_{e}\!\otimes\!\mathbf{E}^{T})\mathbf{\Sigma}_{ge,k}^{1/2*}]\!\right\} ^{*}\nonumber \\
& \overset{(m)}{=} \textrm{Tr}\text{[\ensuremath{\mathbf{\Sigma}_{he,k}^{1/2}\mathbf{\Xi}_{e}\mathbf{\Sigma}_{he,k}^{1/2}}]}+\textrm{Tr}[\mathbf{\Sigma}_{ge,k}^{1/2H}(\mathbf{\Xi}_{e}^{T}\otimes\mathbf{E})\mathbf{\Sigma}_{ge,k}^{1/2}]\nonumber \\
&  \overset{(n)}{=}\!\textrm{Tr}\text{[\ensuremath{\mathbf{\Sigma}_{he,k}^{1/2}\mathbf{\Xi}_{e}\mathbf{\Sigma}_{he,k}^{1/2}}]}\!+\!\bm{\phi}^{T}\!\mathbf{\Sigma}_{gx,k}(\mathbf{I}_{MN_{t}}\!\otimes\!\mathbf{\Xi}_{e})\mathbf{\Sigma}_{gx,k}^{H}\bm{\phi}^{*},\label{eq:TrAkquadraticFull}
\end{alignat}
where the second term of $(m)$ is the same as the $\textrm{Tr}[\mathbf{A}_{e,k}]$
under the partial CSI error, and the second term of $(n)$
is obtained in the same way of \eqref{eq:TrAkquadratic}. The second
term in $(n)$ of \eqref{eq:TrAkquadraticFull} is equivalently
transformed to be concave w.r.t. $\bm{\phi}$ as in \eqref{eq:TrAkConcave}.
By substituting \eqref{eq:TrAkConcave} into \eqref{eq:TrAkquadraticFull},
we reformulate \eqref{eq:TrAkquadraticFull} into a concave form as
\begin{alignat}{1}
&\textrm{Tr}[\tilde{\mathbf{A}}_{e,k}]  =\textrm{Tr}\text{[\ensuremath{\mathbf{\Sigma}_{he,k}^{1/2}\mathbf{\Xi}_{e}\mathbf{\Sigma}_{he,k}^{1/2}}]}+\bm{\phi}^{T}[\mathbf{\Sigma}_{gx,k}(\mathbf{I}_{MN_{t}}\otimes\mathbf{\Xi}_{e})\mathbf{\Sigma}_{gx,k}^{H}\!-\!c_{t,k}\mathbf{I}_{M}]\bm{\phi}^{*}\!+\!Mc_{t,k}.\label{eq:TrAkConcaveFull}
\end{alignat}

By substituting \eqref{ck_partial_Concave} and \eqref{eq:TrAkConcaveFull}
into the constraint in \eqref{eq:BITOCRPMfull_LMI}, we can obtain a
convex constraint equivalent to \eqref{eq:BITOCRPMfull_LMI} as
\begin{alignat}{1}
\eqref{eq:BITOCRPMfull_LMI}\Leftrightarrow & c_{tc,k}^{\textrm{full}}(\bm{\phi})\geq0,\label{findfaiFullLMIconvx}
\end{alignat}
which is convex, and
\begin{alignat}{1}
c_{tc,k}^{\textrm{full}}(\bm{\phi})  \triangleq&\textrm{Tr}\text{[\ensuremath{\mathbf{\Sigma}_{he,k}^{1/2}\mathbf{\Xi}_{e}\mathbf{\Sigma}_{he,k}^{1/2}}]}\!+\!\bm{\phi}^{T}[\mathbf{\Sigma}_{gx,k}(\mathbf{I}_{MN_{t}}\otimes\mathbf{\Xi}_{e})\mathbf{\Sigma}_{gx,k}^{H}\!-\!c_{t,k}\mathbf{I}_{M}]\bm{\phi}^{*}+Mc_{t,k}\!\!-\!\!\sqrt{\!-2\ln(\rho_{k})}\cdot x_{k}\nonumber \\
 &+\ln(\rho_{k})\cdot y_{k}+\bm{\phi}^{T}(\mathbf{\bar{\mathbf{G}}}_{ce,k}\mathbf{\Xi}_{e}\mathbf{\bar{\mathbf{G}}}_{ce,k}^{H}\!-\!c_{c,k}\mathbf{I}_{M})\bm{\phi}^{*}\!+\!2\textrm{Re}\{\bm{\phi}^{T}\mathbf{\bar{\mathbf{G}}}_{ce,k}\mathbf{\Xi}_{e}\mathbf{h}_{ae,k}\}\nonumber \\
 & +\mathbf{h}_{ae,k}^{H}\mathbf{\Xi}_{e}\mathbf{h}_{ae,k}+Mc_{c,k}+\tilde{\sigma}_{e,k}^{2}-\delta_{k}.\label{findfaiFullLMIconvxExprssn}
\end{alignat}

\vspace{-0.2cm}\textbf{Step2}: Transform the constraint in \eqref{eq:BITOCRPMfull_SOCP} to
be convex.

We can find an upper bound of the left hand side of constraint \eqref{eq:BITOCRPMfull_SOCP}
as follows:
\vspace{-0.1cm}\begin{alignat}{1}
&\left\Vert \textrm{vec}(\tilde{\mathbf{A}}_{e,k})\right\Vert ^{2}  =\left\Vert \mathbf{\Sigma}_{he,k}^{1/2}\mathbf{\Xi}_{e}\mathbf{\Sigma}_{he,k}^{1/2}\right\Vert ^{2}\!+\!2\left\Vert \mathbf{\Sigma}_{ge,k}^{1/2T}(\mathbf{\Xi}_{e}\otimes\bm{\phi})\mathbf{\Sigma}_{he,k}^{1/2}\right\Vert ^{2}\!+\!\left\Vert \mathbf{\Sigma}_{ge,k}^{1/2T}(\mathbf{\Xi}_{e}\otimes\mathbf{E}^{T})\mathbf{\Sigma}_{ge,k}^{1/2*}\right\Vert ^{2}\nonumber \\
 & \overset{(o)}{=}\left\Vert (\mathbf{\Sigma}_{he,k}^{1/2T}\otimes\mathbf{\Sigma}_{he,k}^{1/2})\textrm{vec}(\mathbf{\Xi}_{e})\right\Vert ^{2}+2\left\Vert (\mathbf{\Sigma}_{he,k}^{1/2T}\otimes\mathbf{\Sigma}_{ge,k}^{1/2T})\textrm{vec}(\mathbf{\Xi}_{e}\otimes\bm{\phi})\right\Vert ^{2}\nonumber \\
 & \quad+\left\Vert (\mathbf{\Sigma}_{ge,k}^{1/2}\otimes\mathbf{\Sigma}_{ge,k}^{1/2T})\textrm{vec}(\mathbf{\Xi}_{e}\otimes\mathbf{E}^{T})\right\Vert ^{2} \label{stepo}\\
\Rightarrow & \overset{(p)}{\leq}\sigma_{\textrm{max}}^{2}(\mathbf{\Sigma}_{he,k}^{1/2T}\!\otimes\!\mathbf{\Sigma}_{he,k}^{1/2})\!\left\Vert \mathbf{\Xi}_{e}\right\Vert _{F}^{2}\!+\!2\sigma_{\textrm{max}}^{2}(\mathbf{\Sigma}_{he,k}^{1/2T}\!\otimes\!\mathbf{\Sigma}_{ge,k}^{1/2T})M\!\left\Vert \mathbf{\Xi}_{e}\right\Vert _{F}^{2}\!+\!\sigma_{\textrm{max}}^{2}(\mathbf{\Sigma}_{ge,k}^{1/2}\!\otimes\!\mathbf{\Sigma}_{ge,k}^{1/2T})M^{2}\!\left\Vert \mathbf{\Xi}_{e}\right\Vert _{F}^{2}\nonumber \\
 & \overset{(q)}{=}[\lambda_{\textrm{max}}(\mathbf{\Sigma}_{he,k})+\lambda_{\textrm{max}}(\mathbf{\Sigma}_{ge,k})M]^{2}\left\Vert \mathbf{\Xi}_{e}\right\Vert _{F}^{2}\label{Fullnorminequalitypre}
\end{alignat}
\begin{alignat}{1}
&\left\Vert \tilde{\mathbf{u}}_{e,k}\right\Vert ^{2}  =\left\Vert \begin{array}{c}
\mathbf{\Sigma}_{he,k}^{1/2}\mathbf{\Xi}_{e}(\mathbf{h}_{ae,k}+\mathbf{\bar{\mathbf{G}}}_{ce,k}^{H}\bm{\phi}^{*})\\
\mathbf{\Sigma}_{ge,k}^{1/2T}(\mathbf{\Xi}_{e}\otimes\mathbf{I}_{M})((\mathbf{h}_{ae,k}+\mathbf{\bar{\mathbf{G}}}_{ce,k}^{H}\bm{\phi}^{*})\otimes\bm{\phi})
\end{array}\right\Vert ^{2}\nonumber \\
 & =\left\Vert \mathbf{\Sigma}_{he,k}^{1/2}\mathbf{\Xi}_{e}(\mathbf{h}_{ae,k}+\mathbf{\bar{\mathbf{G}}}_{ce,k}^{H}\bm{\phi}^{*})\right\Vert ^{2}+\left\Vert \mathbf{\Sigma}_{ge,k}^{1/2T}(\mathbf{\Xi}_{e}\otimes\mathbf{I}_{M})((\mathbf{h}_{ae,k}+\mathbf{\bar{\mathbf{G}}}_{ce,k}^{H}\bm{\phi}^{*})\otimes\bm{\phi})\right\Vert ^{2}\nonumber \\
 & \overset{(r)}{\leq}\sigma_{\textrm{max}}^{2}(\mathbf{\Sigma}_{he,k}^{1/2}\mathbf{\Xi}_{e})\left\Vert \mathbf{h}_{ae,k}+\mathbf{\bar{\mathbf{G}}}_{ce,k}^{H}\bm{\phi}^{*}\right\Vert ^{2}+\sigma_{\textrm{max}}^{2}(\mathbf{\Sigma}_{ge,k}^{1/2T}(\mathbf{\Xi}_{e}\otimes\mathbf{I}_{M}))M\left\Vert \mathbf{h}_{ae,k}+\mathbf{\bar{\mathbf{G}}}_{ce,k}^{H}\bm{\phi}^{*}\right\Vert ^{2}\nonumber \\
 & =[\lambda_{\textrm{max}}^{2}(\mathbf{\Sigma}_{he,k}^{1/2}\mathbf{\Xi}_{e})+\lambda_{\textrm{max}}^{2}(\mathbf{\Sigma}_{ge,k}^{1/2T}(\mathbf{\Xi}_{e}\otimes\mathbf{I}_{M}))M]\cdot\left\Vert \mathbf{h}_{ae,k}+\mathbf{\bar{\mathbf{G}}}_{ce,k}^{H}\bm{\phi}^{*}\right\Vert ^{2},\label{Fullnorminequalityaftr}
\end{alignat}
where the step in $(o)$ is obtained by using Lemma \ref{Lemma_VectorNorm} as follows.
\begin{lemma}\label{Lemma_VectorNorm}
For any three matrices $\mathbf{O}$, $\mathbf{P}$, $\mathbf{Q}$,
the equality $\left\Vert \mathbf{P}\mathbf{O}\mathbf{Q}\right\Vert ^{2} =\left\Vert (\mathbf{Q}^{T}\otimes\mathbf{P})\textrm{vec}(\mathbf{O})\right\Vert ^{2}$ holds.
\end{lemma}
\begin{proof}[\textbf{Proof}]
Please refer to Appendix \ref{Appendix_B}.
\end{proof}
The $(p)$ is obtained by using $\left\Vert \mathbf{A}\mathbf{x}\right\Vert _{2}\leq\left\Vert \mathbf{A}\right\Vert _{2}\left\Vert \mathbf{x}\right\Vert _{2}$,
$\left\Vert \mathbf{A}\right\Vert _{2}=\sigma_{\max}\{\mathbf{A}\}$
and the property in \eqref{vecKronMproprty}. The $(q)$ is obtained
due to $\textrm{eig}(\mathbf{A}\otimes\mathbf{B})=\textrm{eig}(\mathbf{A})\otimes\textrm{eig}(\mathbf{B})$
and $\sigma_{\max}\{\mathbf{A}\}=\lambda_{\max}\{\mathbf{A}\}$ when
$\mathbf{A}\in\mathbb{H}^{n}$. The $(r)$ is obtained due to Lemma \ref{Lemma_KronckrNorm}.
By using the upper bounds in \eqref{Fullnorminequalitypre} and
\eqref{Fullnorminequalityaftr}, we obtain the convex constraint \eqref{FindfaiFullSOCconvx} instead of the original constraint \eqref{eq:BITOCRPMfull_SOCP} as
\begin{alignat}{1}
	\left\Vert \begin{array}{c}
		[\lambda_{\textrm{max}}(\mathbf{\Sigma}_{he,k})+\lambda_{\textrm{max}}(\mathbf{\Sigma}_{ge,k})M]\textrm{vec}(\mathbf{\Xi}_{e})\\
		\sqrt{2[\lambda_{\textrm{max}}^{2}(\mathbf{\Sigma}_{he,k}^{1/2}\mathbf{\Xi}_{e})+\lambda_{\textrm{max}}^{2}(\mathbf{\Sigma}_{ge,k}^{1/2T}(\mathbf{\Xi}_{e}\otimes\mathbf{I}_{M}))M]}(\mathbf{h}_{ae,k}+\mathbf{\bar{\mathbf{G}}}_{ce,k}^{H}\bm{\phi})
	\end{array}\right\Vert  & \leq\tilde{x}_{k}.\label{FindfaiFullSOCconvx}
\end{alignat}
It is readily to check that when \eqref{FindfaiFullSOCconvx} holds, the \eqref{eq:BITOCRPMfull_SOCP} always holds.

\textbf{Step3}: Transform the constraint in \eqref{eq:BITOCRPMfull_EIG} to
be convex.

The transformation is similar as the process in \eqref{FindfaiEIGconvx}.
By moving $\mathbf{\tilde{A}}_{k}$ in \eqref{eq:BITOCRPMfull_EIG}
from the left hand side to the right hand side of the inequality, we have
\vspace{-0.1cm}\begin{equation}
\tilde{y}_{k}\mathbf{I}_{MN_{t}+N_{t}}\succeq-\mathbf{\tilde{A}}_{k}.\label{eq:BITOCRPMfull_EIG_MOVE}
\end{equation}

Then, we define a matrix $\mathbf{B}_{e,k}$ as
\vspace{-0.2cm}\begin{align}
\!\!\!\!\mathbf{B}_{e,k}\!\!\triangleq\!\!\!\left[\!\!\begin{array}{cc}
\!\!\mathbf{\Sigma}_{he,k}^{1/2}\!(\!c_{g}\mathbf{I}_{N_{t}}\!)\mathbf{\Sigma}_{he,k}^{1/2}\!\!\!\!\!\! & \mathbf{\Sigma}_{he,k}^{1/2}\!(\!c_{g}\mathbf{I}_{N_{t}}\!\!\otimes\!\bm{\phi}^{H}\!)\mathbf{\Sigma}_{ge,k}^{1/2*}\!\!\\
\!\!\mathbf{\Sigma}_{ge,k}^{1/2T}\!(\!c_{g}\mathbf{I}_{N_{t}}\!\!\otimes\!\bm{\phi}\!)\mathbf{\Sigma}_{he,k}^{1/2}\!\!\!\!\!\! & \mathbf{\Sigma}_{ge,k}^{1/2T}\!(\!c_{g}\mathbf{I}_{N_{t}}\!\!\otimes\!\mathbf{E}^{T}\!)\mathbf{\Sigma}_{ge,k}^{1/2*}\!\!
\end{array}\!\!\!\right].
\end{align}

By adding the matrix $\mathbf{B}_{e,k}$ on both sides of the inequality
in \eqref{eq:BITOCRPMfull_EIG_MOVE}, we have the following equivalent inequality in \eqref{FindfaiFullEIGconvx}.
\begin{alignat}{1}
	y_{k}\mathbf{I}_{(N_{t}M+N_{t})}+\mathbf{B}_{e,k}\succeq & \left[\begin{array}{cc}
		\mathbf{\Sigma}_{he,k}^{1/2}(c_{g}\mathbf{I}_{N_{t}})\mathbf{\Sigma}_{he,k}^{1/2} & \mathbf{\Sigma}_{he,k}^{1/2}(c_{g}\mathbf{I}_{N_{t}}\otimes\bm{\phi}^{H})\mathbf{\Sigma}_{ge,k}^{1/2*}\\
		\mathbf{\Sigma}_{ge,k}^{1/2T}(c_{g}\mathbf{I}_{N_{t}}\otimes\bm{\phi})\mathbf{\Sigma}_{he,k}^{1/2} & \mathbf{\Sigma}_{ge,k}^{1/2T}(c_{g}\mathbf{I}_{N_{t}}\otimes\mathbf{E}^{T})\mathbf{\Sigma}_{ge,k}^{1/2*}
	\end{array}\right]-\mathbf{\tilde{A}}_{k}\nonumber \\
	\Leftrightarrow y_{k}\mathbf{I}_{(N_{t}M+N_{t})}+\mathbf{B}_{e,k}\succeq & \left[\begin{array}{cc}
		\mathbf{\Sigma}_{he,k}^{1/2}(\mathbf{V}_{wgz}\mathbf{V}_{wgz}^{H})\mathbf{\Sigma}_{he,k}^{1/2} & \mathbf{\Sigma}_{he,k}^{1/2}((\mathbf{V}_{wgz}\mathbf{V}_{wgz}^{H})\otimes\bm{\phi}^{H})\mathbf{\Sigma}_{ge,k}^{1/2*}\\
		\mathbf{\Sigma}_{ge,k}^{1/2T}((\mathbf{V}_{wgz}\mathbf{V}_{wgz}^{H})\otimes\bm{\phi})\mathbf{\Sigma}_{he,k}^{1/2} & \mathbf{\Sigma}_{ge,k}^{1/2T}((\mathbf{V}_{wgz}\mathbf{V}_{wgz}^{H})\otimes\mathbf{E}^{T})\mathbf{\Sigma}_{ge,k}^{1/2*}
	\end{array}\right]\nonumber \\
	\Leftrightarrow y_{k}\mathbf{I}_{(N_{t}M+N_{t})}+\mathbf{B}_{e,k}\succeq & \left[\begin{array}{c}
		\mathbf{\Sigma}_{he,k}^{1/2}\mathbf{V}_{wgz}\\
		\mathbf{\Sigma}_{ge,k}^{1/2T}(\mathbf{V}_{wgz}\otimes\bm{\phi})
	\end{array}\right]\left[\begin{array}{cc}
		\mathbf{V}_{wgz}^{H}\mathbf{\Sigma}_{he,k}^{1/2} & \;(\mathbf{V}_{wgz}^{H}\otimes\bm{\phi}^{H})\mathbf{\Sigma}_{ge,k}^{1/2*}\end{array}\right].\label{FindfaiFullEIGconvx}
\end{alignat}
Similarly, we find that $\lambda_{\textrm{min}}\{\mathbf{B}_{e,k}\}=0$,
which is proved in Appendix \ref{Appendix_D}. Thus, we have the convex
linear matrix inequality (LMI) constraint as
\begin{alignat}{1}
\eqref{eq:BITOCRPMfull_EIG}\Leftrightarrow & \sqrt{\tilde{y}_{k}}\overset{(s)}{\geq}\left\Vert \begin{array}{c}
\mathbf{\Sigma}_{he,k}^{1/2}\mathbf{V}_{wgz}\\
\mathbf{\Sigma}_{ge,k}^{1/2T}(\mathbf{V}_{wgz}\otimes\bm{\phi})
\end{array}\right\Vert _{2},\tilde{y}_{k}\geq0,\label{FindfaiFullEIG_CVX}
\end{alignat}
where $(s)$ is obtained due to $\left\Vert \mathbf{A}\right\Vert _{2}=\sqrt{\max\textrm{eig}(\mathbf{A}^{H}\mathbf{A})}$.

Finally, based on the mathematical manipulations of \textbf{Steps 1}-\textbf{3} above and
by using the CCP method, the problem for optimizing $\bm{\phi}$ can be reformulated as \vspace{-0.2cm}\begin{subequations}\label{FindfaiConvxCCPFull}
\begin{alignat}{1}
\underset{\bm{\phi},\bm{\delta},\mathbf{x},\mathbf{y}}{\textrm{max}} & \ \ \sum_{k=0}^K{{\delta}_k}-\lambda^{(n+1)}\sum_{l=1}^{2M}{{b}_l}\\
\textrm{s.t.}\;\, & \eqref{findfaiFullLMIconvx},\eqref{FindfaiFullSOCconvx},\eqref{FindfaiFullEIG_CVX},\eqref{BobdatarateRelaxConvx},\eqref{Findfainonconvx_delta}, \eqref{FindfaiConvxCCPunitmodulus_apprx},\eqref{FindfaiConvxCCPunitmodulus_noapprx},\eqref{FindfaiConvxCCPunitmodulus_bvctr}.\label{FindfaiConvxCCPFull_b}
\end{alignat}
\end{subequations}

The same techniques under the partial CSI error can be utilized here to solve Problem \eqref{FindfaiConvxCCPFull}, which is omitted to make the paper compact.
\subsection{Overall Algorithm, Convergence and Complexity Analysis}
The overall AO algorithm proposed under partial/full CSI errors is summarized in Algorithm \ref{AO}, where the expression of ``Problem (A)/Problem (B)'' is utilized to denote that the Problem (A) under partial CSI errors can be replaced by the Problem (B) under full CSI errors. By iteratively solving Problem \eqref{BITOCRPMsimplfyWZxy}/ Problem \eqref{OCR_PMfull_WZ}
and Problem \eqref{FindfaiConvxCCP}/ Problem \eqref{FindfaiConvxCCPFull} optimally in
Step 3 and Step 4 in Algorithm \ref{AO}, the transmit power can be monotonically
reduced with guaranteed convergence. We prove the convergence of the proposed AO algorithm as follows.

\begin{algorithm}
	\caption{Alternating Optimization Algorithm}
	\label{AO} \begin{algorithmic}[1] \STATE Parameter Setting:
		Set the maximum number of iterations $n_{{\rm {max}}}$ and the first
		iterative number $n=1$; Give the penalty factor $\kappa$ and error
		tolerance $\varepsilon$;
		
		\STATE Initialize the variables ${\bf {w}}^{(1)}$, ${\bf {Z}}^{(1)}$
		and ${\bm{\phi}}^{(1)}$ in the feasible region; Compute the OF value
		of Problem \eqref{BITOCRPMsimplfy}/ Problem \eqref{BITOCRPMfull} (i.e., the transmit power) as $p({{\bf {w}}^{(1)}},{{\bf {Z}}^{(1)}}{\rm {)}}$;
		
		\STATE Solve Problem \eqref{BITOCRPMsimplfyWZxy}/ Problem \eqref{OCR_PMfull_WZ} to obtain the ${\bf {w}}^{(n)},{{\bf {Z}}^{(n)}}$
		by fixing ${\bm{\phi}}^{(n)}$; Calculate the OF value of Problem
		\eqref{BITOCRPMsimplfyWZxy}/ Problem \eqref{OCR_PMfull_WZ} as $p({{\bf {w}}^{(n)}},{{\bf {Z}}^{(n)}}{\rm {)}}$;
		
		\STATE Solve Problem \eqref{FindfaiConvxCCP}/ Problem \eqref{FindfaiConvxCCPFull} to obtain ${\bm{\phi}}^{(n+1)}$ by fixing ${\bf {w}}^{(n)},{{\bf {Z}}^{(n)}}$; 
		
		\STATE If ${{\left|{p({\bf {w}}^{(n+1)},{{\bf {Z}}^{(n+1)}})\!-\!p({\bf {w}}^{(n)},{{\bf {Z}}^{(n)}})}\right|}/{p({\bf {w}}^{(n)},{{\bf {Z}}^{(n)}})}}\!<\varepsilon$
		or $n\geq n_{{\rm {max}}}$, terminate. Otherwise, update $n\leftarrow n+1$
		and jump to Step 3. \end{algorithmic}
\end{algorithm}

Denote the OF value of Problem \eqref{BITOCRPMsimplfy}/ Problem \eqref{BITOCRPMfull} and Problem \eqref{BITOCRPMsimplfyWZxy}/ Problem \eqref{OCR_PMfull_WZ} (i.e., the transmit
power) with a feasible solution $(\bm{\phi},\mathbf{w},\mathbf{Z})$
as $p(\bm{\phi},\mathbf{w},\mathbf{Z})$. As shown in step 4 of
Algorithm \ref{AO}, if there exists a feasible solution to Problem \eqref{FindfaiConvxCCP}/ Problem \eqref{FindfaiConvxCCPFull},
i.e., $(\bm{\phi}^{(n+1)},\mathbf{w}^{(n)},\mathbf{Z}^{(n)})$
exists, it is also feasible to Problem \eqref{BITOCRPMsimplfyWZxy}/ Problem \eqref{OCR_PMfull_WZ}. Then, $(\bm{\phi}^{(n+1)},\mathbf{w}^{(n+1)},\mathbf{Z}^{(n+1)})$
and $(\bm{\phi}^{(n)},\mathbf{w}^{(n)},\mathbf{Z}^{(n)})$ in
step 3 are the feasible solutions to Problem \eqref{BITOCRPMsimplfyWZxy}/ Problem \eqref{OCR_PMfull_WZ} in the $(n+1)$th
and $(n)$th iterations, respectively. It then follows that $p(\bm{\phi}^{(n+1)},\mathbf{w}^{(n+1)},\mathbf{Z}^{(n+1)})\overset{(u)}{\leq}p(\bm{\phi}^{(n+1)},\mathbf{w}^{(n)},\mathbf{Z}^{(n)})\overset{(v)}{=}p(\bm{\phi}^{(n)},\mathbf{w}^{(n)},\mathbf{Z}^{(n)})$,
where $(u)$ holds because for given $\bm{\phi}^{(n+1)}$ in step
3 of Algorithm \ref{AO}, $\mathbf{w}^{(n+1)},\mathbf{Z}^{(n+1)}$ is the
optimal solution to Problem \eqref{BITOCRPMsimplfyWZxy}/ Problem \eqref{OCR_PMfull_WZ}; and $(v)$ holds because the OF of Problem \eqref{BITOCRPMsimplfyWZxy}/ Problem \eqref{OCR_PMfull_WZ} is regardless of $\bm{\phi}$ and only
depends on $\mathbf{w},\mathbf{Z}$.

The computational complexity of all the resulted convex problems in Algorithm \ref{AO} can be measured in terms of their worst-case runtime by counting the complexity of LMI and second-order cone (SOC) constraints and ignoring the complexity of linear constraints, and the general expression for complexity has been given in \cite{zhou2020framework}. For Problem \eqref{BITOCRPMsimplfy}, the number of variables is ${{n}_{1}}=2{{N}_{t}}^{2}$. The number of LMIs in \eqref{eq:BITOCRPMsimplfyeig} is $K$ with the size of $M{{N}_{t}}$. The number of LMIs in \eqref{OCR_PM_Partialorgn_Z0} and \eqref{OCR_PM_Partialorgn_W0} is 2 with the size of ${{N}_{t}}$. The number of SOC in \eqref{eq:BITOCRPMsimplfySOC} is $K$ with the size of ${{M}^{2}}{{N}_{t}}^{2}+M{{N}_{t}}$. Thus, the approximate complexity of Problem \eqref{BITOCRPMsimplfy} is
	\begin{align}\hspace{-1cm}
		{{C}_{\mathbf{WZ}}}=&\mathcal{O}\left\{ {{\left[ KM{{N}_{t}}+2{{N}_{t}}+2K \right]}^{1/2}}{{n}_{1}}\left[ {{n}_{1}}^{2}+{{n}_{1}}\left( K{{M}^{2}}+2 \right){{N}_{t}}^{2}  \left( K{{M}^{3}}+2 \right){{N}_{t}}^{3}\right.\right.\nonumber\\ &\quad +\left.\left.{{n}_{1}}K{{\left( {{M}^{2}}{{N}_{t}}^{2}+M{{N}_{t}} \right)}^{2}} \right] \right\}.
\end{align}
For Problem \eqref{FindfaiConvxCCP}, the number of variables is ${{n}_{2}}=M$. The number of LMIs in \eqref{FindfaiSOCconvx} is $K$ with the size of ${{N}_{t}}^{2}+{{N}_{t}}$. The number of SOC in \eqref{FindfaiEIG_CVX} is $K$ with the size of $M{{N}_{t}}^{2}$. Thus, the approximate complexity of Problem \eqref{FindfaiConvxCCP} is
\begin{align}\hspace{-1cm}
	{{C}_{\bm{\phi} }}=\mathcal{O}\left\{ {{\left( 4K \right)}^{1/2}}{{n}_{2}}\left[ {{n}_{2}}^{2}+{{n}_{2}}\left( K{{\left( {{N}_{t}}^{2}+{{N}_{t}} \right)}^{2}}+K{{M}^{2}}{{N}_{t}}^{4} \right) \right] \right\}. 
\end{align}	
Altogether, the approximate computational complexity in each iteration under partial CSI errors is ${{C}_{\mathbf{WZ}}}+{{C}_{\bm{\phi} }}$.

Similarly, for Problem \eqref{OCR_PMfull_WZ}, the number of variables is ${\bar{n}_{1}}=2{{N}_{t}}^{2}$. The number of SOC in \eqref{eq:BITOCRPMfull_SOCP} is $K$ with the size of ${{\left( M{{N}_{t}}+{{N}_{t}} \right)}^{2}}+M{{N}_{t}}+{{N}_{t}}$. The number of LMIs in \eqref{eq:BITOCRPMfull_EIG} is $K$ with the size of $M{{N}_{t}}+{{N}_{t}}$. The number of LMIs in \eqref{OCR_PM_Partialorgn_Z0} and \eqref{OCR_PM_Partialorgn_W0} is 2 with the size of ${{N}_{t}}$. Thus, the approximate complexity of Problem \eqref{OCR_PMfull_WZ} is
\begin{align}\hspace{-1cm}
	{{\tilde{C}}_{\mathbf{WZ}}}=&\mathcal{O}\left\{ {{\left[ K\left( M{{N}_{t}}+{{N}_{t}} \right)+2{{N}_{t}}+2K \right]}^{1/2}}{\bar{n}_{1}}\left[ {\bar{n}_{1}}^{2}+{\bar{n}_{1}}\left( K{{\left( M{{N}_{t}}+{{N}_{t}} \right)}^{2}}+2{{N}_{t}}^{2} \right)\right.\right.\nonumber\\ &\quad +\left.\left.\left( K{{\left( M{{N}_{t}}+{{N}_{t}} \right)}^{3}}+2{{N}_{t}}^{3} \right)+{\bar{n}_{1}}K{{\left( {{\left( M{{N}_{t}}+{{N}_{t}} \right)}^{2}}+M{{N}_{t}}+{{N}_{t}} \right)}^{2}} \right] \right\}.	
\end{align}
For Problem \eqref{FindfaiConvxCCPFull}, the number of variables is ${\bar{n}_{2}}=M$. The number of SOC in \eqref{FindfaiFullSOCconvx} is $K$ with the size of ${{N}_{t}}^{2}+{{N}_{t}}$. The number of LMIs in \eqref{FindfaiFullEIG_CVX} is $K$ with the size of ${{N}_{t}}^{2}+{{N}_{t}}$. Thus, the approximate complexity of Problem \eqref{FindfaiConvxCCPFull} is
\begin{align}\hspace{-1cm}
	{{\tilde{C}}_{\bm{\phi} }}=\mathcal{O}\left\{ {{\left( 2K+2K \right)}^{1/2}}{\bar{n}_{2}}\left[ {\bar{n}_{2}}^{2}+{\bar{n}_{2}}\left( K{{\left( {{N}_{t}}^{2}+{{N}_{t}} \right)}^{2}}+K{{\left( M{{N}_{t}}^{2}+{{N}_{t}}^{2} \right)}^{2}} \right) \right] \right\}.
\end{align}
Altogether, the approximate computational complexity in each iteration under full CSI Errors is ${{\tilde{C}}_{\mathbf{WZ}}}+{{\tilde{C}}_{\bm{\phi} }}$.
\subsection{Extensions to the Multiple-Bob Case}
Consider a multi-Bob extension where $K$ legitimate users
(Bobs) are served by Alice, and are eavesdropped
by $K$ Eves. Specifically, the $k$th Bob
is eavesdropped by the $k$th Eve \cite{chai_secure_2023}. Then, the transmit signal is revised into\begin{align}
	{\bf {x}}=\bar{{\bf {W}}}\bar{\mathbf{s}}{\rm {+}}{\bf {z}}=\sum\limits _{k=1}^{K}{{{\bf {w}}_{k}}{s_{k}}+{\bf {z}}},\label{e1}
\end{align}
where $\bar{\mathbf{s}}=[s_{1},s_{2},\cdots,s_{K}]^{T}$, and $s_{k}$
is the data symbol intended for the $k$th Bob. $\bar{{\bf {W}}}=[{{\bf {w}}_{1}},{{\bf {w}}_{2}},\cdots,{{\bf {w}}_{K}}]$
contains all transmit beamforming vectors, and ${\bf {w}}_{k}$ is the beamforming vector
for the $k$th Bob.
We denote the channel vectors spanning from the BS to the $k$th Bob
and from the RIS to the $k$th Bob by ${{\bf {h}}_{ab,k}}\in{^{{N_{t}}\times1}}$
and ${{\bf {h}}_{rb,k}}\in{^{{N_{t}}\times1}}$ , respectively. The
received signal at the $k$th Bob are expressed as 
\begin{align}
	{y_{b,k}}={\hat {\bf{h}}}_{b,k}^{H}{\bf {x}}+{n_{b,k}}=({\bf {h}}_{ab,k}^{H}+{\bf {h}}_{rb,k}^{H}{\bf {\Phi}}{{\bf {G}}_{ar}}){\bf {x}}+{n_{b,k}},\forall k\in\mathcal{K}\label{e2}
\end{align}
where $\hat{\mathbf{h}}_{b,k}^{H}\triangleq\mathbf{h}_{ab,k}^{H}+{{\bm{\phi}}^{T}}{{\mathbf{G}}_{cb,k}}$
is the overall channel from Alice to Bob, and ${{\mathbf{G}}_{cb,k}}\triangleq\text{diag}(\mathbf{h}_{rb,k}^{H}){{\mathbf{G}}_{ar}}$, ${{\mathbf{G}}_{cb,k}}\in{{\mathbb{C}}^{M\times{{N}_{t}}}}$
is the cascaded channel from Alice to Bob via IRS. The $n_{b,k}\sim\mathcal{CN}(0,\sigma_{b,k}^{2})$ is the additive white Gaussian
noise (AWGN) received at the $k$th Bob. 

For security provisioning, we make a worst-case assumption regarding
the capabilities of the potential eavesdroppers. Specifically, we
assume that the Eves are able to cancel all multiuser interference
before decoding the information transmitted to a given Bob \cite{chai_secure_2023}. Then the achievable data rate of the $k$th
Eve by neglecting the multiuser interference can be written as
\begin{align}
	{{C}_{e,k}}({{\mathbf{W}}_{k}},\mathbf{Z},\mathbf{\Phi})=\log\left(1+\frac{{{\left|{\hat {\bf{h}}}_{e,k}^{H}{{\mathbf{w}}_{k}}\right|}^{2}}}{\sigma_{e,k}^{2}+{{\left|\hat{\mathbf{h}}_{e,k}^{H}\mathbf{z}\right|}^{2}}}\right),\label{e6}
\end{align}
where ${\mathbf{W}}_{k}={\mathbf{w}}_{k}{\mathbf{w}}_{k}^{H}$. To
ensure a fair condition for Bob and Eve, we assume that the legitimate
users can also cancel the multiuser interference, thus the achievable
data rate of the $k$th Bob is
\begin{align}
	{{C}_{b,k}}({{\mathbf{W}}_{k}},\mathbf{Z},\mathbf{\Phi})=\log\left(1+\frac{{{\left|{\hat {\bf{h}}}_{b,k}^{H}{{\mathbf{w}}_{k}}\right|}^{2}}}{\sigma_{b,k}^{2}+{{\left|\hat{\mathbf{h}}_{b,k}^{H}\mathbf{z}\right|}^{2}}}\right).\label{e4}
\end{align}

\subsubsection{Scenario 1: Partial CSI Errors}
By using \eqref{e6} and \eqref{e4}, the OCR-PM problem with partial
CSI errors is formulated as
\begin{subequations}\label{OCR-PMorgn_MultiBob_PartialCSI}\hspace{-1cm} 
	\begin{alignat}{1}
		& \underset{\{{{\mathbf{W}}_{k}}\}_{k=1}^{K},\mathbf{Z},\mathbf{\Phi}}{\mathop{\min}}\,\sum\limits _{k=1}^{K}{\text{Tr}({{\mathbf{W}}_{k}})+\text{Tr}(\mathbf{Z})}\\
		& \text{s}.\text{t}.\text{ }{{C}_{b,k}}({{\mathbf{W}}_{k}},\mathbf{Z},\mathbf{\Phi})\ge\log\gamma_{k},\text{ }\forall k\in\mathcal{K},\label{OCR-PMorgn_MultiBob_PartialCSI_a}\\
		& \quad\,\,\,\text{P}{{\text{r}}_{{{\mathbf{g}}_{ge,k}}}}\left\{ {{C}_{e,k}}({{\mathbf{W}}_{k}},\mathbf{Z},\mathbf{\Phi})\le\log\beta_{k}\right\} \ge1-{{\rho}_{k}},\text{ }\forall k\in\mathcal{K},\label{OCR-PMorgn_MultiBob_PartialCSI_b}\\
		& \quad\,\,\,\mathbf{Z}\succeq0,\label{OCR-PMorgn_MultiBob_PartialCSI_c}\\
		& \quad\,\,\,{{\mathbf{W}}_{k}}\succeq0,\text{ }\forall k\in\mathcal{K},\label{OCR-PMorgn_MultiBob_PartialCSI_d}\\
		& \quad\,\,\,\text{rank}({{\mathbf{W}}_{k}})=1,\text{ }\forall k\in\mathcal{K},\label{OCR-PMorgn_MultiBob_PartialCSI_e}\\
		& \quad\,\,\,\left|{{\bm{\phi}}_{m}}\right|=1,m=1,\cdots,M,\label{OCR-PMorgn_MultiBob_PartialCSI_f}
	\end{alignat}
\end{subequations} where $\gamma_{k}\ge1$ and $\beta_{k}\ge1$ are
constant values, and $\gamma_{k}\ge\beta_{k}$ is imposed to ensure
a positive secrecy rate. Obviously, the constraints of \eqref{OCR-PMorgn_MultiBob_PartialCSI_a}-\eqref{OCR-PMorgn_MultiBob_PartialCSI_f}
above are similar to the constraints of \eqref{OCR_PM_Partialorgn_bob}-\eqref{OCR_PM_Partialorgn_fai1} above. The only difference is that the beamforming vector to be optimized
is extended from $\mathbf{w}$ to $\{\mathbf{w}_{k}\}_{k=1}^{K}$. Thus the problem reformulation and problem solving can be extended directly from those in Section \ref{OCR-Partial} above. 

By using the BTI in a similar manner as the case of single-Bob, Problem \eqref{OCR-PMorgn_MultiBob_PartialCSI}
can be reformulated as
\begin{subequations}\label{OCR-PM_BTI_MultiBob_PartialCSI}\hspace{-1cm} 
	\begin{alignat}{1}
		& \underset{\{{{\mathbf{W}}_{k}}\}_{k=1}^{K},\mathbf{Z},\bm{\phi},\mathbf{x},\mathbf{y}}{\mathop{\min}}\,\sum\limits _{k=1}^{K}{\text{Tr}({{\mathbf{W}}_{k}})+\text{Tr}(\mathbf{Z})}\\
		& \text{s}.\text{t}.\ \ \text{Tr}(\mathbf{\Sigma}_{ge,k}^{1/2}(\mathbf{\Xi}_{e,k}^{T}\otimes\mathbf{E})\mathbf{\Sigma}_{ge,k}^{1/2})-\sqrt{-2\ln({{\rho}_{k}})}\cdot{{x}_{k}}+\ln({{\rho}_{k}})\cdot{{y}_{k}}+c_{k}^{\text{Partial}}\ge0,\forall k\in\mathcal{K},\label{OCR-PM_BTI_MultiBob_PartialCSI_c1}\\
		& \quad\,\,\,{{\left\Vert \left[\begin{matrix}(\mathbf{\Sigma}_{ge,k}^{1/2T}\otimes\mathbf{\Sigma}_{ge,k}^{1/2})\text{vec}(\mathbf{\Xi}_{e,k}^{T}\otimes\mathbf{E})\\
					\sqrt{2}\mathbf{\Sigma}_{ge,k}^{1/2}(\mathbf{\Xi}_{e,k}^{*}\otimes{{\mathbf{I}}_{M}})\text{vec}({{\bm{\phi}}^{*}}\mathbf{h}_{ae,k}^{H}+\mathbf{E}{{\bar{\mathbf{G}}}_{ce,k}})
				\end{matrix}\right]\right\Vert }_{2}}\le{{x}_{k}},\forall k\in\mathcal{K},\label{OCR-PM_BTI_MultiBob_PartialCSI_c2}\\
		& \quad\,\,\,{{y}_{k}}{{\mathbf{I}}_{M{{N}_{t}}}}+\mathbf{\Sigma}_{ge,k}^{1/2}(\mathbf{\Xi}_{e,k}^{T}\otimes\mathbf{E})\mathbf{\Sigma}_{ge,k}^{1/2}\succeq\mathbf{0},{{y}_{k}}\ge0,\forall k\in\mathcal{K},\label{OCR-PM_BTI_MultiBob_PartialCSI_c3}\\
		& \quad\,\,\,\hat{\mathbf{h}}_{b,k}^{H}[{\mathbf{W}}_{k}-(\gamma_{k}-1){\mathbf{Z}}]\hat{\mathbf{h}}_{b,k}\geq(\gamma_{k}-1)\sigma_{b,k}^{2},\forall k\in\mathcal{K},\label{OCR-PM_BTI_MultiBob_PartialCSI_c4}\\
		& \quad\,\,\,\eqref{OCR-PMorgn_MultiBob_PartialCSI_c},\eqref{OCR-PMorgn_MultiBob_PartialCSI_d},\eqref{OCR-PMorgn_MultiBob_PartialCSI_e},\eqref{OCR-PMorgn_MultiBob_PartialCSI_f},\label{OCR-PM_BTI_MultiBob_PartialCSI_c5}
	\end{alignat}
\end{subequations}  where ${{\mathbf{\Xi}}_{e,k}}\triangleq(\beta_{k}-1)\mathbf{Z}-{{\mathbf{W}}_{k}}$, $\tilde{\sigma}_{e,k}^{2}\triangleq(\beta_{k}-1)\sigma_{e,k}^{2}$, and 
\begin{alignat}{1}
	c_{k}^{\text{Partial}} & =(\mathbf{h}_{ae,k}^{H}+\bm{\bm{\phi}}^{T}\bar{\mathbf{\mathbf{G}}}_{ce,k})\mathbf{\Xi}_{e,k}(\mathbf{h}_{ae,k}^{H}+\bm{\bm{\phi}}^{T}\bar{\mathbf{\mathbf{G}}}_{ce,k})^{H}+\tilde{\sigma}_{e,k}^{2}.
\end{alignat}

Then the AO is utilized to estimate the beamformers
$\{\mathbf{W}_{k}\}_{k=1}^{K}$ and $\mathbf{Z}$ alternately.

\textit{(1) Optimization of Transmit Beamforming and AN:} For given phase shifts $\bm{\phi}$, Problem \eqref{OCR-PM_BTI_MultiBob_PartialCSI} cannot be solved
efficiently due to the nonconvexity of $\text{rank}({{\mathbf{W}}_{k}})=1$
in \eqref{OCR-PMorgn_MultiBob_PartialCSI_e}. By removing \eqref{OCR-PMorgn_MultiBob_PartialCSI_e}, the $\left\{ \{{{\mathbf{W}}_{k}}\}_{k=1}^{K},\mathbf{Z}\right\}$ can
be solved by SDR, and the corresponding optimization problem is
\begin{subequations}\label{WZsdr_MultiBob_PartialCSI}\hspace{-1cm} 
	\begin{alignat}{1}
		& \underset{\{{{\mathbf{W}}_{k}}\}_{k=1}^{K},\mathbf{Z},\mathbf{x},\mathbf{y}}{\mathop{\min}}\,\sum\limits _{k=1}^{K}{\text{Tr}({{\mathbf{W}}_{k}})+\text{Tr}(\mathbf{Z})}\ \\
		& \text{s}.\text{t}.\ \eqref{OCR-PM_BTI_MultiBob_PartialCSI_c1},\eqref{OCR-PM_BTI_MultiBob_PartialCSI_c2},\eqref{OCR-PM_BTI_MultiBob_PartialCSI_c3},\eqref{OCR-PM_BTI_MultiBob_PartialCSI_c4},\eqref{OCR-PMorgn_MultiBob_PartialCSI_c},\eqref{OCR-PMorgn_MultiBob_PartialCSI_d}.\label{WZsdr_MultiBob_PartialCSI_C1}
	\end{alignat}
\end{subequations}

Similarly, the tightness of the
SDR method for Problem \eqref{WZsdr_MultiBob_PartialCSI} can be guaranteed. Then the beamforming vector $\mathbf{w}_{k}$
can be recovered from ${\mathbf{W}}_{k}={\mathbf{w}}_{k}{\mathbf{w}}_{k}^{H}$
by performing the Cholesky decomposition.

\textit{(2) Optimization of Phase Shifts at the IRS:} For given $\{{{\mathbf{W}}_{k}}\}_{k=1}^{K}$
and $\mathbf{Z}$, the objective function of \eqref{OCR-PM_BTI_MultiBob_PartialCSI} is irrelevant
with $\bm{\phi}$. To achieve better convergence, slack variables $\bm{\delta}={{[{{\delta}_{1}},{{\delta}_{2}},\cdots,{{\delta}_{2K}}]}^{T}}$
are introduced. Similar to the manipulation from \eqref{ck_partial_Concave} to \eqref{BobdatarateRelaxConvxExprssn} in Section \ref{OCR-Partial-Optphi}, the constraints in \eqref{OCR-PM_BTI_MultiBob_PartialCSI_c1}-\eqref{OCR-PM_BTI_MultiBob_PartialCSI_c3}
can be transformed into convex constraints. As for the constraint in \eqref{OCR-PM_BTI_MultiBob_PartialCSI_c4}, it can
be rewritten as 
\begin{align}
	& (\mathbf{h}_{ab,k}^{H}+{{\bm{\phi}}^{T}}{{\mathbf{G}}_{cb,k}})[{{\mathbf{W}}_{k}}-(\gamma_{k}-1)\mathbf{Z}]({{\mathbf{h}}_{ab,k}}+\mathbf{G}_{cb,k}^{H}{{\bm{\phi}}^{*}})\ge\tilde{\sigma}_{b,k}^{2}+{{\delta}_{K+k}}\Leftrightarrow\nonumber \\
	& {{\bm{\phi}}^{T}}\left[{{\mathbf{G}}_{cb,k}}{{\mathbf{\Xi}}_{b,k}}\mathbf{G}_{cb,k}^{H}-{{c}_{b,k}}{{\mathbf{I}}_{M}}\right]{{\bm{\phi}}^{*}}+2\text{Re}\{{{\bm{\phi}}^{T}}{{\mathbf{G}}_{cb,k}}{{\mathbf{\Xi}}_{b,k}}{{\mathbf{h}}_{ab,k}}\} \nonumber \\ & +\mathbf{h}_{ab,k}^{H}{{\mathbf{\Xi}}_{b,k}}{{\mathbf{h}}_{ab,k}}-\tilde{\sigma}_{b,k}^{2}-{{\delta}_{K+k}}+M{{c}_{b,k}}\ge0,\label{Bob_phi_concave_MultiBob_PartialCSI}
\end{align}
where $\tilde{\sigma}_{b,k}^{2}\triangleq(\gamma_{k}-1)\sigma_{b,k}^{2}$,
${{\mathbf{\Xi}}_{b,k}}={{\mathbf{W}}_{k}}-(\gamma_{k}-1)\mathbf{Z}$,
${{c}_{b,k}}={{\lambda}_{\text{max}}}\{{{\mathbf{G}}_{cb,k}}{{\mathbf{W}}_{k}}\mathbf{G}_{cb,k}^{H}\}$.
Obviously, the constraint in \eqref{Bob_phi_concave_MultiBob_PartialCSI} is concave w.r.t.
${\bm{\phi}}$. By using \eqref{Bob_phi_concave_MultiBob_PartialCSI}, the optimization problem
for ${\bm{\phi}}$ becomes \begin{subequations}\label{Optphi_Vrsn1_MultiBob_PartialCSI}\hspace{-1cm} 
	\begin{alignat}{1}
		& \underset{\bm{\phi},\bm{\delta},\mathbf{x},\mathbf{y}}{\mathop{\text{max}}}\,\ \ \sum\limits _{k=1}^{2K}{{\delta}_{k}}\\
		& \text{s}.\text{t}.\ c_{tc,k}^{\text{partial}}(\bm{\phi})\ge0,\forall k\in\mathcal{K},\label{Optphi_Vrsn1_MultiBob_PartialCSI_C1}\\
		& \quad\,\,\,\left\Vert \begin{matrix}{{\lambda}_{\text{max}}}(\mathbf{\Sigma}_{ge,k}^{1/2T}\otimes\mathbf{\Sigma}_{ge,k}^{1/2})M\text{vec}({{\mathbf{\Xi}}_{e,k}})\\
			\sqrt{2M}{{\lambda}_{\text{max}}}(\mathbf{\Sigma}_{ge,k}^{1/2}(\mathbf{\Xi}_{e,k}^{*}\otimes{{\mathbf{I}}_{M}}))(\mathbf{h}_{ae,k}^{*}+\bar{\mathbf{G}}_{ce,k}^{T}\bm{\phi})
		\end{matrix}\right\Vert \le{{x}_{k}},\forall k\in\mathcal{K},\label{Optphi_Vrsn1_MultiBob_PartialCSI_C2}\\
		& \begin{array}{l}
			\quad\,\,\,{{\left\Vert (\mathbf{\Sigma}_{ge,k}^{1/2}(\mathbf{V}_{wgz,k}^{*}\otimes{{\bm{\phi}}^{*}}))\right\Vert }_{2}}-\sqrt{{{y}_{k}}}\le0,\forall k\in\mathcal{K},\end{array}\label{Optphi_Vrsn1_MultiBob_PartialCSI_C3}\\
		&\quad\,\,\,\eqref{Bob_phi_concave_MultiBob_PartialCSI},\label{Optphi_Vrsn1_MultiBob_PartialCSI_C4}\\
		& \quad\,\,\,\bm{\delta}\succeq\mathbf{0},\label{Optphi_Vrsn1_MultiBob_PartialCSI_C5}\\
		& \quad\,\,\,\left|{{\bm{\phi}}_{m}}\right|=1,m=1,\cdots,M,\label{Optphi_Vrsn1_MultiBob_PartialCSI_C6}
	\end{alignat}
\end{subequations} where $\mathbf{V}_{wgz,k}\mathbf{V}_{wgz,k}^{H}=\mathbf{W}_{k}+\mathbf{Z}_{g,k}$, $\mathbf{Z}_{g,k}\triangleq c_{g,k}\mathbf{I}_{N_{t}}-(\beta_{k}-1)\mathbf{Z}$, $c_{g,k}=\lambda_{\textrm{max}}\{(\beta_{k}-1)\mathbf{Z}\}\geq0$  and
\begin{subequations}
	\begin{align}
		c_{tc,k}^{\text{partial}}(\bm{\phi}) & \triangleq{{\bm{\phi}}^{T}}[{{\mathbf{\Sigma}}_{gx,k}}({{\mathbf{I}}_{M{{N}_{t}}}}\otimes{{\mathbf{\Xi}}_{e,k}})\mathbf{\Sigma}_{gx,k}^{H}-{{c}_{t,k}}{{\mathbf{I}}_{M}}]{{\bm{\phi}}^{*}}+M{{c}_{t,k}}-\sqrt{-2\ln({{\rho}_{k}})}\cdot{{x}_{k}}\nonumber \\
		& \quad\,\,\,+\ln({{\rho}_{k}})\cdot{{y}_{k}}+{{\bm{\phi}}^{T}}({{\bar{\mathbf{G}}}_{ce,k}}{{\mathbf{\Xi}}_{e,k}}\bar{\mathbf{G}}_{ce,k}^{H}-{{c}_{c,k}}{{\mathbf{I}}_{M}}){{\bm{\phi}}^{*}}+2\text{Re}\{{{\bm{\phi}}^{T}}{{\bar{\mathbf{G}}}_{ce,k}}{{\mathbf{\Xi}}_{e,k}}{{\mathbf{h}}_{ae,k}}\}\nonumber \\
		& \quad\,\,\,+\mathbf{h}_{ae,k}^{H}{{\mathbf{\Xi}}_{e,k}}{{\mathbf{h}}_{ae,k}}+M{{c}_{c,k}}+\tilde{\sigma}_{e,k}^{2}-{{\delta}_{k}},\\
		{{c}_{t,k}} & ={{\lambda}_{\max}}\{\mathbf{\Sigma}_{gx,k}({{\mathbf{I}}_{M{{N}_{t}}}}\otimes((\beta_{k}-1)\mathbf{Z}))\mathbf{\Sigma}_{gx,k}^{H}\},\\
		{{c}_{c,k}} & ={{\lambda}_{\max}}\{{{\bar{\mathbf{G}}}_{ce,k}}(\beta_{k}-1)\mathbf{Z}\bar{\mathbf{G}}_{ce,k}^{H}\}.
	\end{align}
\end{subequations}

To handle the only nonconvex constraints \eqref{Optphi_Vrsn1_MultiBob_PartialCSI_C6}, the first-order
Taylor expansion is exploited, and Problem \eqref{Optphi_Vrsn1_MultiBob_PartialCSI} is transformed into an iterative optimization process, and the optimization problem at the $(n+1)$th iteration is \begin{subequations}\label{Optphi_ccp_MultiBob_PartialCSI}\hspace{-1cm}
	\begin{alignat}{1}
		& \underset{\bm{\phi},\bm{\delta},\mathbf{b},\mathbf{x},\mathbf{y}}{\mathop{\text{max}}}\,\ \ \sum\limits _{k=1}^{2K}{{\delta}_{k}}-{{\lambda}^{(n+1)}}\sum\limits _{l=1}^{2M}{{b}_{l}}\\
		& \text{s}.\text{t}.\ \text{ }\eqref{Optphi_Vrsn1_MultiBob_PartialCSI_C1},\eqref{Optphi_Vrsn1_MultiBob_PartialCSI_C2},\eqref{Optphi_Vrsn1_MultiBob_PartialCSI_C3},\eqref{Optphi_Vrsn1_MultiBob_PartialCSI_C4},\eqref{Optphi_Vrsn1_MultiBob_PartialCSI_C5},\\
		& \quad\,\,\,{{\left|\bm{\phi}_{m}^{(n)}\right|}^{2}}-2\text{Re}\{\bm{\phi}_{m}^{*}\bm{\phi}_{m}^{(n)}\}\le{{b}_{m}}-1,m=1,\cdots,M,\\
		& \quad\,\,\,{{\left|{{\bm{\phi}}_{m}}\right|}^{2}}\le1+{{b}_{M+m}},m=1,\cdots,M,\\
		& \quad\,\,\,\mathbf{b}\succeq\mathbf{0},
	\end{alignat}
\end{subequations} where $\bm{\phi}_{m}^{(n)}$ is the phase shift of last
iteration ($n$th iteration) and $\mathbf{b}={{[{{b}_{1}},\cdots,{{b}_{2M}}]}^{T}}$
are the introduced slack variables. The $\sum\limits _{l=1}^{2M}{{b}_{l}}$
is added into the objective function as a penalty term and ${{\lambda}^{(n+1)}}$
is employed as the regularization factor to control the feasibility
of the constraints. Similarly, the penalty CCP algorithm
is leveraged to solve Problem \eqref{Optphi_ccp_MultiBob_PartialCSI}. 
\subsubsection{Scenario 2: Full CSI Errors}
In the scenario of full CSI errors, the OCR-PM problem is formulated as
\begin{subequations}\label{OCR-PMorgn_MultiBob_FullCSI}\hspace{-1cm} 
	\begin{alignat}{1}
		& \underset{\{{{\mathbf{W}}_{k}}\}_{k=1}^{K},\mathbf{Z},\mathbf{\Phi}}{\mathop{\min}}\,\sum\limits _{k=1}^{K}{\text{Tr}({{\mathbf{W}}_{k}})+\text{Tr}(\mathbf{Z})}\\
	    & \quad\,\,\,\textrm{Pr}_{\mathbf{g}_{he,k},\mathbf{g}_{ge,k}}\left\{ {{C}_{e,k}}({{\mathbf{W}}_{k}},\mathbf{Z},\mathbf{\Phi})\le\log\beta_{k}\right\} \ge1-{{\rho}_{k}},\text{ }\forall k\in\mathcal{K},\label{OCR-PMorgn_MultiBob_FullCSI_b}\\
		& \eqref{OCR-PMorgn_MultiBob_PartialCSI_a},\eqref{OCR-PMorgn_MultiBob_PartialCSI_c},\eqref{OCR-PMorgn_MultiBob_PartialCSI_d},\eqref{OCR-PMorgn_MultiBob_PartialCSI_e},\eqref{OCR-PMorgn_MultiBob_PartialCSI_f}
	\end{alignat}
\end{subequations}
Similarly, the reformulation and solving of Problem \eqref{OCR-PMorgn_MultiBob_FullCSI} can be extended directly from those in Section \ref{OCR-Full}. By applying the BTI similarly, the OCR-PM problem with full CSI errors is formulated
as\begin{subequations}\label{OCR-PM_BTI_MultiBob_FullCSI}\hspace{-1cm} 
	\begin{alignat}{1}
		& \underset{\{{{\mathbf{W}}_{k}}\}_{k=1}^{K},\mathbf{Z},\bm{\phi},\mathbf{x},\mathbf{y}}{\mathop{\min}}\,\sum\limits _{k=1}^{K}{\text{Tr}({{\mathbf{W}}_{k}})+\text{Tr}(\mathbf{Z})}\\
		& \text{s}.\text{t}.\ \ \text{Tr}({\tilde{\mathbf{{A}}}_{e,k}})-\sqrt{-2\ln({{\rho}_{k}})}\cdot{{\tilde{x}}_{k}}+\ln({{\rho}_{k}})\cdot{{\tilde{y}}_{k}}+c_{k}^{\text{full}}\ge0,\forall k\in\mathcal{K},\label{OCR-PM_BTI_MultiBob_FullCSI_C1}\\
		& \quad\,\,\,{{\left\Vert \left[\begin{matrix}\text{vec}({{\tilde{\mathbf{A}}}_{e,k}})\\
					\sqrt{2}{{\tilde{\mathbf{u}}}_{e,k}}
				\end{matrix}\right]\right\Vert }_{2}}\le{{\tilde{x}}_{k}},\text{ }\forall k\in\mathcal{K},\label{OCR-PM_BTI_MultiBob_FullCSI_C2}\\
		& \quad\,\,\,{{\tilde{y}}_{k}}{{\mathbf{I}}_{M{{N}_{t}}+{{N}_{t}}}}+{{\tilde{\mathbf{A}}}_{e,k}}\succeq\mathbf{0},{{\tilde{y}}_{k}}\ge0,\text{ }\forall k\in\mathcal{K},\label{OCR-PM_BTI_MultiBob_FullCSI_C3}\\
		& \quad\,\,\,\hat{\mathbf{h}}_{b,k}^{H}[{\mathbf{W}}_{k}-(\gamma_{k}-1){\mathbf{Z}}]\hat{\mathbf{h}}_{b,k}\geq(\gamma_{k}-1)\sigma_{b,k}^{2},\forall k\in\mathcal{K},\label{OCR-PM_BTI_MultiBob_FullCSI_C4}\\
		& \quad\,\,\, \eqref{OCR-PMorgn_MultiBob_PartialCSI_c},\eqref{OCR-PMorgn_MultiBob_PartialCSI_d},\eqref{OCR-PMorgn_MultiBob_PartialCSI_e},\eqref{OCR-PMorgn_MultiBob_PartialCSI_f},\label{OCR-PM_BTI_MultiBob_FullCSI_C5}
	\end{alignat}
\end{subequations} where 
\begin{subequations}
	\begin{alignat}{1}
		\tilde{\mathbf{{A}}}_{e,k} & \triangleq\left[\begin{matrix}\mathbf{\Sigma}_{he,k}^{1/2}{{\mathbf{\Xi}}_{e,k}}\mathbf{\Sigma}_{he,k}^{1/2} & \mathbf{\Sigma}_{he,k}^{1/2}({{\mathbf{\Xi}}_{e,k}}\otimes{{\bm{\phi}}^{H}})\mathbf{\Sigma}_{ge,k}^{1/2*}\\
			\mathbf{\Sigma}_{ge,k}^{1/2T}({{\mathbf{\Xi}}_{e,k}}\otimes\bm{\phi})\mathbf{\Sigma}_{he,k}^{1/2} & \mathbf{\Sigma}_{ge,k}^{1/2T}({{\mathbf{\Xi}}_{e,k}}\otimes{{\mathbf{E}}^{T}})\mathbf{\Sigma}_{ge,k}^{1/2*}
		\end{matrix}\right],\\
		{\tilde{\mathbf{u}}}_{e,k} & =\left[\begin{matrix}\mathbf{\Sigma}_{he,k}^{1/2}{{\mathbf{\Xi}}_{e,k}}({{\bar{\mathbf{h}}}_{ae,k}}+\bar{\mathbf{G}}_{ce,k}^{H}{{\bm{\phi}}^{*}})\\
			\mathbf{\Sigma}_{ge,k}^{1/2T}({{\mathbf{\Xi}}_{e,k}}\otimes{{\mathbf{I}}_{M}})(({{\bar{\mathbf{h}}}_{ae,k}}+\bar{\mathbf{G}}_{ce,k}^{H}{{\bm{\phi}}^{*}})\otimes\bm{\phi})
		\end{matrix}\right],\\
		c_{k}^{\textrm{full}} & =(\bar{\mathbf{h}}_{ae,k}^{H}+{{\bm{\phi}}^{T}}{{\bar{\mathbf{G}}}_{ce,k}}){{\mathbf{\Xi}}_{e,k}}{{(\bar{\mathbf{h}}_{ae,k}^{H}+{{\bm{\phi}}^{T}}{{\bar{\mathbf{G}}}_{ce,k}})}^{H}}+\tilde{\sigma}_{e,k}^{2}.
	\end{alignat}
\end{subequations}

\textit{(1) Optimization of Transmit Beamforming and AN:} For given phase shifts $\bm{\phi}$, Problem \eqref{OCR-PM_BTI_MultiBob_FullCSI} cannot be solved
efficiently only due to the nonconvexity of $\text{rank}({{\mathbf{W}}_{k}})=1$
in \eqref{OCR-PMorgn_MultiBob_PartialCSI_e}. By removing \eqref{OCR-PMorgn_MultiBob_PartialCSI_e}, the $\left\{ \{{{\mathbf{W}}_{k}}\}_{k=1}^{K},\mathbf{Z}\right\}$ can
be solved by SDR, and the corresponding optimization problem is
\begin{subequations}\label{WZsdr_MultiBob_FullCSI}\hspace{-1cm} 
	\begin{alignat}{1}
		& \underset{\{{{\mathbf{W}}_{k}}\}_{k=1}^{K},\mathbf{Z},\mathbf{x},\mathbf{y}}{\mathop{\min}}\,\sum\limits _{k=1}^{K}{\text{Tr}({{\mathbf{W}}_{k}})+\text{Tr}(\mathbf{Z})}\\
		& \text{s}.\text{t}.\ \ \eqref{OCR-PM_BTI_MultiBob_FullCSI_C1},\eqref{OCR-PM_BTI_MultiBob_FullCSI_C2},\eqref{OCR-PM_BTI_MultiBob_FullCSI_C3},\eqref{OCR-PM_BTI_MultiBob_FullCSI_C4},\eqref{OCR-PMorgn_MultiBob_PartialCSI_c},\eqref{OCR-PMorgn_MultiBob_PartialCSI_d},\eqref{OCR-PMorgn_MultiBob_PartialCSI_f}.
	\end{alignat}
\end{subequations}The SDR in Problem \eqref{WZsdr_MultiBob_FullCSI} is tight, and the beamforming vector $\mathbf{w}_{k}$ can be
recovered from ${\mathbf{W}}_{k}={\mathbf{w}}_{k}{\mathbf{w}}_{k}^{H}$
by performing the Cholesky decomposition.

\textit{(2) Optimization of Phase Shifts at the IRS:} When $\{{{\mathbf{W}}_{k}}\}_{k=1}^{K}$ and $\mathbf{Z}$ are fixed,
the slack variables $\bm{\delta}={{[{{\delta}_{1}},{{\delta}_{2}},\cdots,{{\delta}_{2K}}]}^{T}}$
are introduced to improve the convergence of the AO algorithm. Similar
to the manipulation from \eqref{eq:TrAkquadraticFull} to \eqref{FindfaiFullEIG_CVX} in Section \ref{OCR-Full-Optphi},
the constraints in \eqref{OCR-PM_BTI_MultiBob_FullCSI_C1}-\eqref{OCR-PM_BTI_MultiBob_FullCSI_C3} can be transformed
into convex constraints. Thus, by using \eqref{Bob_phi_concave_MultiBob_PartialCSI},
the optimization problem for ${\bm{\phi}}$ becomes
\begin{subequations}\label{Optphi_Vrsn1_MultiBob_FullCSI}\hspace{-1cm} 
	\begin{alignat}{1}
		& \underset{\bm{\phi},\bm{\delta},\mathbf{x},\mathbf{y}}{\mathop{\text{max}}}\,\ \ \sum\limits _{k=1}^{2K}{{\delta}_{k}}\\
		& \text{s}.\text{t}.\; c_{tc,k}^{\text{full}}(\bm{\phi})\ge0,\forall k\in\mathcal{K},\label{Optphi_Vrsn1_MultiBob_FullCSI_C1}\\
		& \quad\,\,\,\left\Vert \begin{matrix}[{{\lambda}_{\text{max}}}({{\mathbf{\Sigma}}_{he,k}})+{{\lambda}_{\text{max}}}({{\mathbf{\Sigma}}_{ge,k}})M]\text{vec}({{\mathbf{\Xi}}_{e,k}})\\
			\sqrt{2[\lambda_{\text{max}}^{2}(\mathbf{\Sigma}_{he,k}^{1/2}{{\mathbf{\Xi}}_{e,k}})+\lambda_{\text{max}}^{2}(\mathbf{\Sigma}_{ge,k}^{1/2T}({{\mathbf{\Xi}}_{e}}\otimes{{\mathbf{I}}_{M}}))M]}({{\bar{\mathbf{h}}}_{ae,k}}+\bar{\mathbf{G}}_{ce,k}^{H}\bm{\phi})
		\end{matrix}\right\Vert \le{{\tilde{x}}_{k}},\text{ }\forall k\in\mathcal{K},\label{Optphi_Vrsn1_MultiBob_FullCSI_C2}\\
		& \begin{array}{l}
			\quad\,\,\,\sqrt{{{\tilde{y}}_{k}}}\overset{{}}{\mathop{\ge}}\,{{\left\Vert \begin{matrix}\mathbf{\Sigma}_{he,k}^{1/2}{{\mathbf{V}}_{wgz,k}}\\
						\mathbf{\Sigma}_{ge,k}^{1/2T}({{\mathbf{V}}_{wgz,k}}\otimes\bm{\phi})
					\end{matrix}\right\Vert }_{2}},\text{ }{{\tilde{y}}_{k}}\ge0,\text{ }\forall k\in\mathcal{K},\end{array}\label{Optphi_Vrsn1_MultiBob_FullCSI_C3}\\
		& \quad\,\,\,\eqref{Bob_phi_concave_MultiBob_PartialCSI}, \eqref{Optphi_Vrsn1_MultiBob_PartialCSI_C5}, \eqref{Optphi_Vrsn1_MultiBob_PartialCSI_C6}, \label{Optphi_Vrsn1_MultiBob_FullCSI_C4}
	\end{alignat}
\end{subequations} where 
\begin{align}
	c_{tc,k}^{\text{full}}(\bm{\phi})\triangleq & \text{Tr}[\mathbf{\Sigma}_{he,k}^{1/2}{{\mathbf{\Xi}}_{e,k}}\mathbf{\Sigma}_{he,k}^{1/2}]+{{\bm{\phi}}^{T}}[{{\mathbf{\Sigma}}_{gx,k}}({{\mathbf{I}}_{M{{N}_{t}}}}\otimes{{\mathbf{\Xi}}_{e,k}})\mathbf{\Sigma}_{gx,k}^{H}-{{c}_{t,k}}{{\mathbf{I}}_{M}}]{{\bm{\phi}}^{\text{*}}}+M{{c}_{t,k}}\nonumber \\
	{} &
	-\sqrt{{-2}\ln({{\rho}_{k}})}\cdot{{x}_{k}}
	+\ln({{\rho}_{k}})\cdot{{y}_{k}}+{{\bm{\phi}}^{T}}({{\bar{\mathbf{G}}}_{ce,k}}{{\mathbf{\Xi}}_{e,k}}\bar{\mathbf{G}}_{ce,k}^{H}-{{c}_{c,k}}{{\mathbf{I}}_{M}}){{\bm{\phi}}^{*}}\nonumber \\
	{} &
	+2\text{Re}\{{{\bm{\phi}}^{T}}{{\bar{\mathbf{G}}}_{ce,k}}{{\mathbf{\Xi}}_{e,k}}{{\bar{\mathbf{h}}}_{ae,k}}\} +\bar{\mathbf{h}}_{ae,k}^{H}{{\mathbf{\Xi}}_{e,k}}{{\bar{\mathbf{h}}}_{ae,k}}+M{{c}_{c,k}}+\tilde{\sigma}_{e,k}^{2}-{{\delta}_{k}}.
\end{align}

To handle the only nonconvex constraints  \eqref{Optphi_Vrsn1_MultiBob_PartialCSI_C6},
the first-order Taylor expansion is exploited. Then Problem \eqref{Optphi_Vrsn1_MultiBob_FullCSI}
is transformed into an iterative optimization process,
and the optimization problem at the $(n+1)$th iteration is\begin{subequations}\label{Optphi_ccp_MultiBob_FullCSI}\hspace{-1cm}
	\begin{alignat}{1}
		& \underset{\bm{\phi},\bm{\delta},\mathbf{b},\mathbf{x},\mathbf{y}}{\mathop{\text{max}}}\,\ \ \sum\limits _{k=1}^{2K}{{\delta}_{k}}-{{\lambda}^{(n+1)}}\sum\limits _{l=1}^{2M}{{b}_{l}}\\
		& \text{s}.\text{t}.\ \eqref{Optphi_Vrsn1_MultiBob_FullCSI_C1},\eqref{Optphi_Vrsn1_MultiBob_FullCSI_C2},\eqref{Optphi_Vrsn1_MultiBob_FullCSI_C3},\eqref{Optphi_Vrsn1_MultiBob_FullCSI_C4},\eqref{Bob_phi_concave_MultiBob_PartialCSI}, \eqref{Optphi_Vrsn1_MultiBob_PartialCSI_C5},\\
		& \quad\,\,\,{{\left|\bm{\phi}_{m}^{(n)}\right|}^{2}}-2\text{Re}\{\bm{\phi}_{m}^{*}\bm{\phi}_{m}^{(n)}\}\le{{b}_{m}}-1,m=1,\cdots,M,\\
		& \quad\,\,\,{{\left|{{\bm{\phi}}_{m}}\right|}^{2}}\le1+{{b}_{M+m}},m=1,\cdots,M,\\
		& \quad\,\,\,\mathbf{b}\succeq\mathbf{0}.
	\end{alignat}
\end{subequations}
Similarly, the penalty CCP algorithm
is leveraged to solve Problem \eqref{Optphi_ccp_MultiBob_FullCSI}.

\vspace{-0.3cm}\section{Simulation Results}
\begin{figure}
	\begin{minipage}[t]{0.495\linewidth}
		\centering
		\includegraphics[width=2.6in]{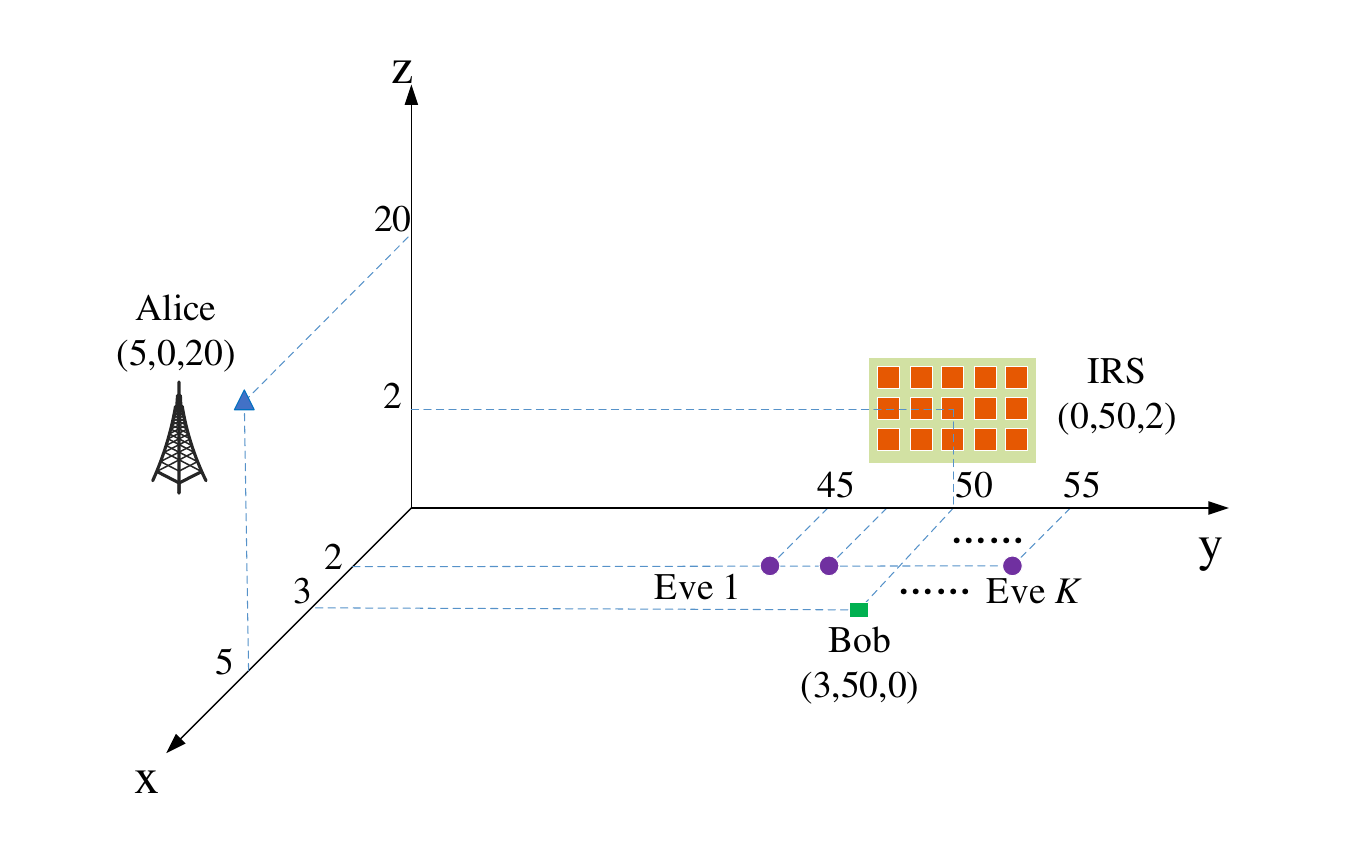}\vspace{-0.6cm}
		\caption{Simulation setup of the considered IRS-aided secure communication}
		\label{figsimulscenario}
	\end{minipage}%
	\hfill
	\begin{minipage}[t]{0.495\linewidth}
		\centering
		\includegraphics[width=2.6in]{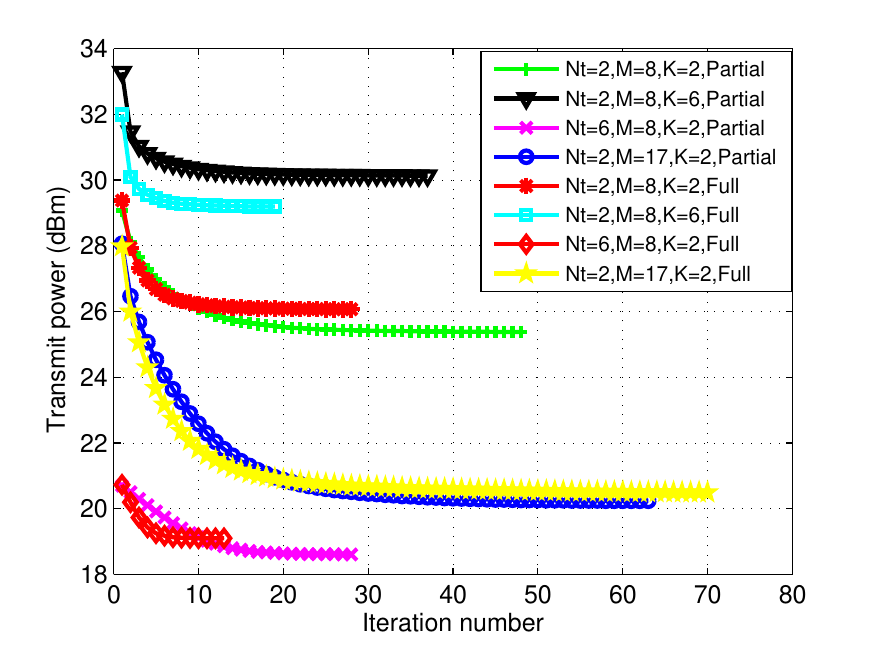}\vspace{-0.6cm}
		\caption{Convergence performance.}
		\label{figPowvsIterPCUandFCU}
	\end{minipage}\vspace{-0.7cm}
\end{figure}

Fig. \ref{figsimulscenario} describes the considered IRS-aided secure communication system, where the Alice, IRS, and Bob are located \cite{guan2020intelligent} at (5,0,20) m, (0,50,2) m, and (3,50,0) m respectively. The Eves are randomly distributed on the line from (2,45,0) m to (2,55,0) m. We assume that the channel from Alice to the IRS is Rician fading and can be modeled as
\begin{alignat}{1}
\!\!\mathbf{\mathbf{G}}_{ar} & \!=\!\sqrt{\!L_{0}d_{ar}^{-\alpha_{ar}}}\!\!\left(\!\!\sqrt{\!\frac{\beta_{ar}}{1\!+\!\beta_{ar}}}\mathbf{G}_{ar}^{LOS}\!\!+\!\!\sqrt{\!\frac{1}{1\!+\!\beta_{ar}}}\mathbf{G}_{ar}^{NLOS}\!\!\right),
\end{alignat}
where the pathloss at the reference distance is set to be
$L_{0}=-40$ $\textrm{dB}$  based on the 3GPP UMi model \cite{2017technical}. The distance from Alice to the IRS is denoted
by $d_{ar}$, and the path loss exponent of the Alice-IRS link is denoted by $\alpha_{ar}$. $\beta_{ar}$ is the corresponding Rician factor. The channel $\mathbf{\mathbf{G}}_{ar}$ contains the line of sight (LoS) $\mathbf{G}_{ar}^{LOS}$ and the Rayleigh fading non-LoS (NLoS) $\mathbf{G}_{ar}^{NLOS}$ components. Other channels are modeled similarly.

The path loss exponent for Alice-IRS channel is $\alpha_{ar}=2.2$ \cite{zhou2020framework}. The path loss exponents for IRS-Bob channel $\alpha_{rb}$ and IRS-Eve channel $\alpha_{re}$ are 2 \cite{zhou2020framework}. A more scattering environment is assumed for the direct links, thus the path loss exponents for Alice-Bob channel and Alice-Eve channel are $\alpha_{ab}=\alpha_{ae}=3.5$ \cite{feng2021physical}. The Rician factors for the Alice-IRS channel $\beta_{ar}$, IRS-Bob channel $\beta_{rb}$, IRS-Eve channel $\beta_{re}$, Alice-Bob channel $\beta_{ab}$, and Alice-Eve channel $\beta_{ae}$ are 5 \cite{yu2020robust}. The Eves' outage probability is $\rho_{k}=0.05$ \cite{zhou2020framework}. The noise power at Bob and Eves is set as $-85$ $\textrm{dBm}$. The threshold for convergence is $\varepsilon=10^{-3}$ \cite{yu2020robust}.

For comparision, we exploit the following baseline schemes.
\begin{itemize}
\item MRT-isoAN-randIRS: The maximum ratio transmission (MRT) based beamforming is performed, where $\mathbf{w}=\sqrt{p_{w}}\frac{\hat{\mathbf{h}}_{b}}{\left\Vert \hat{\mathbf{h}}_{b}\right\Vert }$, and $p_{w}$ is the power allocated to Bob. The isotropic AN \cite{liao2010qos} is generated, where the AN covariance matrix
is $\mathbf{Z}=p_{z}\mathbf{P}_{\hat{\mathbf{h}}_{b}}^{\perp}$, $\mathbf{P}_{\hat{\mathbf{h}}_{b}}^{\perp}=\mathbf{I}_{N_{t}}-\hat{\mathbf{h}}_{b}\hat{\mathbf{h}}_{b}^{H}/\left\Vert \hat{\mathbf{h}}_{b}\right\Vert ^{2}$,
and $p_{z}$ is the power invested on AN. We assume that the $\mathbf{\Phi}$ in $\hat{\mathbf{h}}_{b}=\mathbf{h}_{ab}+\mathbf{G}_{ar}^{H}\mathbf{\mathbf{\Phi}}^{H}\mathbf{h}_{rb}$ are randomly chosen. The allocated
power $p_{w}$ and $p_{z}$ is optimized in Problem
\eqref{BITOCRPMsimplfyWZxy} or Problem \eqref{OCR_PMfull_WZ}.
\item MRT-isoAN-optIRS: The difference of the MRT-isoAN-optIRS scheme from the MRT-isoAN-randIRS scheme is that the
utilized $\mathbf{\Phi}$ at the IRS is optimized.
\item randIRS: The difference of the Random-IRS scheme from the proposed algorithm is that the utilized $\mathbf{\Phi}$ at the IRS is random.
\end{itemize}
\vspace{-0.5cm}\subsection{The Case of Uncorrelated CSI Errors}
Firstly, we consider the uncorrelated CSI errors, where the variance matrix of $\mathbf{g}_{ge,k}$
is defined as $\mathbf{\Sigma}_{ge,k}=\varepsilon_{g,k}^{2}\mathbf{I}$,
where $\varepsilon_{g,k}^{2}=\delta_{g,k}^{2}\left\Vert \textrm{vec}(\mathbf{\bar{\mathbf{G}}}_{ce,k})\right\Vert _{2}^{2}$, while the variance matrix of $\mathbf{g}_{he,k}$ is defined as $\mathbf{\Sigma}_{he,k}=\varepsilon_{h,k}^{2}\mathbf{I}$, where $\varepsilon_{h,k}^{2}=\delta_{h,k}^{2}\left\Vert \bar{\mathbf{h}}_{ae,k}\right\Vert _{2}^{2}$.
$\delta_{g,k}, \delta_{h,k}\in\left[0,1\right)$ are the normalized CSI errors,
which measure the relative amount of CSI errors. Unless specified, the normalized CSI error for partial Eves' CSI error is $\delta_{g,k}=0.01$, $\forall k$, and the normalized CSI error for full Eves' CSI error is $\delta_{g,k}=\delta_{h,k}=0.01$, $\forall k$ \cite{zhou2020framework}.
\subsubsection{Convergence Performance}
The convergence performance of the proposed method is investigated in Fig. \ref{figPowvsIterPCUandFCU}. It is observed that the proposed algorithm monotonically converges with different values of $N_t$, $M$, and $K$ for both the partial and full CSI errors. The convergence speed decreases with a larger number of IRS reflection units $M$ and a larger number of Eves $K$, while increases with a larger number of the transmit antennas $N_t$. The number of iterations required for convergence is more sensitive to $M$ than $N_t$ and $K$. As observed in Fig. \ref{figPowvsIterPCUandFCU}, the convergence speed with partial CSI error is generally slower than that of full CSI error, except the case when the $M$ is relatively large. Since the dimensions of the search space in the optimization subproblems increase with $N_t$ and $M$, and the number of constraints in the optimization subproblems increases with $K$, the running time required by the proposed robust design grows with $N_t$, $M$, and $K$.
\subsubsection{Transmit Power vs the Minimum Data Rate of Bob}
\begin{figure}
	\begin{minipage}[t]{0.495\linewidth}
		\centering
		\includegraphics[width=2.6in]{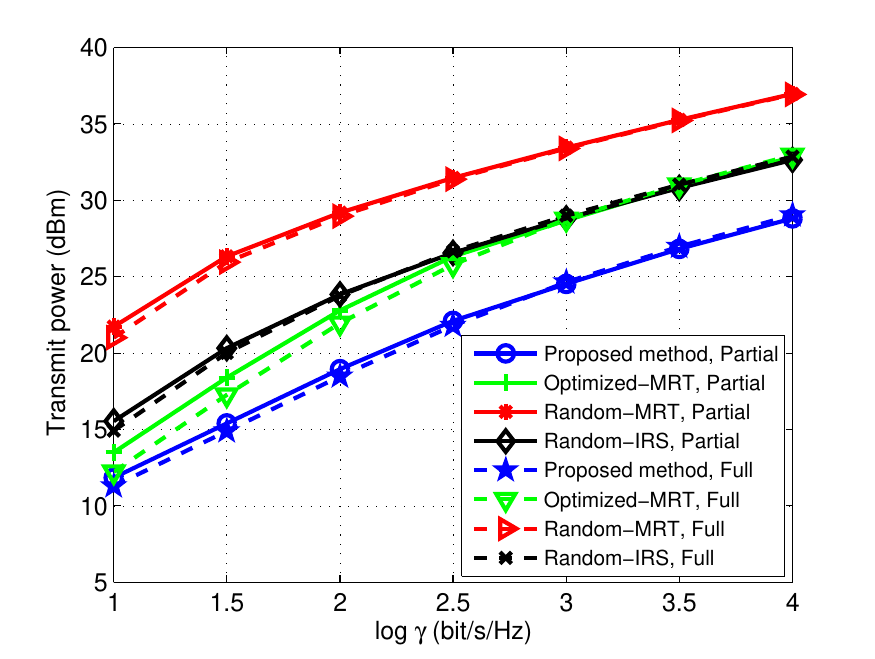}\vspace{-0.6cm}
		\caption{Required transmit power at each minimum data rate $\log\gamma$ of
			Bob, where $N_{t}=2$, $M=15$, $K=2$, and $\log\beta=0.5$ $\textrm{bit/s/Hz}$.}
		\label{figPowvsBobGamma}
	\end{minipage}%
	\hfill
	\begin{minipage}[t]{0.495\linewidth}
		\centering
		\includegraphics[width=2.6in]{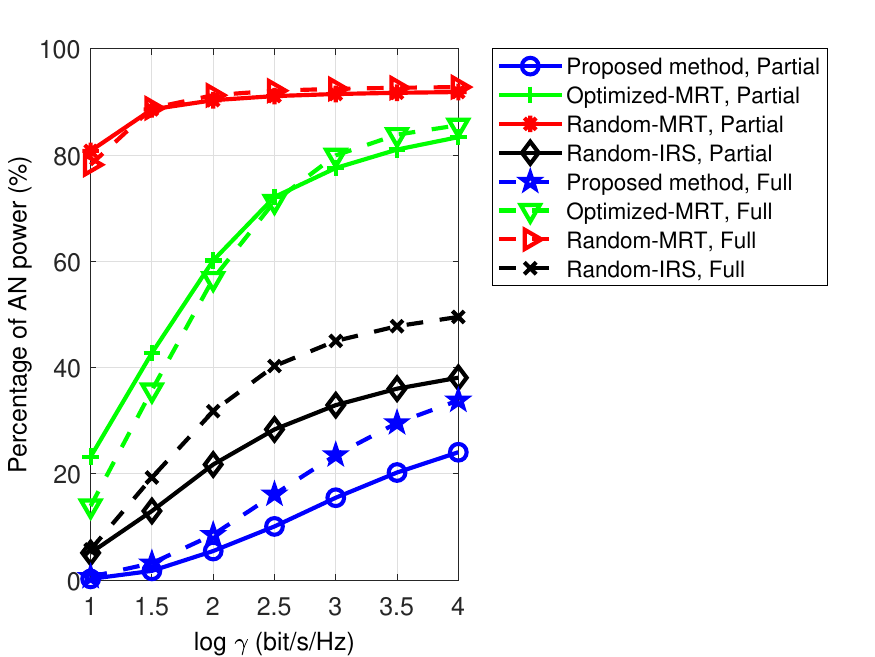}\vspace{-0.6cm}
		\caption{Percentage of invested AN power at each minimum data rate $\log\gamma$ of Bob, where $N_{t}=2$, $M=15$, $K=2$, and $\log\beta=0.5$ $\textrm{bit/s/Hz}$.}
		\label{figPercentANvsBobGamma}
	\end{minipage}\vspace{-0.7cm}
\end{figure}

Fig. \ref{figPowvsBobGamma} describes the transmit power at different values of $\log\gamma$, which is the
minimum data rate of Bob. It is observed that the transmit power increases monotonically with $\log\gamma$. This means that Alice has to transmit more power to ensure a larger data rate of Bob.


Here, the Eves' data rates can be limited by three factors, which are the CSI errors of Eves' channels, the AN impairment, and the reconfiguration on wireless propagation by the IRS. For the MRT-isoAN-randIRS and MRT-isoAN-optIRS schemes, the AN is isotropically distributed on the orthogonal complement subspace of the equivalent channel of Bob, thus the AN power directed to Eves is relatively low. For the proposed algorithm and the randIRS scheme, the AN is steered towards Eves' direction, thus the AN power invested on Eves is spatially focused and relatively high.

When $\log\gamma$ is relatively small, the transmit power required by the MRT-isoAN-optIRS is close to that required by the proposed algorithm, and lower than that of the randIRS scheme. This means that when $\log\gamma$ is small, the isotropically distributed AN is low but enough to confuse the Eves, and the optimization of the IRS phase shifts seems more effective than the AN focusing. As the $\log\gamma$ increases, the performance of the MRT-isoAN-optIRS scheme degrades into that of the randIRS scheme, and their required power is higher than that of the proposed algorithm. This signifies that when $\log\gamma$ is large, the optimization on IRS phase shifts is not enough to interfere Eves, and the AN should be designed and focused on Eves.

It is also observed that when $\log\gamma$ is small, the transmit power of all these schemes with full CSI errors is smaller than that with partial CSI errors. This is because when $\log\gamma$ is small, the full CSI errors can impair Eves with larger errors than the partial CSI errors. When $\log\gamma$ is large, neither the full or the partial CSI errors are enough for impairing Eves, thus more AN power is required. Since it is more difficult to transmit AN through channels with worse quality, the robust design with full CSI errors require more transmit power.

The AN power invested from the total transmit power at different values of $\log\gamma$ is shown in
Fig. \ref{figPercentANvsBobGamma}. As observed, the percentage of AN power increases with $\log\gamma$ for all schemes. A higher percentage of AN power is required by the MRT-isoAN strategy than the proposed algorithm and randIRS scheme. That's because the AN in the MRT-isoAN strategy is isotropically distributed, while the AN in the proposed algorithm and randIRS scheme is spatially focused.
\subsubsection{Transmit Power vs the Maximum Data Rate of Eves}
\begin{figure}
	\begin{minipage}[t]{0.495\linewidth}
		\centering
		\includegraphics[width=2.6in]{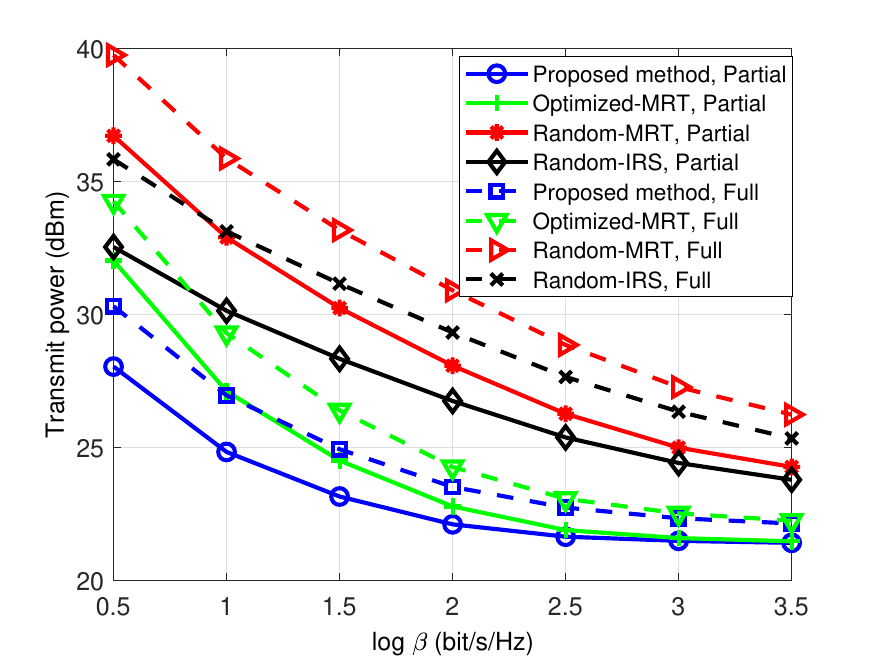}\vspace{-0.6cm}
		\caption{Required transmit power at each maximum data rate $\log\beta$
			of Eves, where $N_{t}=2$, $M=15$, $K=2$, and $\log\gamma=4$ $\textrm{bit/s/Hz}$.}
		\label{figPvsEveBeita}
	\end{minipage}%
	\hfill
	\begin{minipage}[t]{0.495\linewidth}
		\centering
		\includegraphics[width=2.6in]{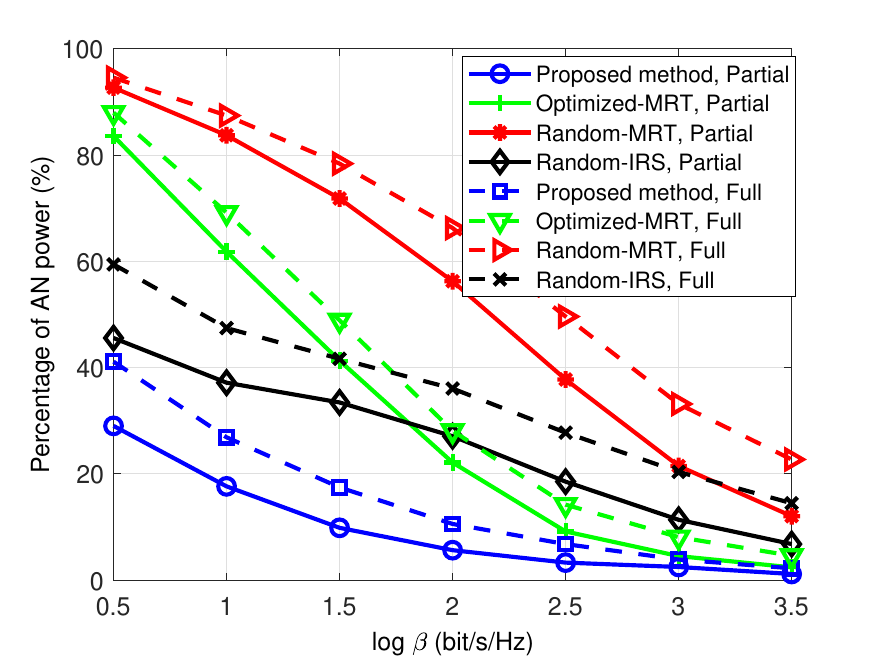}\vspace{-0.6cm}
		\caption{Percentage of invested AN power at each maximum data rate $\log\beta$
			of Eves, where $N_{t}=2$, $M=15$, $K=2$, and $\log\gamma=4$ $\textrm{bit/s/Hz}$.}
		\label{figPercentANvsEveBeita}
	\end{minipage}\vspace{-0.7cm}
\end{figure}

The transmit power at different values of $\log\beta$ is depicted in Fig. \ref{figPvsEveBeita}, where $\log\beta$ denotes the maximum data rate of Eves. As observed, the transmit power monotonically decreases with $\log\beta$, and the decreasing speed slows down with increased $\log\beta$. Increasing $\log\beta$ means that the limitation on the information leakage for Eves is relaxed, thus less AN power and transmit power are required. With a relatively large $\log\beta$, the transmit power required by the MRT-isoAN-optIRS scheme becomes close to that of the proposed algorithm, and the MRT-isoAN-randIRS scheme becomes close to the randIRS scheme. This means that it is enough to transmit almost isotropically distributed AN by MRT strategy to interfere Eves when the data rate limitation for Eves is relaxed. The transmit power required by the MRT-isoAN-optIRS and the proposed algorithm is much lower than the MRT-isoAN-randIRS and randIRS schemes, which demonstrates the necessity to optimize the phase shifts of the IRS. It is also found that the transmit power with full CSI errors is higher than that with partial CSI errors, and the gap is narrowed gradually with $\log\beta$.

The invested AN power from the total transmit power at different values of $\log\beta$ is shown in Fig. \ref{figPercentANvsEveBeita}. As observed, the AN power decreases with $\log\beta$. The tighter the limitation for information leakage is, the more AN power is required. The AN power of the proposed algorithm and Random-IRS scheme is less than that of the Optimized-MRT and Random-MRT, which means that the idea of making AN spatially focused is beneficial for reducing the transmit power.
\subsubsection{The Impact of the Number of IRS Reflection Units}
\begin{figure}
	\begin{minipage}[t]{0.495\linewidth}
		\centering
		\includegraphics[width=2.6in]{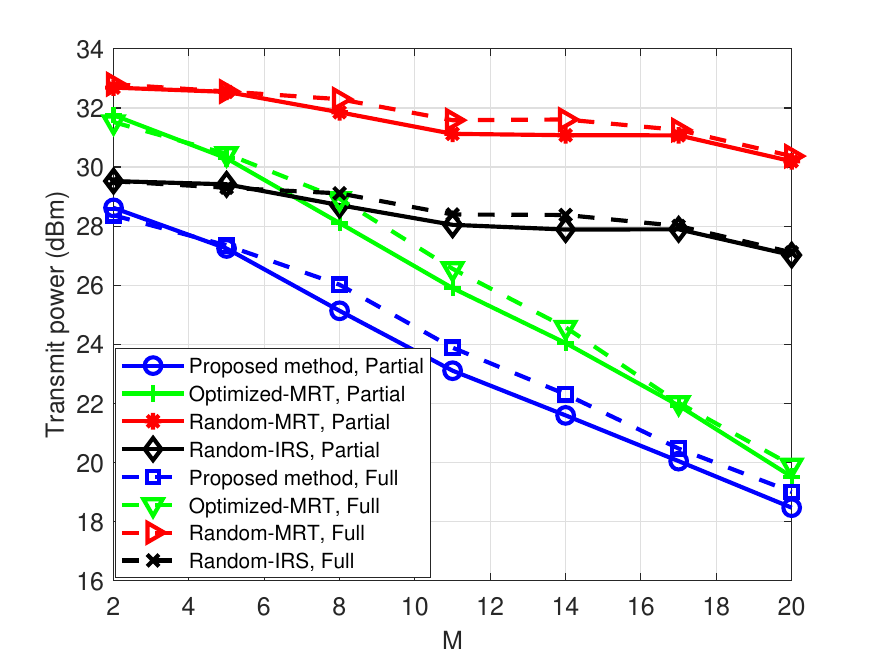}\vspace{-0.6cm}
		\caption{Required transmit power with different unit numbers $M$ of the IRS, where $N_{t}=2$, $K=2$, $\log\beta=1$ $\textrm{bit/s/Hz}$, and $\log\gamma=3$ $\textrm{bit/s/Hz}$.}
		\label{figPvsM}
	\end{minipage}%
	\hfill
	\begin{minipage}[t]{0.495\linewidth}
		\centering
		\includegraphics[width=2.6in]{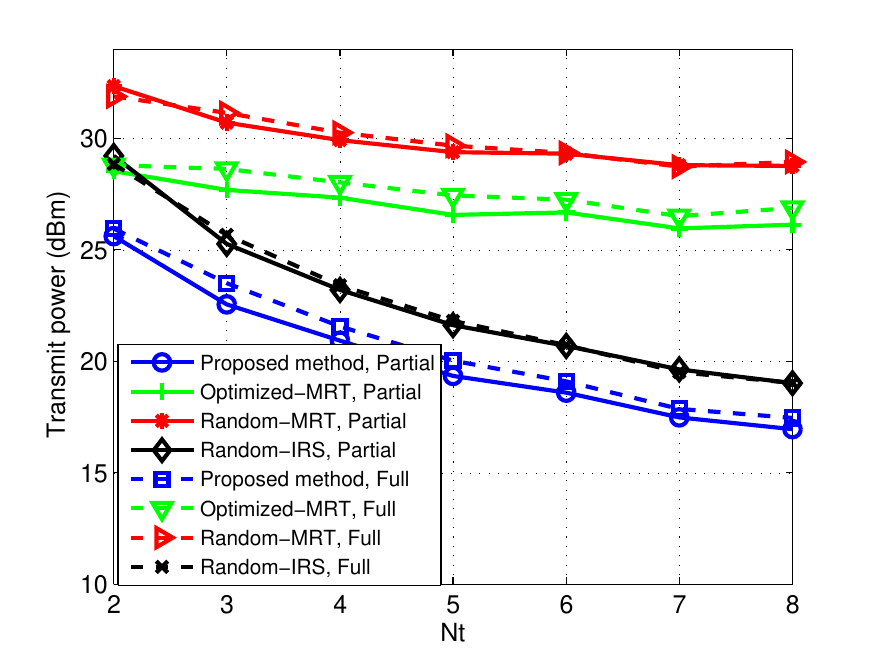}\vspace{-0.6cm}
		\caption{Required transmit power with different transmit antenna numbers $N_{t}$, where $M=8$, $K=2$, $\log\beta=1$ $\textrm{bit/s/Hz}$, and $\log\gamma=3$ $\textrm{bit/s/Hz}$.}
		\label{figPvsNt}
	\end{minipage}\vspace{-0.7cm}
\end{figure}

Fig. \ref{figPvsM} investigates the impact of the unit number of IRS on reducing the transmit power. It is observed that the transmit power can be reduced with more IRS units for all these schemes. The robust transmission design of the proposed algorithm outperforms the other schemes in terms of the lowest transmit power. The transmit power consumption of the proposed algorithm and the MRT-isoAN-optIRS scheme drops much faster than that of the randIRS and MRT-isoAN-randIRS schemes, since the phase shifts of the IRS are optimized for them. When $M$ becomes large, the power consumption of the MRT-isoAN-optIRS tends to approach that of the proposed algorithm. This is because larger $M$ brings more cascaded CSI errors, which enhance the capability of impairing Eves by the isoAN strategy. The transmit power with the full CSI errors is slightly larger than that with the partial CSI errors.
\subsubsection{The Impact of the Number of Transmit Antennas}
Fig. \ref{figPvsNt} investigates the impact of the number of Alice's antennas $N_{t}$ on the transmit power consumption. As observed, the transmit power decreases with $N_{t}$ for all schemes, which means that more transmit antennas can reduce the transmit power in the robust design. The proposed algorithm and the randIRS scheme are more sensitive to $N_{t}$ than the MRT-isoAN-optIRS scheme and MRT-isoAN-randIRS scheme. This is because a larger $N_{t}$ is helpful for increasing the Bob's data rate, while is not conductive to decrease the Eves' data rate. The AN power for the isoAN strategy is almost isotropically distributed, which is less effective to reduce the transmit power when $N_{t}$ is large. Thus, the transmit power required by the MRT-isoAN-optIRS and MRT-isoAN-randIRS schemes reduces gently. The transmit power with full CSI errors is slightly higher than that with partial CSI errors.
\subsubsection{The Impact of Channel Errors}
\begin{figure}
	\begin{minipage}[t]{0.495\linewidth}
		\centering
		\includegraphics[width=2.6in]{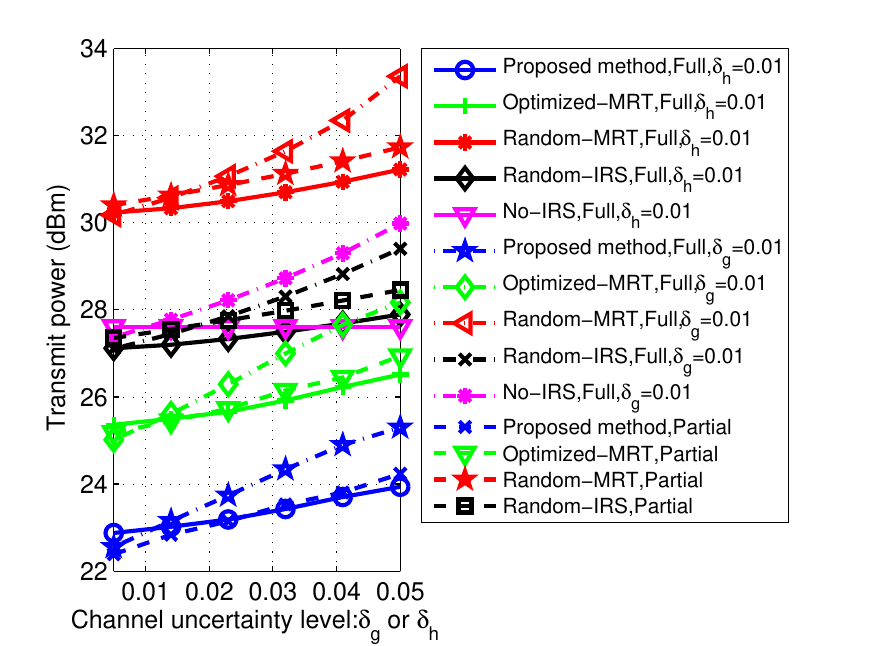}\vspace{-0.6cm}
		\caption{Required transmit power with different channel uncertainty levels $\delta$, where $N_t=2$, $M=12$, $K=3$, $\log\beta=1$ $\textrm{bit/s/Hz}$, and $\log\gamma=3$ $\textrm{bit/s/Hz}$. }
		\label{figPvsCSIerr}
	\end{minipage}%
	\hfill
	\begin{minipage}[t]{0.495\linewidth}
		\centering
		\includegraphics[width=2.6in]{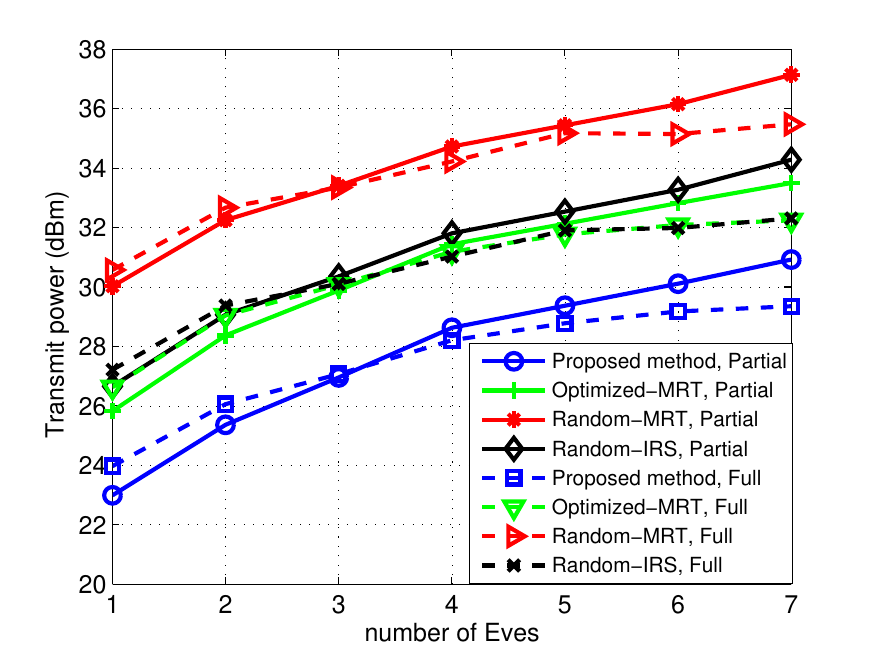}\vspace{-0.6cm}
		\caption{Required transmit power with different numbers of Eves $K$, where $N_t=2$, $M=8$, $\log\beta=1$ $\textrm{bit/s/Hz}$, and $\log\gamma=3$ $\textrm{bit/s/Hz}$.}
		\label{figPowvsKenum}
	\end{minipage}\vspace{-0.7cm}
\end{figure}

The impact of channel uncertainty on the transmit power is investigated in Fig. \ref{figPvsCSIerr}, where the channel uncertainty is measured by the normalized
CSI error $\delta$ of Eves' channels, and we assume that $\delta_{g,k}=\delta_{h,k}=\delta$, $\forall k$. It is found that the transmit power consumption increases with the normalized CSI errors $\delta$, which means that more power has to be transmitted when the channel quality degrades. Under a relatively large CSI error, the transmit power with full CSI errors is slightly lower than that with partial CSI errors. This means that when the CSI error is large, the full CSI errors are preferred to impair the Eves. It is also observed that when full CSI errors exist for Eves, the transmit power changing with the channel error of the direct link is higher than that changing with the channel error of the reflecting link. This signifies that the robust transmission design is more sensitive to the CSI error of the direct link. Among all these schemes, the proposed algorithm requires the lowest transmit power. By comparing with the No-IRS scheme, it demonstrates that the security performance of the IRS-aided communication system is superior to the no-IRS system even with CSI errors of both the direct link and IRS reflecting link.
\subsubsection{The Impact of the Number of Eves}
Fig. \ref{figPowvsKenum} investigates the impact of the number of Eves on transmit power. As observed, the transmit power increases with the number of Eves for all schemes. When the number of Eves is relatively small, the transmit power with partial CSI errors is slightly lower than that with full CSI errors, which means that the partial CSI error is sufficient to impair Eves. When more Eves exist, the transmit power with partial CSI errors is slightly higher than that with full CSI errors, which means the full CSI error is preferred to interfere more Eves.
\subsubsection{The Impact of the Initial Values on the Performance	of the Proposed Algorithm}
\begin{figure}[h]
	\vspace{-0.6cm}
	\centering
	\includegraphics[width=3.5in]{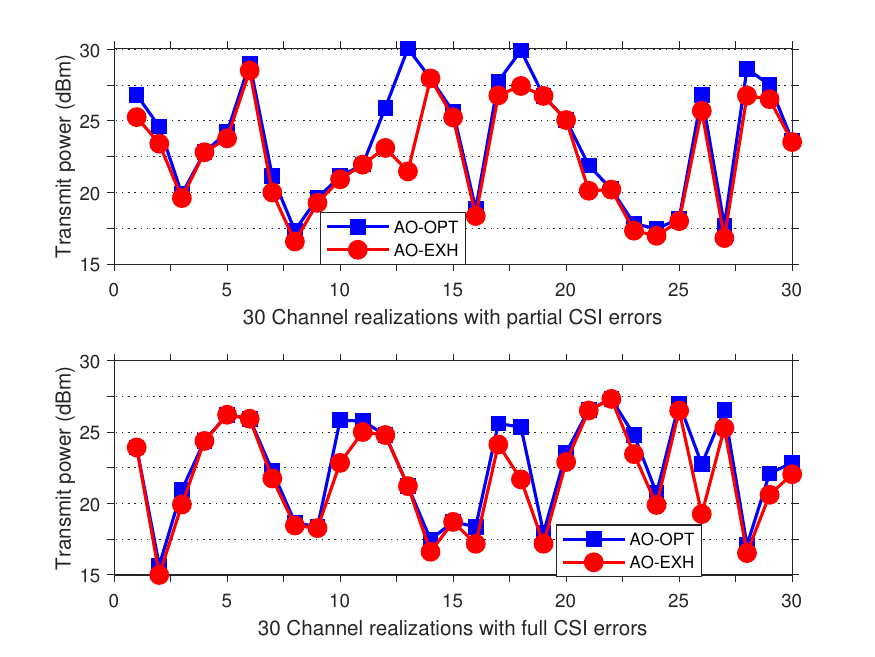}
	\caption{Performance comparison of AO-OPT and AO-EXH, where $N_{t}=2$, $M=15$, $K=2$, $\log\gamma=2.5$ $\textrm{bit/s/Hz}$, and $\log\beta=0.5$ $\textrm{bit/s/Hz}$.}\vspace{-0.5cm}
	\label{optimality}
	\vspace{-0.7cm}
\end{figure}

Due to the nonconvexity of Problem \eqref{OCR_PM_Partialextnd},
different initial values may
result in different locally optimal solutions obtained by the
proposed algorithm. It is difficult to find a good initial value for $\bm{\phi}$ since the a good $\bm{\phi}$ should enhance the Alice-Bob channel and deteriorate the Alice-Eve channel. Thus, the initial $\bm{\phi}$ is chosen randomly, and the initial $\mathbf{w}$ and $\mathbf{Z}$ are generated by solving Problem \eqref{BITOCRPMsimplfyWZxy} or Problem \eqref{OCR_PMfull_WZ}. To study the impact of the initialization of $\mathbf{w}$ and $\mathbf{Z}$ on the performance of the proposed algorithm,
we test 30 randomly generated channels shown in Fig. \ref{optimality}, where
AO-OPT refers to using optimized $\mathbf{w}$ and $\mathbf{Z}$ as
initial values and AO-EXH refers to that $\mathbf{w}$ and $\mathbf{Z}$ are initialized by the best of 1000 feasible random
normalized vectors and matrices for each channel realization. It can be seen
that the total power consumption of AO-OPT is almost the
same as that of AO-EXH, implying that the optimized $\mathbf{w}$ and $\mathbf{Z}$ is a good option for the initialization.

\subsubsection{The Multiple-Bob Case}
We further investigate the effectiveness of the proposed robust transmission design in the multiple-Bob case. It is noted that the baseline scheme of MRT and isotropic AN was proposed aiming for the single-Bob case \cite{liao2010qos}. If we extend this baseline scheme into the multiple-Bob case directly, the $k$th beamformer becomes $\mathbf{w}_{k}=\sqrt{p_{k}}\frac{\hat{\mathbf{h}}_{b,k}}{\left\Vert \hat{\mathbf{h}}_{b,k}\right\Vert }$, and the AN covariance
matrix becomes $\mathbf{Z}=p_{z}\mathbf{P}_{\check{\mathbf{H}}_{b}}^{\perp}$,
$\mathbf{P}_{\check{\mathbf{H}}_{b}}^{\perp}=\mathbf{I}_{N_{t}}-\check{\mathbf{H}}_{b}\check{\mathbf{H}}_{b}^{H}/\left\Vert \check{\mathbf{H}}_{b}\right\Vert _{\textrm{F}}^{2}$,
$\check{\mathbf{H}}_{b}=[\hat{\mathbf{h}}_{b,1},\hat{\mathbf{h}}_{b,2},\cdots,\hat{\mathbf{h}}_{b,K}]$, where the $p_{k}$ and $p_{z}$ are power scale factors to be optimized. However, we find the infeasibility rate of this baseline scheme increase greatly, because this form of $\mathbf{w}_{k}$ and $\mathbf{Z}$ makes the data rate of each Bob stay at a low level and cannot ensure a large secrecy rate.
\begin{figure}[h]
		\begin{minipage}[t]{0.495\linewidth}
			\centering
			\renewcommand\thefigure{\arabic{figure}}
			\includegraphics[width=2.6in]{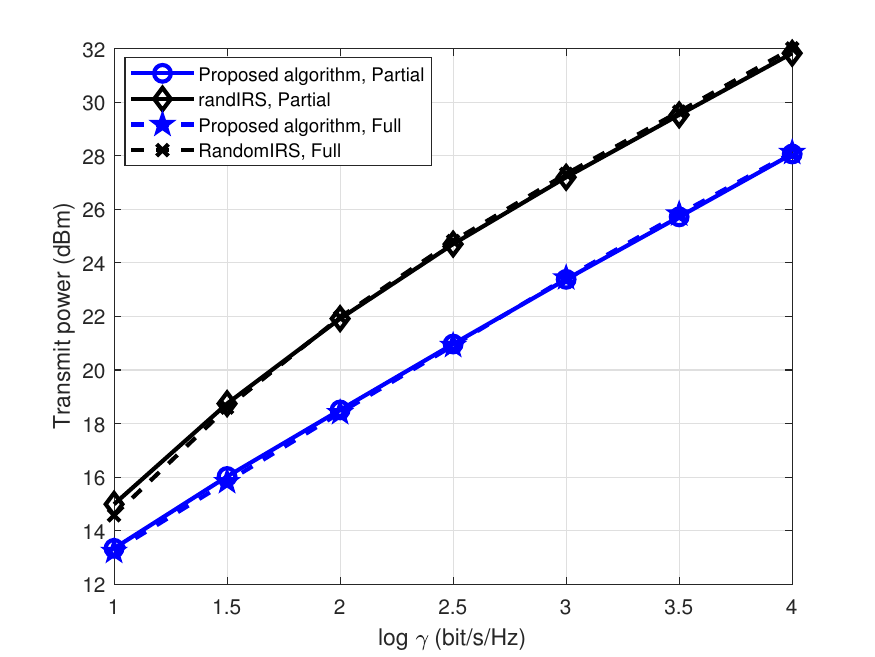}\vspace{-0.6cm}
			\caption{Required transmit power at each minimum data rate $\log\gamma$ of
				multiple Bobs, where $N_{t}=2$, $M=15$, $K=2$, $\log\beta=0.5$ $\textrm{bit/s/Hz}$, ${\gamma}_k=\gamma$, ${\beta}_k=\beta$, $\forall k\in\mathcal{K}$.}
			\label{figPowvsMultiBobGamma}
		\end{minipage}%
		\hfill
		\begin{minipage}[t]{0.495\linewidth}
			\centering
			\renewcommand\thefigure{\arabic{figure}}
			\includegraphics[width=2.6in]{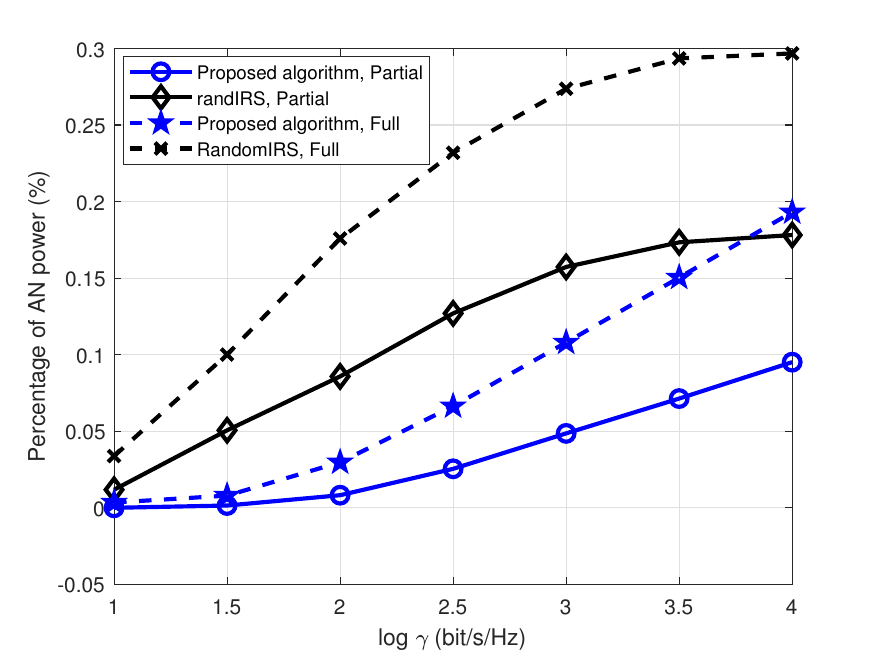}\vspace{-0.6cm}
			\caption{Percentage of invested AN power at each minimum data rate $\log\gamma$ of multiple Bobs, where $N_{t}=2$, $M=15$, $K=2$, $\log\beta=0.5$ $\textrm{bit/s/Hz}$, ${\gamma}_k=\gamma$, ${\beta}_k=\beta$, $\forall k\in\mathcal{K}$.}
			\label{figPercentANvsMultiBobGamma}
		\end{minipage}\vspace{-0.7cm}
\end{figure}

To make the comparison fair, we only compare our proposed algorithm with the randIRS scheme. Fig. \ref{figPowvsMultiBobGamma} describes the transmit power of different schemes at different values of $\log\gamma$ in the multi-Bob case, where two Bobs are located at (3,36,0) m and (3,48,0) m. The percentage of AN power invested from the total transmit power at different values of $\log\gamma$ is shown in Fig. \ref{figPercentANvsMultiBobGamma}. It is observed from Fig. \ref{figPowvsMultiBobGamma} that the transmit power of the proposed algorithm is much lower than the randIRS scheme. It is seen from Fig. \ref{figPercentANvsMultiBobGamma} that the percentage of AN power increases with $\log\gamma$, and invested AN power of the proposed algorithm is much slower than the randIRS scheme. Moreover, as compared with the single-Bob case, the required AN power is greatly reduced in the multi-Bob case and is almost negligible.

\subsection{The Case of Correlated CSI Errors}
In this subsection, we consider the robust transmission design with correlated CSI errors, where $\mathbf{\Sigma}_{ge,k}^{1/2}=\varepsilon_{g,k}\mathbf{S}_{MN_{t}}$,
$\mathbf{\Sigma}_{he,k}^{1/2}=\varepsilon_{h,k}\mathbf{S}_{N_{t}}$,
$[\mathbf{S}_{MN_{t}}]_{m,n},[\mathbf{S}_{N_{t}}]_{m,n}=\eta^{\left|m-n\right|}$,
and $\eta$ is set as 0.9 \cite{Marco2003on}. To make the relative amount of correlated CSI errors
comparable with the case of uncorrelated CSI errors, we assume that
$\varepsilon_{g,k}^{2}=\delta_{g,k}^{2}\left\Vert \textrm{vec}(\mathbf{\bar{\mathbf{G}}}_{ce,k})\right\Vert _{2}^{2}[\left\Vert \mathbf{I}_{MN_{t}}\right\Vert _{\textrm{F}}/\left\Vert \mathbf{S}_{MN_{t}}^{2}\right\Vert _{\textrm{F}}]$,
and $\varepsilon_{h,k}^{2}=\delta_{h,k}^{2}\left\Vert \bar{\mathbf{h}}_{ae,k}\right\Vert _{2}^{2}[[\left\Vert \mathbf{I}_{N_{t}}\right\Vert _{\textrm{F}}/\left\Vert \mathbf{S}_{N_{t}}^{2}\right\Vert _{\textrm{F}}]$.

\begin{figure}
	\begin{minipage}[t]{0.495\linewidth}
		\centering
		\includegraphics[width=2.6in]{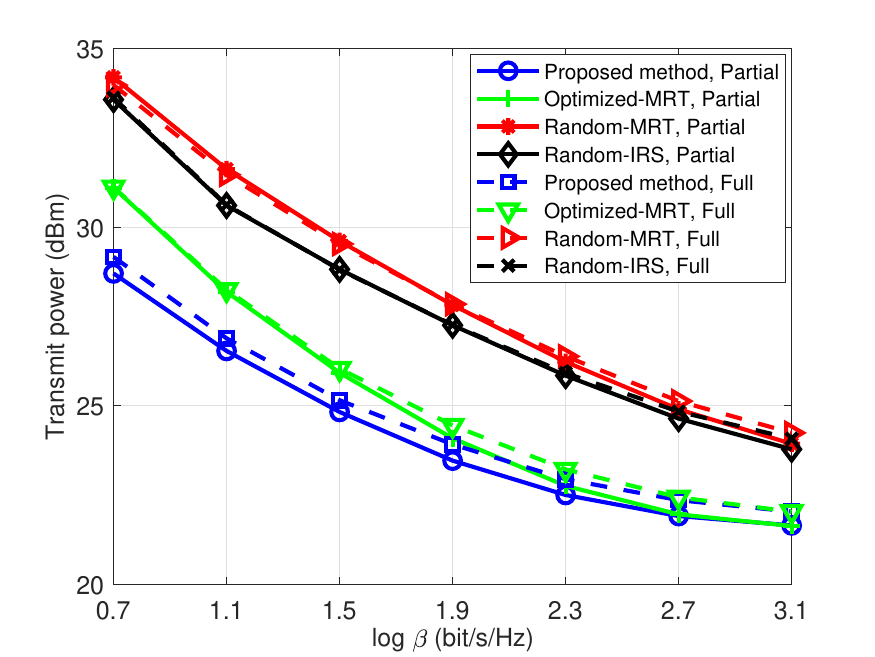}\vspace{-0.6cm}
		\caption{Required transmit power at each $\log\beta$ with correlated CSI errors, where $N_{t}=2$, $M=15$, $K=2$, and $\log\gamma=3.5$ $\textrm{bit/s/Hz}$.}
		\label{figPowvsBeta_CommCSI}
	\end{minipage}%
	\hfill
	\begin{minipage}[t]{0.495\linewidth}
		\centering
		\includegraphics[width=2.6in]{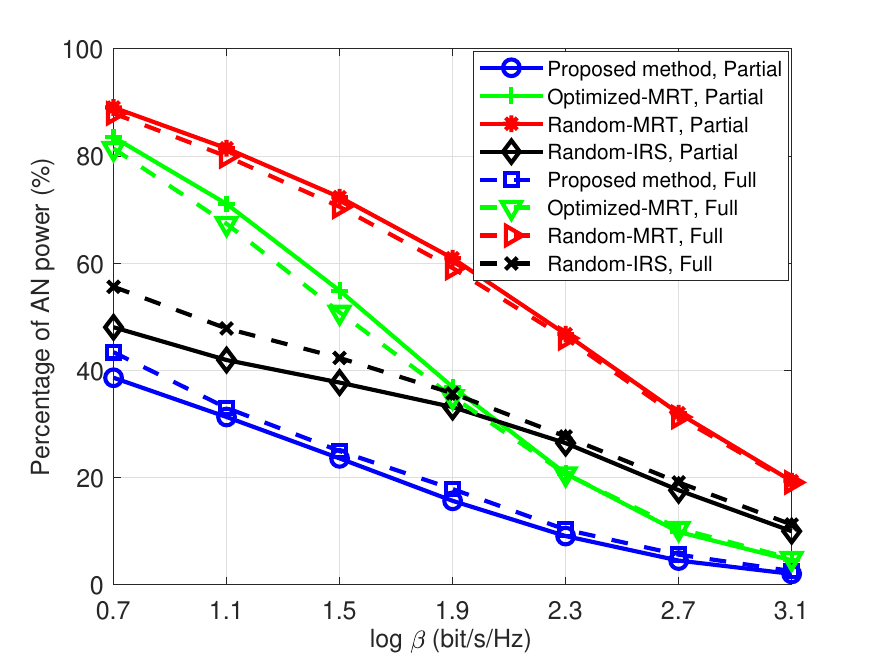}\vspace{-0.6cm}
		\caption{Percentage of invested AN power at each $\log\beta$ with correlated CSI errors, where $N_{t}=2$, $M=15$, $K=2$, and $\log\gamma=3.5$ $\textrm{bit/s/Hz}$.}
		\label{figPercentANvsBeta_CommCSI}
	\end{minipage}\vspace{-0.7cm}
\end{figure}

The transmit power versus the maximum tolerable channel
capacity $\log\beta$ of Eves is shown in Fig. \ref{figPowvsBeta_CommCSI}. It is shown that the transmit power of the proposed algorithm is the lowest among all the schemes, which demonstrates the effectiveness and superiority of the proposed algorithm for the uncorrelated CSI errors. The gap between the proposed algorithm and the Optimized-MRT scheme becomes smaller as $\log\beta$ increases. The transmit power of the full CSI errors is slightly higher than that of the partial CSI errors.

The invested AN power from the total transmit power with different $\log\beta$ is shown in Fig. \ref{figPercentANvsBeta_CommCSI}. The required AN power reduces when the data rate limit of Eves is relaxed. The AN power with the full CSI errors is slightly higher than that with the partial CSI errors for the proposed algorithm.

\section{Conclusion}
An robust transmission strategy was designed for an IRS-aided secure communication system by considering the statistical CSI errors on both direct links and IRS reflecting links to Eves. We proposed an AO algorithm to solve the formulated OCR-PM problem by leveraging the BTI, SDR technique, and CCP method. The proposed algorithm can apply with the uncorrelated and correlated CSI errors, and its superiority over baseline schemes is verified by simulations.

\begin{appendices}
\vspace{-0.5cm}
\section{Proof of Lemma \ref{Lemma_KronckrNorm}}\label{Appendix_C}
We define a matrix $\mathbf{C}=\mathbf{a}\mathbf{b}^{T}$, where $\mathbf{a}$
and $\mathbf{b}$ are vectors. Since we have $\textrm{vec}(\mathbf{C})=\mathbf{b}\otimes\mathbf{a}$,
then we have
\begin{align}
\left\Vert \mathbf{C}\right\Vert _{\textrm{F}} & =\left\Vert \mathbf{b}\otimes\mathbf{a}\right\Vert _{2}=\sqrt{\textrm{Tr}(\mathbf{C}\mathbf{C}^{H})}=\sqrt{\textrm{Tr}(\mathbf{a}\mathbf{b}^{T}\mathbf{b}^{*}\mathbf{a}^{H})} =\sqrt{\mathbf{b}^{T}\mathbf{b}^{*}\mathbf{a}^{H}\mathbf{a}}=\left\Vert \mathbf{b}\right\Vert _{2}\left\Vert \mathbf{a}\right\Vert _{2}.
\end{align}

\section{Proof of \eqref{FindfaiEIGconvx}}\label{Appendix_A}
By utilizing the rank properties $\textrm{rank}\{\mathbf{AB}\}\leq\min\{\textrm{rank}(\mathbf{A}),\textrm{rank}(\mathbf{B})\}$
and $\textrm{rank}(\mathbf{A}\otimes\mathbf{B})=\textrm{rank}(\mathbf{A})\cdot\textrm{rank}(\mathbf{B})$,
we can prove $\lambda_{\textrm{min}}\{\mathbf{\Sigma}_{e,k}^{1/2}((c_{g}\mathbf{I}_{N_{t}})^{T}\otimes\mathbf{E})\mathbf{\Sigma}_{e,k}^{1/2}\}=0$
as follows.

We have the inequality of $\textrm{rank}\{\mathbf{\Sigma}_{ge,k}^{1/2}((c_{g}\mathbf{I}_{N_{t}})^{T}\otimes\mathbf{E})\mathbf{\Sigma}_{ge,k}^{1/2}\}\leq\min\{\textrm{rank}(\mathbf{\Sigma}_{ge,k}^{1/2}),\textrm{rank}((c_{g}\mathbf{I}_{N_{t}})^{T}\otimes\mathbf{E})\}$.
Since $\textrm{rank}((c_{g}\mathbf{I}_{N_{t}})^{T}\otimes\mathbf{E})=\textrm{rank}(c_{g}\mathbf{I}_{N_{t}})\cdot\textrm{rank}(\mathbf{E})=N_{t}$,
we have $\textrm{rank}\{\mathbf{\Sigma}_{ge,k}^{1/2}((c_{g}\mathbf{I}_{N_{t}})^{T}\otimes\mathbf{E})\mathbf{\Sigma}_{ge,k}^{1/2}\}\leq N_{t}$.
Since the matrix $\mathbf{\Sigma}_{ge,k}^{1/2}((c_{g}\mathbf{I})^{T}\otimes\mathbf{E})\mathbf{\Sigma}_{ge,k}^{1/2}$
is an $MN_{t}\times MN_{t}$ dimensional semidefinite matrix, its minimum
eigenvalue is zero as
\begin{alignat}{1}
\lambda_{\textrm{min}}\{\mathbf{\Sigma}_{ge,k}^{1/2}((c_{g}\mathbf{I})^{T}\otimes\mathbf{E})\mathbf{\Sigma}_{ge,k}^{1/2}\}=0.\label{eq:lamdaminzero}
\end{alignat}

\section{Proof of Lemma \ref{Lemma_VectorNorm}}\label{Appendix_B}
For any three matrices $\mathbf{O}$, $\mathbf{P}$, $\mathbf{Q}$,
we have
\begin{align}
\left\Vert \mathbf{P}\mathbf{O}\mathbf{Q}\right\Vert ^{2} & \overset{(m_{1})}{=}\textrm{Tr}[\mathbf{P}\mathbf{O}\mathbf{Q}\mathbf{Q}^{H}\mathbf{O}^{H}\mathbf{P}^{H}]=\textrm{Tr}[\mathbf{O}^{H}\mathbf{P}^{H}\mathbf{P}\mathbf{O}\mathbf{Q}\mathbf{Q}^{H}]\nonumber \\
 & \overset{(m_{2})}{=}\textrm{vec}^{H}(\mathbf{O})[(\mathbf{Q}\mathbf{Q}^{H})^{T}\otimes(\mathbf{P}^{H}\mathbf{P})]\textrm{vec}(\mathbf{O}) =\textrm{vec}^{H}(\mathbf{O})[(\mathbf{Q}^{T}\otimes\mathbf{P})^{H}(\mathbf{Q}^{T}\otimes\mathbf{P})]\textrm{vec}(\mathbf{O})\nonumber \\
 & =\left\Vert (\mathbf{Q}^{T}\otimes\mathbf{P})\textrm{vec}(\mathbf{O})\right\Vert ^{2},
\end{align}
where $(m_{1})$ is due to $\left\Vert \mathbf{A}\right\Vert _{F}^{2}=\textrm{Tr}[\mathbf{A}\mathbf{A}^{H}]$,
and $(m_{2})$ is obtained by invoking the identity $\textrm{Tr}(\mathbf{A}^{H}\mathbf{BCD})=\textrm{vec}^{H}(\mathbf{A})(\mathbf{D}^{T}\otimes\mathbf{B})\textrm{vec}(\mathbf{C})$.

\section{Proof of that $\mathbf{B}_{e,k}$ is rank deficient}\label{Appendix_D}
According to the definition of $\mathbf{B}_{e,k}$, it can be equivalently
expressed as $\mathbf{B}_{e,k}=\mathbf{C}_{e,k}^{H}\mathbf{C}_{e,k}$,
where $\mathbf{C}_{e,k}$ is defined by
\begin{align}
\mathbf{C}_{e,k} & =\left[\begin{array}{cc}
c_{g}^{1/2}\mathbf{I}_{N_{t}}\mathbf{\Sigma}_{he,k}^{1/2} & \;(c_{g}^{1/2}\mathbf{I}_{N_{t}}\otimes\bm{\phi}^{H})\mathbf{\Sigma}_{ge,k}^{1/2*}\end{array}\right]
\end{align}

Then we have $\lambda_{\textrm{min}}\{\mathbf{B}_{e,k}\}=\lambda_{\textrm{min}}\{\mathbf{C}_{e,k}^{H}\mathbf{C}_{e,k}\}$.
We can find
\begin{alignat}{1}
\textrm{rank}\left\{ \mathbf{B}_{e,k}\right\} &\leq  \textrm{rank}\left\{ \mathbf{C}_{e,k}\right\} \leq  \textrm{rank}\left\{ \left[\begin{array}{cc}
c_{g}^{1/2}\mathbf{I}_{N_{t}}\mathbf{\Sigma}_{he,k}^{1/2} & \;(c_{g}^{1/2}\mathbf{I}_{N_{t}}\otimes\bm{\phi}^{H})\mathbf{\Sigma}_{ge,k}^{1/2*}\end{array}\right]\right\} \nonumber \\
&\leq  \textrm{rank}\{c_{g}^{1/2}\mathbf{I}_{N_{t}}\mathbf{\Sigma}_{he,k}^{1/2}\}+\textrm{rank}\{(c_{g}^{1/2}\mathbf{I}_{N_{t}}\otimes\bm{\phi}^{H})\mathbf{\Sigma}_{ge,k}^{1/2*}\}\leq  2N_{t},
\end{alignat}
where the properties $\textrm{rank}(\mathbf{A}\mathbf{B})\leq\min\{\textrm{rank}(\mathbf{A}),\textrm{rank}(\mathbf{B})\}$
and $\textrm{rank}([\mathbf{A},\mathbf{B}])\leq\textrm{rank}(\mathbf{A})+\textrm{rank}(\mathbf{B})$
are utilized. Since the matrix $\mathbf{B}_{e,k}$ is a $(MN_{t}+N_{t})\times(MN_{t}+N_{t})$
dimensional semidefinite matrix, its minimum eigenvalue is zero as
$\lambda_{\textrm{min}}\{\mathbf{B}_{e,k}\}=0$ for any $M\geq2$.
\end{appendices}

\bibliographystyle{IEEEtran}
\bibliography{OC-PM-WD-Ref}

\begin{thebibliography}{10}
\providecommand{\url}[1]{#1}
\csname url@samestyle\endcsname
\providecommand{\newblock}{\relax}
\providecommand{\bibinfo}[2]{#2}
\providecommand{\BIBentrySTDinterwordspacing}{\spaceskip=0pt\relax}
\providecommand{\BIBentryALTinterwordstretchfactor}{4}
\providecommand{\BIBentryALTinterwordspacing}{\spaceskip=\fontdimen2\font plus
\BIBentryALTinterwordstretchfactor\fontdimen3\font minus
  \fontdimen4\font\relax}
\providecommand{\BIBforeignlanguage}[2]{{%
\expandafter\ifx\csname l@#1\endcsname\relax
\typeout{** WARNING: IEEEtran.bst: No hyphenation pattern has been}%
\typeout{** loaded for the language `#1'. Using the pattern for}%
\typeout{** the default language instead.}%
\else
\language=\csname l@#1\endcsname
\fi
#2}}
\providecommand{\BIBdecl}{\relax}
\BIBdecl

\bibitem{wu2019towards}
Q.~Wu and R.~Zhang, ``Towards smart and reconfigurable environment: Intelligent
  reflecting surface aided wireless network,'' \emph{IEEE Communications
  Magazine}, vol.~58, no.~1, pp. 106--112, 2019.

\bibitem{pan2020reconfigurable}
C.~Pan, H.~Ren, K.~Wang, J.~F. Kolb, M.~Elkashlan, M.~Chen, M.~Di~Renzo,
  Y.~Hao, J.~Wang, A.~L. Swindlehurst, X.~You, and L.~Hanzo, ``Reconfigurable
  intelligent surfaces for {6G} systems: Principles, applications, and research
  directions,'' \emph{IEEE Communications Magazine}, vol.~59, no.~6, pp.
  14--20, 2021.

\bibitem{cui2014coding}
T.~J. Cui, M.~Q. Qi, X.~Wan, J.~Zhao, and Q.~Cheng, ``Coding metamaterials,
  digital metamaterials and programmable metamaterials,'' \emph{Light: Science
  \& Applications}, vol.~3, no.~10, pp. 218--218, 2014.

\bibitem{di2020reconfigurable}
M.~Di~Renzo, K.~Ntontin, J.~Song, F.~H. Danufane, X.~Qian, F.~Lazarakis,
  J.~De~Rosny, D.-T. Phan-Huy, O.~Simeone, R.~Zhang \emph{et~al.},
  ``Reconfigurable intelligent surfaces vs. relaying: Differences,
  similarities, and performance comparison,'' \emph{IEEE Open Journal of the
  Communications Society}, vol.~1, pp. 798--807, 2020.

\bibitem{pan2019intelligent2}
C.~{Pan}, H.~{Ren}, K.~{Wang}, W.~{Xu}, M.~{Elkashlan}, A.~{Nallanathan}, and
  L.~{Hanzo}, ``Multicell {MIMO} communications relying on intelligent
  reflecting surfaces,'' \emph{IEEE Transactions on Wireless Communications},
  vol.~19, no.~8, pp. 5218--5233, 2020.

\bibitem{9133107}
T.~{Bai}, C.~{Pan}, Y.~{Deng}, M.~{Elkashlan}, A.~{Nallanathan}, and
  L.~{Hanzo}, ``Latency minimization for intelligent reflecting surface aided
  mobile edge computing,'' \emph{IEEE Journal on Selected Areas in
  Communications}, vol.~38, no.~11, pp. 2666--2682, 2020.

\bibitem{zhou2020intelligent}
G.~{Zhou}, C.~{Pan}, H.~{Ren}, K.~{Wang}, and A.~{Nallanathan}, ``Intelligent
  reflecting surface aided multigroup multicast {MISO} communication systems,''
  \emph{IEEE Transactions on Signal Processing}, vol.~68, pp. 3236--3251, 2020.

\bibitem{zhang2020intelligent}
L.~Zhang, Y.~Wang, W.~Tao, Z.~Jia, T.~Song, and C.~Pan, ``Intelligent
  reflecting surface aided {MIMO} cognitive radio systems,'' \emph{IEEE
  Transactions on Vehicular Technology}, vol.~69, no.~10, pp. 11\,445--11\,457,
  2020.

\bibitem{pan2020intelligent}
C.~{Pan}, H.~{Ren}, K.~{Wang}, M.~{Elkashlan}, A.~{Nallanathan}, J.~{Wang}, and
  L.~{Hanzo}, ``Intelligent reflecting surface aided {MIMO} broadcasting for
  simultaneous wireless information and power transfer,'' \emph{IEEE Journal on
  Selected Areas in Communications}, vol.~38, no.~8, pp. 1719--1734, 2020.

\bibitem{9014322}
X.~{Yu}, D.~{Xu}, and R.~{Schober}, ``Enabling secure wireless communications
  via intelligent reflecting surfaces,'' in \emph{2019 IEEE Global
  Communications Conference (GLOBECOM)}, 2019, pp. 1--6.

\bibitem{chen2019intelligent}
J.~Chen, Y.-C. Liang, Y.~Pei, and H.~Guo, ``Intelligent reflecting surface: A
  programmable wireless environment for physical layer security,'' \emph{IEEE
  Access}, vol.~7, pp. 82\,599--82\,612, 2019.

\bibitem{8972406}
L.~{Dong} and H.~{Wang}, ``Secure {MIMO} transmission via intelligent
  reflecting surface,'' \emph{IEEE Wireless Communications Letters}, vol.~9,
  no.~6, pp. 787--790, 2020.

\bibitem{cui2019secure}
M.~Cui, G.~Zhang, and R.~Zhang, ``Secure wireless communication via intelligent
  reflecting surface,'' \emph{IEEE Wireless Communications Letters}, vol.~8,
  no.~5, pp. 1410--1414, 2019.

\bibitem{guan2020intelligent}
X.~Guan, Q.~Wu, and R.~Zhang, ``Intelligent reflecting surface assisted secrecy
  communication: Is artificial noise helpful or not?'' \emph{IEEE Wireless
  Communications Letters}, vol.~9, no.~6, pp. 778--782, 2020.

\bibitem{9201173}
S.~{Hong}, C.~{Pan}, H.~{Ren}, K.~{Wang}, and A.~{Nallanathan},
  ``Artificial-noise-aided secure {MIMO} wireless communications via
  intelligent reflecting surface,'' \emph{IEEE Transactions on Communications},
  vol.~68, no.~12, pp. 7851--7866, 2020.

\bibitem{chu2019intelligent}
Z.~Chu, W.~Hao, P.~Xiao, and J.~Shi, ``Intelligent reflecting surface aided
  multi-antenna secure transmission,'' \emph{IEEE Wireless Communications
  Letters}, vol.~9, no.~1, pp. 108--112, 2019.

\bibitem{xu2019resource}
D.~Xu, X.~Yu, Y.~Sun, D.~W.~K. Ng, and R.~Schober, ``Resource allocation for
  secure {IRS}-assisted multiuser {MISO} systems,'' in \emph{2019 IEEE Globecom
  Workshops (GC Wkshps)}.\hskip 1em plus 0.5em minus 0.4em\relax IEEE, 2019,
  pp. 1--6.

\bibitem{wei2021channel}
X.~Wei, D.~Shen, and L.~Dai, ``Channel estimation for {RIS} assisted wireless
  communications—{P}art {I}: Fundamentals, solutions, and future
  opportunities,'' \emph{IEEE Communications Letters}, vol.~25, no.~5, pp.
  1398--1402, 2021.

\bibitem{taha2019enabling}
A.~Taha, M.~Alrabeiah, and A.~Alkhateeb, ``Enabling large intelligent surfaces
  with compressive sensing and deep learning,'' \emph{IEEE Access}, vol.~9, pp.
  44\,304--44\,321, 2021.

\bibitem{9241029}
Z.~{Zhou}, N.~{Ge}, Z.~{Wang}, and L.~{Hanzo}, ``Joint transmit precoding and
  reconfigurable intelligent surface phase adjustment: A decomposition-aided
  channel estimation approach,'' \emph{IEEE Transactions on Communications},
  vol.~69, no.~2, pp. 1228--1243, 2021.

\bibitem{wang2020channel}
Z.~{Wang}, L.~{Liu}, and S.~{Cui}, ``Channel estimation for intelligent
  reflecting surface assisted multiuser communications: Framework, algorithms,
  and analysis,'' \emph{IEEE Transactions on Wireless Communications}, vol.~19,
  no.~10, pp. 6607--6620, 2020.

\bibitem{wang2019compressed}
P.~{Wang}, J.~{Fang}, H.~{Duan}, and H.~{Li}, ``Compressed channel estimation
  for intelligent reflecting surface-assisted millimeter wave systems,''
  \emph{IEEE Signal Processing Letters}, vol.~27, pp. 905--909, 2020.

\bibitem{chen2019channel}
J.~Chen, Y.-C. Liang, H.~V. Cheng, and W.~Yu, ``Channel estimation for
  reconfigurable intelligent surface aided multi-user mm{W}ave {MIMO}
  systems,'' \emph{IEEE Transactions on Wireless Communications}, vol. early
  access, pp. 1--17, 2023.

\bibitem{zhang2020capacity}
S.~Zhang and R.~Zhang, ``Capacity characterization for intelligent reflecting
  surface aided {MIMO} communication,'' \emph{IEEE Journal on Selected Areas in
  Communications}, vol.~38, no.~8, pp. 1823--1838, 2020.

\bibitem{zhou2020robust}
G.~{Zhou}, C.~{Pan}, H.~{Ren}, K.~{Wang}, M.~D. {Renzo}, and A.~{Nallanathan},
  ``Robust beamforming design for intelligent reflecting surface aided {MISO}
  communication systems,'' \emph{IEEE Wireless Communications Letters}, vol.~9,
  no.~10, pp. 1658--1662, 2020.

\bibitem{yu2020robust}
X.~{Yu}, D.~{Xu}, Y.~{Sun}, D.~W.~K. {Ng}, and R.~{Schober}, ``Robust and
  secure wireless communications via intelligent reflecting surfaces,''
  \emph{IEEE Journal on Selected Areas in Communications}, vol.~38, no.~11, pp.
  2637--2652, 2020.

\bibitem{zhou2020framework}
G.~{Zhou}, C.~{Pan}, H.~{Ren}, K.~{Wang}, and A.~{Nallanathan}, ``A framework
  of robust transmission design for {IRS}-aided {MISO} communications with
  imperfect cascaded channels,'' \emph{IEEE Transactions on Signal Processing},
  vol.~68, pp. 5092--5106, 2020.

\bibitem{zhang2020robust}
L.~Zhang, C.~Pan, Y.~Wang, H.~Ren, and K.~Wang, ``Robust beamforming design for
  intelligent reflecting surface aided cognitive radio systems with imperfect
  cascaded {CSI},'' \emph{IEEE Transactions on Cognitive Communications and
  Networking}, vol.~8, no.~1, pp. 186--201, 2022.

\bibitem{yu2020irs}
X.~Yu, D.~Xu, D.~W.~K. Ng, and R.~Schober, ``{IRS}-assisted green communication
  systems: Provable convergence and robust optimization,'' \emph{IEEE
  Transactions on Communications}, vol.~69, no.~9, pp. 6313--6329, 2021.

\bibitem{hu_robust_2021}
S.~Hu, Z.~Wei, Y.~Cai, C.~Liu, D.~W.~K. Ng, and J.~Yuan, ``Robust and secure
  sum-rate maximization for multiuser {MISO} downlink systems with
  self-sustainable {IRS},'' \emph{IEEE Transactions on Communications},
  vol.~69, no.~10, pp. 7032--7049, Oct. 2021.

\bibitem{9167248}
T.~A. {Le}, T.~{Van Chien}, and M.~{Di Renzo}, ``Robust
  probabilistic-constrained optimization for {IRS}-aided {MISO} communication
  systems,'' \emph{IEEE Wireless Communications Letters}, vol.~10, no.~1, pp.
  1--5, 2021.

\bibitem{zhao2020outage}
M.-M. Zhao, A.~Liu, and R.~Zhang, ``Outage-constrained robust beamforming for
  intelligent reflecting surface aided wireless communication,'' \emph{IEEE
  Transactions on Signal Processing}, vol.~69, pp. 1301--1316, 2021.

\bibitem{hong2020robust}
S.~Hong, C.~Pan, H.~Ren, K.~Wang, K.~K. Chai, and A.~Nallanathan, ``Robust
  transmission design for intelligent reflecting surface-aided secure
  communication systems with imperfect cascaded {CSI},'' \emph{IEEE
  Transactions on Wireless Communications}, vol.~20, no.~4, pp. 2487--2501,
  2021.

\bibitem{wang2014outage}
K.-Y. Wang, A.~M.-C. So, T.-H. Chang, W.-K. Ma, and C.-Y. Chi, ``Outage
  constrained robust transmit optimization for multiuser {MISO} downlinks:
  Tractable approximations by conic optimization,'' \emph{IEEE Transactions on
  Signal Processing}, vol.~62, no.~21, pp. 5690--5705, 2014.

\bibitem{liao2010qos}
W.-C. Liao, T.-H. Chang, W.-K. Ma, and C.-Y. Chi, ``{Q}o{S}-based transmit
  beamforming in the presence of eavesdroppers: An optimized
  artificial-noise-aided approach,'' \emph{IEEE Transactions on Signal
  Processing}, vol.~59, no.~3, pp. 1202--1216, 2010.

\bibitem{chai_secure_2023}
L.~Chai, L.~Bai, T.~Bai, J.~Shi, and A.~Nallanathan, ``Secure {RIS}-aided
  {MISO-NOMA} system design in the presence of active eavesdropping,''
  \emph{IEEE Internet of Things Journal}, vol. early access, pp. 1--16, 2023.

\bibitem{2017technical}
3GPP, ``Technical specification group radio access network; study on {3D}
  channel model for {LTE} (release 12),'' TR 36.873 V12.7.0, Tech. Rep., Dec.
  2017.

\bibitem{feng2021physical}
K.~Feng, X.~Li, Y.~Han, S.~Jin, and Y.~Chen, ``Physical layer security
  enhancement exploiting intelligent reflecting surface,'' \emph{IEEE
  Communications Letters}, vol.~25, no.~3, pp. 734--738, 2021.

\bibitem{Marco2003on}
M.~Chiani, M.~Win, and A.~Zanella, ``On the capacity of spatially correlated
  {MIMO} rayleigh-fading channels,'' \emph{IEEE Transactions on Information
  Theory}, vol.~49, no.~10, pp. 2363--2371, 2003.

\end{thebibliography}

\end{document}